\documentclass[%
        a4paper,
        twocolumn,
        10pt,
        accepted=2024-11-11
    ]{quantumarticle}
\pdfoutput=1

\usepackage[utf8]{inputenc}

\usepackage[
    numbers,
    square,
    comma,
    sort&compress
    ]{natbib}

\usepackage{graphicx}

\usepackage{hyperref}
\hypersetup{
	colorlinks=true,
	linkcolor=blue,
	citecolor=blue,
	urlcolor=blue,
}

\usepackage{physics}
\usepackage{amsmath}
\usepackage{amssymb}
\usepackage{bm} 
\usepackage{bbm} 
\usepackage{mathtools}

\usepackage{xspace}
\usepackage{orcidlink}
\usepackage[normalem]{ulem} 

\newcommand{\Z}{\mathbb{Z}}
\newcommand{\R}{\mathbb{R}}
\newcommand{\C}{\mathbb{C}}
\newcommand{\bigO}[1]{\mathcal{O}\!\left(#1\right)}

\newcommand{\poly}[1]{\text{poly}\!\left(#1\right)}

\renewcommand{\d}{\!\!\text{d}}
\renewcommand{\vec}[1]{\bm{#1}}
\newcommand{\Id}{\mathbbm{1}}
\newcommand{\transpose}{T}
\renewcommand{\ket}[1]{|#1\rangle}
\renewcommand{\bra}[1]{\langle#1|}
\newcommand{\cnot}{\textsc{cnot}\xspace}
\newcommand{\verteq}{\rotatebox{90}{$\,=$}}
\definecolor{nice-red}{HTML}{D62728}

\makeatletter 
    
\renewcommand\onecolumngrid{
    \do@columngrid{one}{\@ne}%
    \def\set@footnotewidth{\onecolumngrid}
}

\renewcommand\twocolumngrid{
    \do@columngrid{mlt}{\tw@}
}%

\makeatother

\begin{document}
\title{Efficient MPS representations and quantum circuits from the Fourier modes of classical image data}

\author{Bernhard Jobst\,\orcidlink{0000-0001-7027-3918}}
\email{bernhard.jobst@tum.de}%
\affiliation{Technical University of Munich, TUM School of Natural Sciences, Physics Department, 85748 Garching, Germany}%
\affiliation{Munich Center for Quantum Science and Technology (MCQST), Schellingstr. 4, 80799 München, Germany}%
\affiliation{BMW Group, 80788 München, Germany}
\orcid{0000-0001-7027-3918}

\author{Kevin Shen\,\orcidlink{0009-0005-2506-4056}}
\affiliation{Technical University of Munich, TUM School of Natural Sciences, Physics Department, 85748 Garching, Germany}%
\affiliation{Munich Center for Quantum Science and Technology (MCQST), Schellingstr. 4, 80799 München, Germany}%
\affiliation{BMW Group, 80788 München, Germany}%
\affiliation{LIACS, Leiden University, Niels Bohrweg 1, 2333 CA, Leiden, The Netherlands}%
\affiliation{$\langle aQa^L \rangle$ Applied Quantum Algorithms, Leiden University, The Netherlands}
\orcid{0009-0005-2506-4056}

\author{Carlos A. Riofr\'{i}o\,\orcidlink{0000-0002-7346-9198}}
\affiliation{BMW Group, 80788 München, Germany}
\orcid{0000-0002-7346-9198}

\author{Elvira Shishenina}
\thanks{Now at Quantinuum, Leopoldstrasse 180, 80804 München, Germany.}
\affiliation{BMW Group, 80788 München, Germany}

\author{Frank Pollmann\,\orcidlink{0000-0003-0320-9304}\,}
\affiliation{Technical University of Munich, TUM School of Natural Sciences, Physics Department, 85748 Garching, Germany}%
\affiliation{Munich Center for Quantum Science and Technology (MCQST), Schellingstr. 4, 80799 München, Germany}
\orcid{0000-0003-0320-9304}

\begin{abstract}
    Machine learning tasks are an exciting application for quantum computers, as it has been proven that they can learn certain problems more efficiently than classical ones. Applying quantum machine learning algorithms to classical data can have many important applications, as qubits allow for dealing with exponentially more data than classical bits. However, preparing the corresponding quantum states usually requires an exponential number of gates and therefore may ruin any potential quantum speedups. Here, we show that classical data with a sufficiently quickly decaying Fourier spectrum after being mapped to a quantum state can be well-approximated by states with a small Schmidt rank (i.e., matrix-product states) and we derive explicit error bounds. These approximated states can, in turn, be prepared on a quantum computer with a linear number of nearest-neighbor two-qubit gates. We confirm our results numerically on a set of $\bm{1024\times1024}$-pixel images taken from the `Imagenette' and DIV2K datasets. Additionally, we consider different variational circuit ansätze and demonstrate numerically that one-dimensional sequential circuits achieve the same compression quality as more powerful ansätze.
\end{abstract}

\maketitle

\section{Introduction}
Quantum computers provide a powerful new platform in the context of machine learning~\cite{Biamonte2017, Schuld2018}, which---in certain instances---can yield a rigorously proven exponential speedup compared to classical machine learning methods~\cite{Liu2021, Sweke2021, Huang2021, Pirnay2023, Gyurik2023}. Quantum machine learning (QML) algorithms can be applied to quantum data~\cite{Wiebe2012, Biamonte2017, Cong2019, Liu2023, Lake2023}, where the input data are already given as a quantum state, or to classical data~\cite{Rebentrost2014, Biamonte2017, Schuld2018, Havlek2019, Schuld2021}, where the input data first need to be mapped to a quantum state, which is then prepared as an initial state on the quantum device. This preparation of the initial state is generally exponentially costly~\cite{Plesch2011, Iten2016, Amankwah2022}, and thus constitutes a significant bottleneck: even if the subsequent QML algorithm is efficient, the exponential speedup is already lost at the stage of loading the input data~\cite{Aaronson2015, Biamonte2017, Liu2021}.

Typical strategies for mapping the classical data to a quantum state leverage the superposition of basis states to encode an exponential amount of data in a linear number of qubits~\cite{Rebentrost2014, Schuld2018, Amankwah2022}. Thus, intuitively, it is not surprising that the number of gates required to prepare the state scales exponentially in the number of qubits: to fix the exponentially many amplitudes in the quantum state generically exponentially many operations are needed~\cite{Plesch2011}. The scaling can be reduced if it is sufficient to only approximately prepare the state up to a small error. For this, there are several proposals in the literature.
Ref.~\cite{Moosa2023} proposes to approximate the state by a truncated Fourier series and prepare it by loading the lowest-frequency components into a subset of qubits, before extending it to a sparse state on the whole qubit register and running an inverse quantum Fourier transform. If the number of included Fourier modes is fixed, this leads to a circuit with a depth that scales linearly with the number of qubits and a number of \cnot gates that scales quadratically in the number of qubits. However, some long-range two-qubit gates are needed and a priori it is not clear how many Fourier coefficients need to be retained.
Another strategy proposed in Refs.~\cite{Garcia-Ripoll2021, Holmes2020, Dilip2022, Iaconis2023} is to first approximate the data classically by a matrix-product state (MPS), which is argued to be possible due to the observation that states encoding a probability distribution with a bounded derivative have a small entanglement entropy. These MPS can then be prepared with linear-depth quantum circuits on a quantum computer~\cite{Schön2005, Schön2007, Lubasch2020, Smith2022, Lin2021, Barratt2021}. Alternatively, as shown in Ref.~\cite{Dilip2022}, one can directly variationally optimize hardware-efficient linear-depth circuits to approximate the target state, which yields circuits that can be faithfully run on current quantum hardware~\cite{Iaconis2023, Shen2024}. In both cases, however, it is not clear exactly which bond dimension or circuit depth is needed to accurately capture the target state, and whether or how this has to scale with increasing system size.
Other approaches, as proposed in Refs.~\cite{Park2019, Wang2009, Matteo2020, Ashhab2022}, assume a certain structure of the target state to be efficient, i.e., either the number of basis states in the superposition being small~\cite{Park2019}, all states included in the superposition having a given number of zeros and ones~\cite{Wang2009, Matteo2020} or all weights in the superposition being of a roughly equal magnitude~\cite{Ashhab2022}. Additionally, a few of the approaches above use mid-circuit measurements which can be challenging to implement on some hardware realizations, with error rates possibly being an order of magnitude worse than those of \cnot gates~\cite{Gaebler2021, Rudinger2022, Graham2023}.

In this work, we focus the discussion on classical image data with sufficiently quickly decaying Fourier coefficients, i.e., those where the coefficients of the Fourier transform as a function of the frequencies $k_x$ and $k_y$ decay faster than $\bigO{|k_x|^{-1}|k_y|^{-1}}$, which we will numerically demonstrate includes a wide range of naturally occurring images (see also Refs.~\cite{Burton1987, Field1987, Tolhurst1992, Field1993, Ruderman1994, vanderSchaaf1996, Ruderman1997}). However, the results are also applicable to any two-dimensional classical data with Fourier modes decaying in the same way~\cite{Lubasch2018, Lubasch2020, Gourianov2022, Nunez-Fernandez2022, Ritter2024}. We show that, using different mappings from the classical data to a quantum state, the resulting states can be efficiently described by MPS, in the sense that the bond dimension of the MPS does not need to grow with increasing image resolution to remain below a certain error threshold. This generalizes the findings of Refs.~\cite{Garcia-Ripoll2021, Holmes2020} that discretized real-valued functions with a bounded derivative are well-approximated by MPS as the resolution of the discretization grid increases, allowing now also for complex-valued functions and diverging derivatives as long as the Fourier coefficients still decay sufficiently fast. We verify our results numerically on $1024\times1024$-pixel images taken from the `Imagenette' dataset~\cite{imagenette} (a subset of the well-known ImageNet dataset~\cite{ImageNet}) and the DIV2K dataset~\cite{DIV2K}, which contain a variety of images depicting humans, animals and objects, and thus should be reasonably representative of the general features of generic images. The MPS approximation of the quantum state then naturally leads to a linear-depth quantum circuit that only requires a linear number of nearest-neighbor \cnot gates and does not require any mid-circuit measurements~\cite{Schön2005, Schön2007, Lubasch2020, Smith2022, Lin2021, Barratt2021}. Additionally, we try out several different circuit ansätze for variationally compressing images in the Fashion-MNIST dataset~\cite{FashionMNIST} (similarly to Ref.~\cite{Dilip2022}), and find that sequential circuits inspired by MPS work just as well as more powerful ansätze with the same number of parameters.

This work is structured as follows: in Sec.~\ref{sec:image_encodings} we introduce the different mappings from classical data to quantum states that we consider. In Sec.~\ref{sec:image_compression_MPS}, we first present numerical data showing that states encoding classical image data have a small entanglement entropy and are well-approximated by MPS compared to states with randomly sampled amplitudes. Then, we derive explicit bounds on the error of the MPS approximation, depending on the decay of the Fourier modes. Notably, for a fixed bond dimension these bounds do not scale with the image resolution. In Sec.~\ref{sec:image_compression_circuits} we compare the numerical results for variationally optimizing different circuit ansätze to approximate states encoding classical image data. Finally, we conclude in Sec.~\ref{sec:conclusion} with a discussion of our results.

\section{\label{sec:image_encodings}Image encodings}
\begin{figure*}[t]
    \begin{tabular}{lclclc}
        \hspace{-0.5em}{\sffamily(a)}\hspace{-0.8em} && {\sffamily(b)}\hspace{-0.8em} && {\sffamily(c)}\hspace{-0.8em} & \\[-1.1em]
        &  \includegraphics[width=0.28\linewidth]{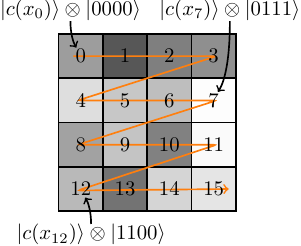}%
        && \includegraphics[width=0.28\linewidth]{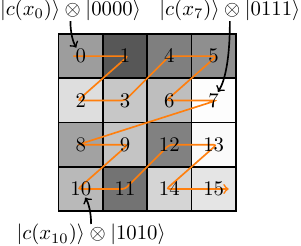}%
        && \includegraphics[width=0.28\linewidth]{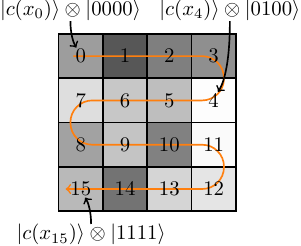}%
    \end{tabular}
    \caption{\label{fig:pixel_indexing}%
    \textbf{Different indexing variants for labeling the pixels in an image during the mapping to a quantum state.} (a)~The pixels are labeled row by row from top to bottom and from left to right~\cite{Le2011}. This is equivalent to having a separate index for the $y$- and the $x$-coordinate and concatenating their binary representation to obtain a single binary integer labeling all pixels. (b)~The pixels are labeled in a hierarchical fashion, where the first two bits of the binary integer denote the quadrant of the pixel, the next two bits label the subquadrant, and so on. See also Refs.~\cite{Latorre2005, Le2011}. (c)~The pixels are labeled in a snake pattern that traverses every other row in the opposite direction (used in Ref.~\cite{Dilip2022}).}
\end{figure*}

Several different strategies for mapping a classical ${2^n\times2^n}$-pixel image to a quantum state have been proposed~\cite{Latorre2005, Le2011, Le2011_2, Khan2019, Zhang2013, Jiang2014, Sun2011, Sun2013, Sang2016, Su2021, Amankwah2022}. While we will focus the discussion on images, in principle one could also apply the same mappings to other classical data. The quantum states resulting from the different mappings all share a similar structure, given by~\cite{Amankwah2022}
\begin{equation}
    \ket{\psi} = \frac{1}{2^n} \sum_{j=0}^{2^{2n}-1} \ket{c(x_j)} \otimes \ket{j}.
    \label{eq:state_general_structure}
\end{equation}
Each term in the sum is a tensor product of $\ket{c(x_j)}$ and $\ket{j}$. The state $\ket{c(x_j)}$ is a mapping from the data value $x_j$ of the $j$th pixel to a quantum state; the number of qubits needed, and whether color or only grayscale images can be taken into account, depend on the details of the chosen encoding. As the corresponding qubits store the information of the pixel values, they are often referred to as \emph{color qubits}. The state $\ket{j}$ lives on $2n$ so-called \emph{address qubits}, and labels each of the $2^{2n}$ pixels with a binary integer. While the order of this labeling can in principle be chosen arbitrarily, different orders can have an effect on the entanglement entropy of the state. Common choices of the indexing are shown in Fig.~\ref{fig:pixel_indexing}. Fig.~\ref{fig:pixel_indexing}a shows a row-by-row indexing of the pixels, as it was e.g. proposed in Ref.~\cite{Le2011}; Fig.~\ref{fig:pixel_indexing}b shows a hierarchical indexing, where the first two qubits label the quadrant of the image in which a pixel lies, the next two qubits label the subquadrant, and so on~\cite{Latorre2005, Le2011}; and Fig.~\ref{fig:pixel_indexing}c shows a snake indexing, where every other row is traversed in the reverse direction, which was used in Ref.~\cite{Dilip2022}.

In the following, we will present the three most common encodings in some detail. While there exist variants of these that accommodate color images, we will focus on grayscale images.

\textbf{Amplitude encoding.}
The simplest strategy for encoding data is \emph{amplitude encoding}. Instead of using another state $\ket{c(x_j)}$ to encode the information of the pixel values, we just use the state amplitudes to encode the information~\cite{Schuld2018, Latorre2005, Rebentrost2014}. The state is then given by
\begin{equation}
    \ket{\psi} = \frac{1}{\sqrt{\sum_{j=0}^{2^{2n}-1} |x_j|^2}}\sum_{j=0}^{2^{2n}-1}x_j\,\ket{j}.
    \label{eq:state_amplitude_encoding}
\end{equation}
This way, we only need $2n$ qubits to store the information of $2^{2n}$ grayscale pixel values, which is an exponential reduction compared to the classical case. However, some information about the contrast of the image, which is present in the classical data, gets lost during this mapping due to the normalization of the quantum state.
Before being able to construct such a state on a quantum computer for a given set of pixel values $x_j$, we first need to decompose the $2n$-qubit unitary preparing such a state into the native gate set, e.g., \cnot gates and single-qubit gates. While this can be done in principle, the number of gates will generically scale exponentially as $\bigO{2^{2n}}$~\cite{Plesch2011, Iten2016}, and the calculation of the decomposition introduces a classical overhead, which can be computationally costly.

\textbf{Flexible representation of quantum images.}
Some of the drawbacks of the amplitude encoding are addressed by the so-called \emph{flexible representation of quantum images (FRQI)}~\cite{Le2011, Le2011_2}. It uses the general form of the state in Eq.~\eqref{eq:state_general_structure}, and the value $x_j$ of the $j$th pixel, normalized to lie between $0$ and $1$, is encoded in the state $\ket{c(x_j)}$ of a single color qubit given by
\begin{equation}
    \ket{c(x_j)} = \cos(\frac{\pi}{2} x_j)\ket{0} + \sin(\frac{\pi}{2} x_j)\ket{1}.
    \label{eq:state_FRQI}
\end{equation}
This encodes the $2^{2n}$ values of a $2^n\times2^n$-pixel grayscale image into only $2n+1$ qubits. (Color images can be treated similarly using several color qubits, see Refs.~\cite{Sun2011, Sun2013}.) Since the state $\ket{c(x_j)}$ is already properly normalized, there is no loss of information after normalizing the final state, in contrast to the amplitude encoding. Additionally, there is a prescription for constructing these states in terms of \cnot gates and single-qubit rotations on a quantum computer, so there is no need for a classical overhead. The state can be constructed exactly using $\bigO{2^{4n}}$ gates~\cite{Le2011}, or if $2n-2$ auxiliary qubits are present, using $\bigO{n2^{2n}}$ gates~\cite{Khan2019}. If one does allow for classical preprocessing with $\bigO{n2^{2n}}$ operations, the number of gates can be further reduced to $\bigO{2^{2n}}$ without using auxiliary qubits~\cite{Amankwah2022}.
Using the FRQI, some image processing operations such as pattern detection~\cite{Schützhold2003}, edge detection~\cite{Zhang2015} and certain color transformations~\cite{Le2011_3} can be implemented efficiently.

\textbf{Novel enhanced quantum representation of digital images.}
The pixel values of digital images are usually not given as continuous variables, but rather as discrete $8$-bit integers. An encoding based on this is the \emph{novel enhanced quantum representation of digital images (NEQR)}~\cite{Zhang2013}. Given a grayscale $2^n\times2^n$-pixel image, where each pixel value $x_j$ can be written as a $q$-bit binary integer $x_j = x_{0,j} x_{1,j} \ldots x_{q-1,j}$ describing one of $2^q$ different grayscale values, the NEQR maps the image to a quantum state of the form in Eq.~\eqref{eq:state_general_structure}, with the pixel value $x_j$ being mapped to a $q$-qubit quantum state given by
\begin{equation}
    \ket{c(x_j)} = \ket{x_{0,j} x_{1,j} \ldots x_{q-1,j}}.
    \label{eq:state_NEQR}
\end{equation}
This encodes the $2^{2n}$ pixel values of a grayscale image into $2n+q$ qubits. (Color images can also be represented in this framework using more color qubits, see Refs.~\cite{Sang2016, Su2021}.) The state can be prepared exactly on a quantum computer using $\bigO{qn2^{2n}}$ \cnot and single-qubit gates if one utilizes $2n-2$ auxiliary qubits~\cite{Zhang2013}. If one allows for $\bigO{qn2^{2n}}$ operations of classical preprocessing, the state can be prepared using $\bigO{q2^{2n}}$ gates without using any auxiliary qubits~\cite{Amankwah2022}. Compared to the FRQI, more color transformations can be implemented efficiently~\cite{Zhang2013}, and algorithms for feature extraction and image scaling have been proposed~\cite{Zhang2015_2, Jiang2015}.
\medskip

For all three encodings, the asymptotic scaling of the number of gates needed to prepare the state is exponential in the number of qubits. The exponential cost for preparing the state exactly seems to be unavoidable in a general setting, since we need to load exponentially many pixel values into the quantum state.

\section{\label{sec:image_compression_MPS}Image compression with MPS}
For machine learning tasks, it is usually not necessary to prepare the exact quantum state representing the image, a good approximation of the state is sufficient. The idea is that as long as we are close enough to the target state, we can view the input state as the exact training data with some noise added, and the classifier should still be able to recognize relevant features and assign the correct label. In fact, in classical machine learning it was observed that adding random noise to the training data can act as a kind of regularization and actually improve the generalization of the classifier~\cite{Sietsma1991, Bishop1995, An1996}. Similarly, for QML algorithms applied to quantum data, Ref.~\cite{Liu2023} found that adding perturbations to quantum states during training helps the classifier to learn to distinguish different quantum phases of matter, while for QML algorithms applied to classical data, Ref.~\cite{Dilip2022} found numerically that the classifier is still able to learn from the quantum states in the training set if they are only prepared approximately.

If we relax the constraint of preparing the target state exactly and allow for a small approximation error, we can overcome the exponential scaling of the number of gates needed to prepare the state. For small images, Ref.~\cite{Dilip2022} found that matrix-product states (MPS) with a small bond dimension and shallow quantum circuits, i.e., circuits with a number of gates scaling linearly in the number of qubits, suffice to approximate the corresponding quantum states well enough, so that a classifier trained on these compressed states is still able to learn the classification task. The question remains, however, whether this observation also scales to much larger images.

\begin{figure*}[t]
    \begin{tabular}{lclc}
        \hspace{-0.5em}{\sffamily(a)}\hspace{-1em} && {\sffamily(b)}\hspace{-1em} & \\[-1.5em]
        &  \includegraphics[width=0.45\linewidth]{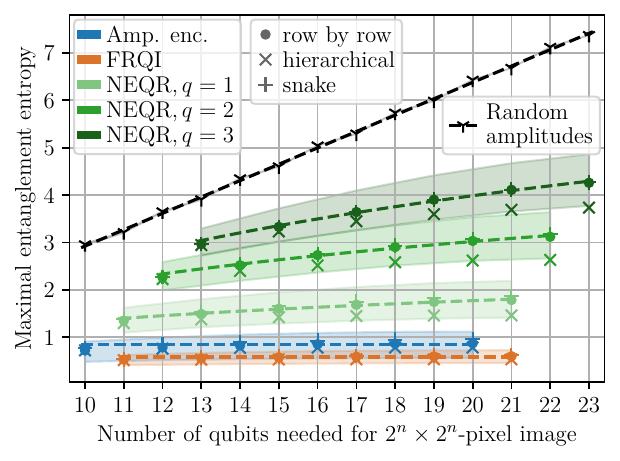}%
        && \includegraphics[width=0.45\linewidth]{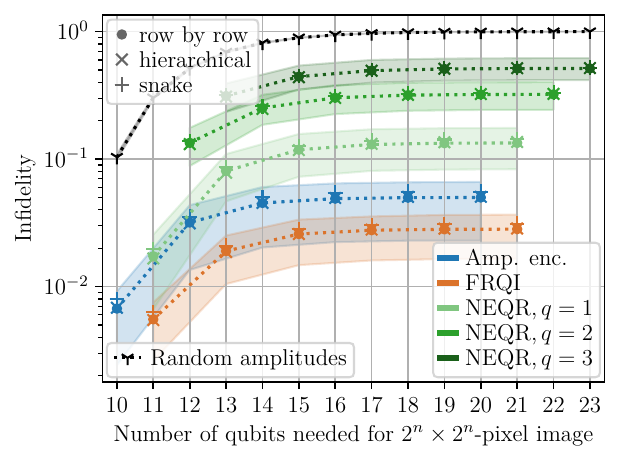}%
    \end{tabular}
    \caption{\label{fig:scaling_resolution}%
    \textbf{Scaling of the entanglement entropy and the approximation error with increasing image resolution.} The quantum states are obtained from encoding the $1058$ highest-resolution images of the `Imagenette'~\cite{imagenette} and DIV2K~\cite{DIV2K} datasets for different levels of resolution. (a)~The plot shows the scaling of the maximal entanglement entropy over any bipartition of the state into two contiguous halves. Blue markers show the results using amplitude encoding, orange markers the results using the FRQI, and the different shades of green show the results using the NEQR with $1$--$3$ color qubits. Note that the three different encodings may use a different number of qubits, so that data points corresponding to the same image resolution do not necessarily align vertically. The marker shapes---dots, crosses and pluses---indicate the different indexing variants---respectively, row-by-row indexing, hierarchical indexing and snake indexing (see also Fig.~\ref{fig:pixel_indexing}). The shaded areas show the $25$th--$75$th percentiles for the different encodings using row-by-row indexing. The dashed blue and orange lines show the average value of the entanglement entropy that is approached with increasing image resolution, the green dashed lines show logarithmic fits. For reference, we also show the average half-chain entanglement entropy for $200$ states whose amplitudes were drawn from a normal distribution before normalization as black three-pointed stars, the black shaded area shows the $25$th--$75$th percentiles. The black dashed line is a linear fit to the data, showing the generic growth of entanglement with system size~\cite{Page1993}. (b)~The plot shows the infidelity (defined as $1 - |\langle\psi_{\text{exact}}|\psi_{\chi=16}\rangle|^2$) between the exact encoded state and its bond dimension $\chi=16$ approximation. The colors and marker shapes denote the same encodings as before.}
\end{figure*}

\subsection{Numerical results}
To test the scaling of compressing images into quantum states, we consider images from the `Imagenette' dataset~\cite{imagenette}, which is a subset of the well-known large-scale ImageNet dataset~\cite{ImageNet}, and the DIV2K dataset~\cite{DIV2K}. These datasets contain a variety of photographs of humans, animals and objects in front of different backgrounds. They should capture a wide range of features that commonly occur in natural images, and thus can be expected to be reasonably representative for drawing conclusions about the properties of generic images. In particular, we select the $1058$ images in the datasets with a resolution higher than $1024\times1024$ pixels ($174$ from the `Imagenette' dataset and $884$ from the DIV2K dataset), as we are mainly interested in the scaling towards high-resolution images. While this comes at the cost of discarding a part of the dataset, this ensures that we do not bias our results from artificially increasing the resolution of originally low-resolution images. In order to ultimately work with a uniformly sized test set, we take the largest square section of each image and resize them to all have the fixed size of $1024\times1024$ pixels using bilinear interpolation. We map the images to quantum states according to the encodings discussed in Sec.~\ref{sec:image_encodings}, and we consider different resolutions of the same images by repeatedly halving the image size, retaining only every other row and column.

First, we look at the entanglement entropy of these states, as low-entangled states are typically well-approximated by MPS~\cite{Verstraete2006, Schuch2008, Hastings2007, Gottesman2010, Eisert2010}. A good MPS approximation also implies a polynomial-depth circuit, as an $n$-qubit MPS with a fixed bond dimension $\chi$ can be prepared using $\bigO{\chi^2 n}$ gates~\cite{Schön2005, Schön2007, Lubasch2020, Smith2022, Lin2021, Barratt2021}.

Fig.~\ref{fig:scaling_resolution}a shows the maximal entanglement entropy over all bipartitions of the state into two contiguous halves, averaged over all images in the dataset, plotted against the number of qubits in the state. The different colors denote the different encodings, i.e., blue shows the data for amplitude encoding, orange the data for the FRQI and the different shades of green the data for the NEQR with $1$--$3$ color qubits. Note that the different encodings may use a different number of qubits for encoding the same resolution image, so data points corresponding to the same image resolution may not align vertically. The marker shapes denote the different ways the pixels are indexed (see Fig.~\ref{fig:pixel_indexing}); dots show the data for row-by-row indexing, crosses the data for hierarchical indexing and pluses the data for snake indexing. The shaded areas show the $25$th--$75$th percentiles for the different image encodings using the row-by-row indexing; we omit the shaded areas for the data points corresponding to the other indexing variants for visual clarity, the results would look similar. As a reference value, we also plot the half-chain entanglement entropy of states with amplitudes that are sampled randomly from a normal distribution with zero mean and a variance of one before normalization. The average of $200$ realizations is marked by black three-pointed stars, the $25$th--$75$th percentiles are shaded in black. The black dashed line shows a linear fit to the data, illustrating the generic case where the entanglement entropy of quantum states grows linearly with the subsystem size~\cite{Page1993}. Conversely, we see that states using amplitude encoding or the FRQI to represent images, independent of the type of indexing used, display a saturation of the entanglement entropy at rather small values with increasing image resolution. The limiting value of the entanglement entropy is roughly reached for image resolutions of $64\times64$ pixels and above; the averages of the corresponding data points are shown as blue and orange dashed lines. In comparison, the entanglement entropy of states using the NEQR is larger but still significantly smaller than that of generic quantum states. For the system sizes at hand it is unclear whether the growth of the entanglement entropy will also eventually saturate for states using the NEQR. It seems plausible that the image resolution is not yet large enough to observe such a saturation, or that the relatively small number of color qubits is insufficient to properly resolve the image features and allow for such a saturation. In the numerically accessible regime, the growth of the entanglement entropy seems roughly logarithmic, as indicated by the green dashed lines, which show the results of fitting a growth logarithmic in the number of qubits to the data.

The saturation of the entanglement entropy at comparably small values for states using amplitude encoding or the FRQI suggests that these states can be well-approximated by MPS with a small bond dimension, while the approximation should be significantly worse for states using the NEQR. Fig.~\ref{fig:scaling_resolution}b shows the average infidelity of the exact encoded states and their MPS approximations with bond dimension $\chi=16$. The MPS approximation is obtained by performing subsequent singular value decompositions (SVDs) of the original state and only keeping the $\chi$ dominant singular values~\cite{Schollwöck2011}. As before, the different colors indicate the different image encodings, and the marker shapes denote the different indexing variants. The shaded areas show the $25$th--$75$th percentiles for different encodings using the row-by-row indexing. For reference, we again show the results for states with randomly sampled amplitudes as black three-pointed stars. We see that the average infidelity of states using amplitude encoding and the FRQI is below that of the NEQR, confirming our expectation. For all three encodings, the average infidelity is far below that of the random states, which cannot be effectively approximated by an MPS with bond dimension $\chi=16$. For states using amplitude encoding and the FRQI, the infidelity seems to saturate with increasing system size at some value, just like the entanglement entropy; for the NEQR, it is again not quite clear whether the error saturates or grows very slowly. The observation that (for certain encodings) the infidelity does not grow asymptotically has significant consequences for the asymptotic scaling of the state preparation. Since the MPS approximation can be prepared using $\bigO{\chi^2 n}$ gates, and we do not need to increase $\chi$ with increasing image resolution, the preparation will only scale linearly in the number of qubits. We present further numerical data showing the scaling of the infidelity with the bond dimension in App.~\ref{app:numerics}.

\begin{figure}[t]
    \hspace{-0.5em}%
    {\setlength{\tabcolsep}{3pt}
    \begin{tabular}{ll}
        \hspace{-3pt}{\sffamily(a) Uncompressed image.}
        &\hspace{-3pt}{\sffamily(b)\hspace{1.5pt}Compression\hspace{1.5pt}with\hspace{1.5pt}%
        $\chi\hspace{-2pt}=\hspace{-2pt}64$.}\\[0.25em]
        \includegraphics[width=0.484\linewidth]{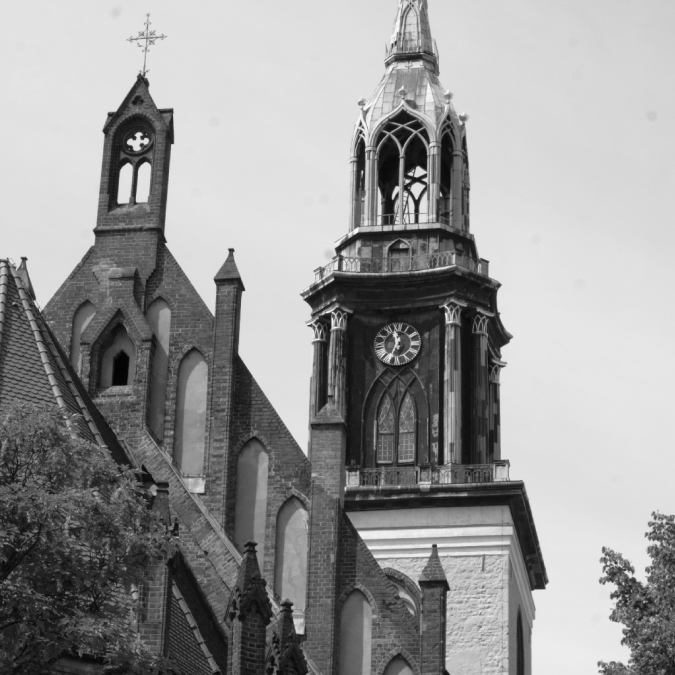}%
        &\includegraphics[width=0.484\linewidth]{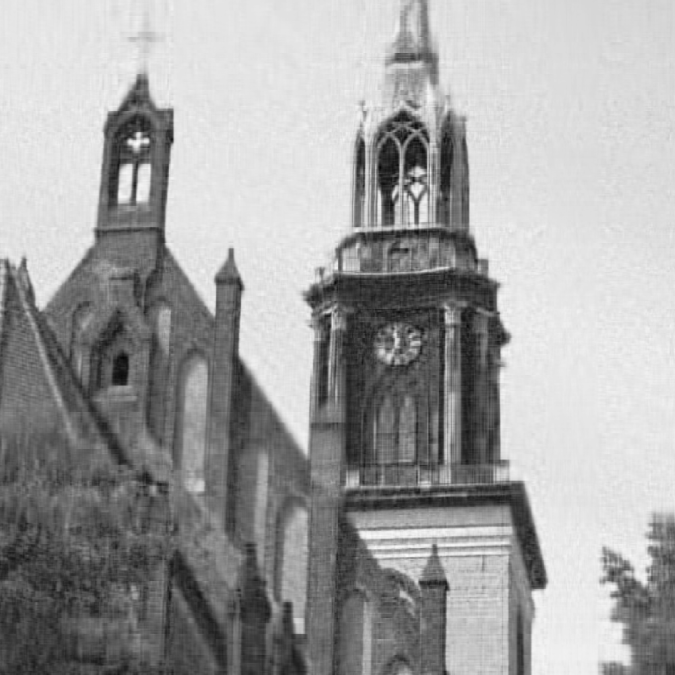}\\[0.5em]
        \hspace{-3pt}{\sffamily(c)\hspace{1.5pt}Compression\hspace{1.5pt}with\hspace{1.5pt}%
        $\chi\hspace{-2pt}=\hspace{-2pt}32$.}
        &\hspace{-3pt}{\sffamily(d)\hspace{1.5pt}Compression\hspace{1.5pt}with\hspace{1.5pt}%
        $\chi\hspace{-2pt}=\hspace{-2pt}16$.}\\[0.25em]
        \includegraphics[width=0.484\linewidth]{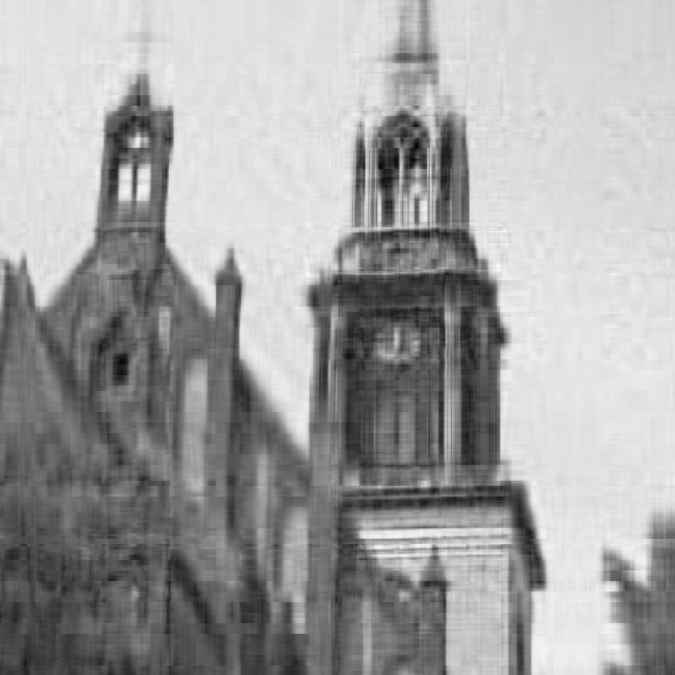}%
        &\includegraphics[width=0.484\linewidth]{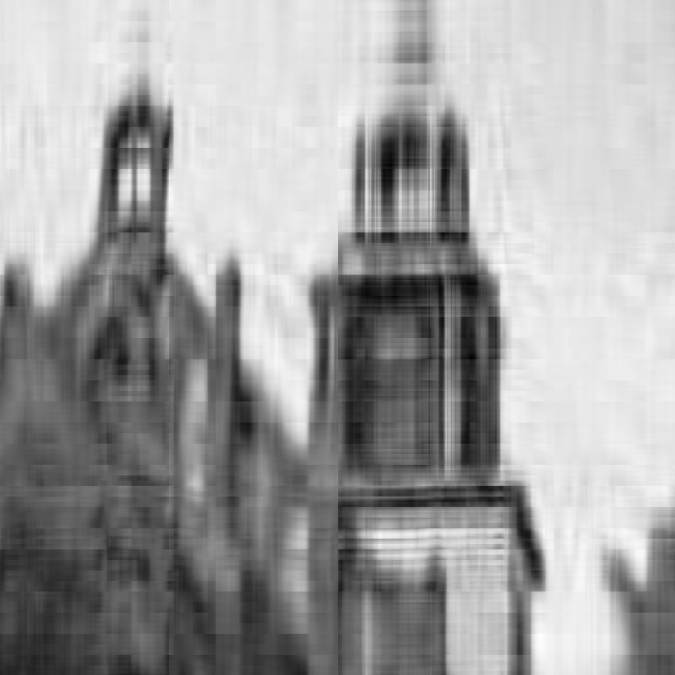}%
    \end{tabular}}
    \caption{\label{fig:example_compression}%
    \textbf{Comparing an (a)~uncompressed image to its MPS approximations with different bond dimensions (b)~$\bm{\chi=64}$, (c)~$\bm{\chi=32}$ and (d)~$\bm{\chi=16}$.} (a)~The original image with $1024\times1024$ pixels is obtained from resizing the raw data of a digital camera, without using any image compression. (b)~Compressing the image with a bond dimension $\chi=64$ yields an infidelity of $0.006$, (c)~a bond dimension $\chi=32$ yields an infidelity of $0.011$, and (d)~a bond dimension $\chi=16$ yields an infidelity of $0.017$. The $\chi=16$ example is representative of the visual quality of the amplitude-encoded images approximated by a $\chi=16$ MPS for which the infidelities are shown in Fig.~\ref{fig:scaling_resolution}b. Note that also the values of the infidelity shown there are similar to the infidelity of the example here.}
\end{figure}

An example for the quality of the compression is shown in Fig.~\ref{fig:example_compression}. The uncompressed image, shown in Fig.~\ref{fig:example_compression}a, is obtained from the raw data of a digital camera, and then resized to $1024\times1024$ pixels. The compressed images are obtained by mapping the original image to a quantum state using amplitude encoding with row-by-row indexing, and constructing an MPS with bond dimension $\chi$ by performing subsequent SVDs and keeping only the dominant $\chi$ singular values. Figs.~\ref{fig:example_compression}b--c show the visual quality of the MPS approximation with decreasing bond dimension. The MPS approximations yield the following infidelities with the original state: $0.006$ for $\chi=64$, $0.011$ for $\chi=32$, and $0.017$ for $\chi=16$. Note that, at least for small systems, it has been demonstrated that preparing input states with infidelities of around $0.1$ is enough to yield classification accuracies comparable with exactly prepared input states for variational QML algorithms~\cite{Dilip2022, Shen2024, West2024}.
In Fig.~\ref{fig:scaling_resolution}b we showed the infidelities of $\chi=16$ MPS approximations; there, the infidelities for $1024\times1024$-pixel images using amplitude encoding saturate at values around $3\cdot10^{-2}$, similar to the infidelity of the $\chi=16$ example in Fig.~\ref{fig:example_compression}d. This example image is representative of the visual quality expected from the amplitude-encoded images corresponding to the data presented in Fig.~\ref{fig:scaling_resolution}b. Note that truncating to a bond dimension of $\chi=16$ is a substantial truncation for a $1024\times1024$-pixel image, keeping less than $2\%$ of all Schmidt values. Still, the main features of the image remain recognizable, although the overall image quality diminishes noticeably at this level of truncation.

\subsection{Bounds on the approximation error}
\begin{figure}[!t]
    \includegraphics[width=\linewidth]{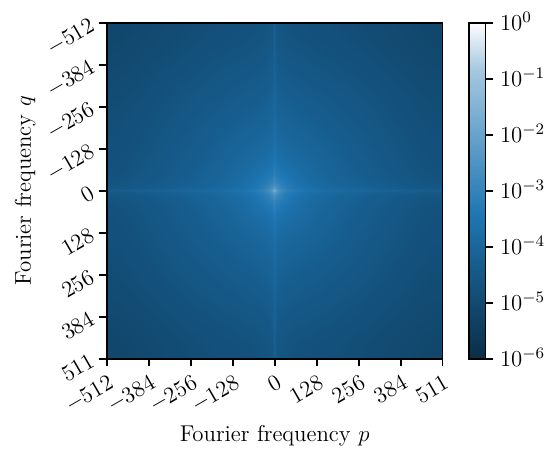}\\[-7.09cm]
    \raggedright{\sffamily(a)}\\[6.2cm]
    \raggedright{\sffamily(b)}\\
    \includegraphics[width=\linewidth]{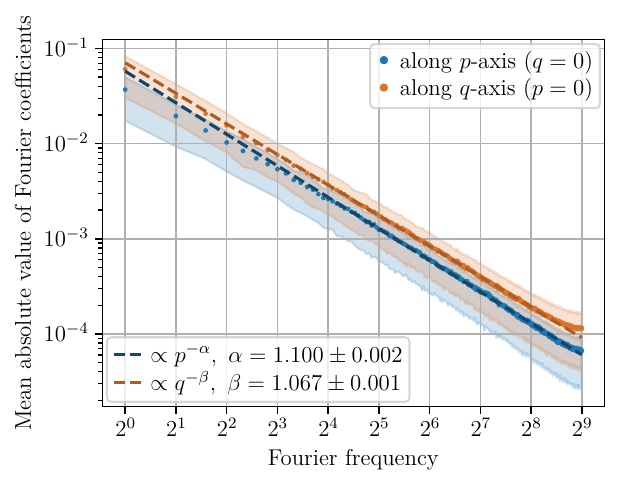}
    \caption{\label{fig:Fourier_coefficients}%
    \textbf{Average decay of the Fourier coefficients of $\bm{1024\times1024}$-pixel images.} (a)~The color plot shows the average absolute values of the Fourier coefficients of the $1058$ highest-resolution images of the `Imagenette'~\cite{imagenette} and DIV2K~\cite{DIV2K} datasets. Most of the weight is concentrated at the bright white spot in the center, and the two axes corresponding to zero-frequency modes in either direction. (b)~The plot below shows the decay of the average absolute value of the Fourier coefficients along the zero-frequency modes---along the $p$-axis where $q=0$ in blue and along the $q$-axis where $p=0$ in orange. The shaded areas show the $25$th--$75$th percentiles. The decay is almost algebraic, as shown by the algebraic fits to the data that are plotted as dashed lines. The deviation from the algebraic decay for large frequencies is due to finite-size effects, see Eq.~\eqref{eq:app_bound_fpq_alg} in the Appendix.}
\end{figure}

In this section we derive error bounds for the MPS approximation of quantum states that encode image data. To be able to make rigorous statements, we need to restrict ourselves to a class of images with well-defined properties. Generally, there is no precise mathematical definition of what constitutes an actual image, as opposed to, e.g., an image made from pixel values drawn from a random distribution. Therefore, we will focus on the subset of images with Fourier coefficients that decay strictly faster than $\bigO{|p|^{-1}|q|^{-1}}$ as a function of the Fourier frequencies $p$ and $q$ in $x$- and $y$-direction. This includes many typical images~\cite{Burton1987, Field1987, Tolhurst1992, Field1993, Ruderman1994, vanderSchaaf1996, Ruderman1997}, where the high-frequency components are small, as high-frequency changes in the pixel values are usually perceived as noise---this assumption also underlies classical image compression methods like the JPEG compression algorithm~\cite{Ahmed1974, Wallace1992}. In practice, we can look at the same $1058$ images from the `Imagenette'~\cite{imagenette} and DIV2K~\cite{DIV2K} datasets we used before and calculate the mean absolute value of the Fourier coefficients; the results are shown in Fig.~\ref{fig:Fourier_coefficients}. The color plot shows the full spectrum, where most of the weight is concentrated at the bright white spot in the center and the two lines where either the frequency in the $x$- or the $y$-direction (labeled by $p$ and $q$) are zero. Outside this region, the weight of the coefficients decreases quickly with increasing frequency. The plot below in the figure shows cuts through the upper color plot along the two white lines, i.e., for either of the two frequencies being zero. In both cases the weight of the Fourier coefficients decays approximately algebraically, indicated by the dashed lines which show algebraic fits to the data, and the decay is faster than $\bigO{1/p}$ or  $\bigO{1/q}$, respectively. The deviation for large frequencies can be understood as a finite-size effect, see the discussion in the following and the derivation in App.~\ref{app:proof_appr_err_bounds}.

To derive the error bounds, we first focus on states using amplitude encoding with row-by-row indexing, i.e., for data given as a $2^n\times2^n$ matrix $f_{ab}$ with $a,b\in\{0,1,\ldots,2^n-1\}$, the quantum state is of the form
\begin{equation}
    \ket{\psi} = \frac{1}{\sqrt{\sum_{a,b=0}^{2^n-1} |f_{ab}|^2}} \sum_{a,b=0}^{2^n-1} f_{ab} \ket{b}\ket{a}.
    \label{eq:example_amp_enc}
\end{equation}
The strategy is then to relate the Fourier coefficients of an $2^n\times2^n$-pixel image to the Fourier coefficients of an `infinitely high-resolution' image. Then, we explain how to construct an MPS from a truncated Fourier series, before showing that a truncated Fourier series is a good approximation if the Fourier coefficients decay fast enough. Finally, we discuss how to extend the derived error bounds to the other indexing and encoding variants.

\textbf{Fourier representation.}
As the goal is to understand the limiting behavior of the quantum states as the number of qubits tends to infinity, we need to specify the limit that the image data approach as we increase the image resolution (and therefore the number of qubits). We can describe this limiting value by a function
\begin{equation}
    F: [0,1)^2\subset\R^2\to\C,~(x,y) \mapsto F(x,y)
    \label{eq:cont_func}
\end{equation}
with continuous inputs $x$ and $y$ defined on the unit square. In principle, this could describe any form of data, but in terms of images, one can restrict to functions with real outputs and think of them as the light intensity distribution of a subject convolved with the point spread function of an imaging system, e.g., one could think of a grayscale image taken by a digital camera with `infinitely high resolution'. From this function with continuous variables, we obtain the $2^n\times2^n$-pixel image $f_{ab}$ by evaluating it at discrete points $(x,y) = \big(\frac{a}{2^n},\frac{b}{2^n}\big)$, with $a,b\in\{0,1,\ldots,2^n-1\}$. To differentiate between the function with continuous inputs describing the limit of infinite pixels and the $2^n\times2^n$-pixel image with discrete indices, we denote the former by an uppercase $F$ and the latter by a lowercase $f$ (and we will also stick to this convention for their Fourier transforms $\hat{F}$ and $\hat{f}$ in the following).
We can write the Fourier series of the function $F$ in Eq.~\eqref{eq:cont_func} as
\begin{equation}
\begin{aligned}
    F(x,y) &= \sum_{k,\ell = -\infty}^{\infty} \!\hat{F}(k,\ell)\;e^{i2\pi kx}\,e^{i2\pi \ell y}\\
    \hat{F}(k,\ell) &= \int_0^1 \d x \int_0^1 \d y~F(x,y)\;e^{-i2\pi kx}\,e^{-i2\pi \ell y}
    \label{eq:Fourier_series}
\end{aligned}
\end{equation}
and relate it to the discrete Fourier transform (DFT) of the pixelated image as
\begin{equation}
\begin{aligned}
    &\qquad f_{ab} \quad = \sum_{p,q = -2^n\!/2}^{2^n\!/2-1} \!\hat{f}_{pq}\;e^{i2\pi pa/2^n}\,e^{i2\pi qb/2^n}\\[-0.6em]
    &\qquad~~\verteq\\[-0.3em]
    &F\!\left(\frac{a}{2^n},\frac{b}{2^n}\right) = \sum_{k,\ell = -\infty}^{\infty} \!\hat{F}(k,\ell)\;e^{i2\pi ka/2^n}\,e^{i2\pi \ell b/2^n},
    \label{eq:relating_DFT_and_FT_sampling}
\end{aligned}
\end{equation}
where $\hat{F}(k,\ell)$ are the coefficients of the Fourier series of the function $F(x,y)$ with continuous arguments as in Eq.~\eqref{eq:Fourier_series} and $\hat{f}_{pq}$ are the coefficients obtained from the DFT of the pixelated image $f_{ab}$. Since the phase factor $e^{i2\pi ka/2^n}$ is invariant under shifts $k\to k+j2^n$ with $j\in\Z$ and the DFT is unique, we can identify
\begin{equation}
    \hat{f}_{pq} = \sum_{i,j\in\Z} \hat{F}(p+i2^n,q+j2^n).
    \label{eq:relating_DFT_and_FT_coeff}
\end{equation}
This allows to relate the Fourier coefficients $\hat{f}_{pq}$ of the ${2^n\times2^n}$-pixel image to those of the sampled function $\hat{F}(k,\ell)$, and allows us to study what happens as the number of pixels is increased.

Consider, for example, a function $F$ with some cutoff $\Lambda$ in the Fourier spectrum, i.e., all Fourier coefficients $\hat{F}(k,\ell)$ are zero if either $|k|>\Lambda$ or $|\ell|>\Lambda$. Then for large enough images, where $2^n>2\Lambda$, we see from Eq.~\eqref{eq:relating_DFT_and_FT_coeff} that the Fourier coefficients remain unchanged when increasing $n$ because shifting the Fourier frequencies by $2^n$ always takes them outside of the cutoff. The image can then be written as
\begin{equation}
    f_{ab} = \sum_{p,q=-\Lambda}^{\Lambda} \!\hat{F}(p,q)\;e^{i2\pi pa/2^n}\,e^{i2\pi qb/2^n},
    \label{eq:Fourier_with_cutoff}
\end{equation}
i.e., as a sum of a finite number of Fourier modes, and now neither the value of the Fourier coefficients nor the number of Fourier modes in the sum depend on $n$.

\textbf{MPS from Fourier modes.}
It turns out that matrices containing only a few Fourier modes, such as for the case above, can be written efficiently as an MPS. An MPS for $n$ qubits has the form~\cite{Fannes1992, Cirac2021, Schollwöck2011}
\begin{equation}
\begin{aligned}
    \ket{\psi} &= \hspace{-0.5em}\sum_{\{\sigma_j\}_{j=0}^{n-1}}\hspace{-0.5em}
    A^{[0]\sigma_0} A^{[1]\sigma_1} \cdots A^{[n-1]\sigma_{n-1}}\\[-1.5em]
    &\hspace{11em}\times\ket{\sigma_0\sigma_1\ldots\sigma_{n-1}},
\end{aligned}
\end{equation}
where for each qubit $j$ there are two $\chi_j\times\chi_{j+1}$ matrices $A^{[j]0}$ and $A^{[j]1}$, and the probability amplitude for a given basis state is obtained by performing the matrix multiplication of the corresponding matrices. For the matrix multiplication to yield a number, the matrix dimensions $\chi_0$ and $\chi_n$ must be one. The maximal matrix dimension $\chi=\max_j \chi_j$ is often referred to as the bond dimension of the MPS, and it controls the amount of entanglement contained in the state. Any state can be written as an MPS, however, generically the bond dimension will need to increase exponentially with the system size $n$---in certain cases, the bond dimension does not scale with the system size and the $2^n$ amplitudes of the quantum state can be expressed using only $\bigO{\chi^2 n}$ parameters.

In one dimension, a sum of $\chi$ Fourier modes can be written as an MPS with bond dimension~$\chi$~\cite{Khoromskij2011, Oseledets2010, Oseledets2011, Oseledets2013, Khoromskij2018}. This can be seen by first considering an {$n$-qubit} state with coefficients given by a single Fourier mode:
\begin{equation}
    \ket{\psi} = \frac{1}{\sqrt{2^n}}\sum_{a=0}^{2^n-1} e^{i2\pi ka/2^n} \ket{a}.
\end{equation}
We can express the summation index $a$ as an $n$-bit binary integer, $a=\sum_{j=0}^{n-1}2^{n-1-j}\sigma_j$ with $\sigma_j\in\{0,1\}$, and find
\begin{equation}
\begin{aligned}
    \hspace{-0.3em}\ket{\psi} &= \!\frac{1}{\sqrt{2^n}}\hspace{-0.65em}\sum_{\{\sigma_j\}_{j=0}^{n-1}}\hspace{-0.65em}
    e^{i2\pi k \sum_{j=0}^{n-1}2^{n-1-j}\sigma_j/2^n}\ket{\sigma_0\sigma_1\ldots \sigma_{n-1}}\hspace{-2em}\\
    &= \bigotimes_{j=0}^{n-1} \left(\frac{1}{\sqrt{2}} \sum_{\sigma_j=0}^1 e^{i2\pi k \sigma_j/2^{j+1}} \ket{\sigma_j}\right),
\end{aligned}
\end{equation}
i.e., a single Fourier mode corresponds to a product state. Hence, a sum of $\chi$ Fourier modes can be written as a bond-dimension-$\chi$ MPS.

\begin{figure}[!t]
    \begin{tabular}{l}
        {\sffamily(a)}\\
        \hspace{-5pt}\includegraphics[width=\linewidth]%
        {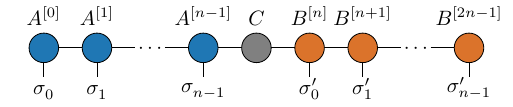}\\[0.8em]
        {\sffamily(b)}\\[-1.4em]
        \hspace{-5pt}\includegraphics[width=\linewidth]%
        {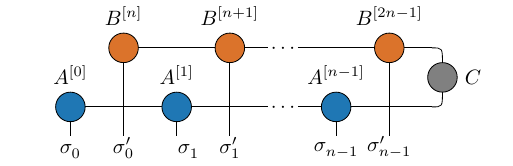}\\
        {\sffamily(c)}\\
        \hspace{-5pt}\includegraphics[width=\linewidth]%
        {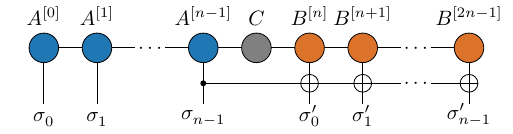}
    \end{tabular}
    \caption{\label{fig:MPS}%
    \textbf{MPS representation of an amplitude encoding state using (a)~row-by-row, (b)~hierarchical and (c)~snake indexing.} Consider a $2n$-qubit amplitude encoding state $\ket{\psi}\propto\sum_{a,b}f_{ab}\ket{b}\ket{a}$, whose amplitudes $f_{ab}$ can be written as a sum of $\chi\times\chi$ Fourier modes. The first $n$ qubits corresponding to the index $b$ (the $y$-coordinate) are labeled $\sigma_j$, the remaining $n$ qubits corresponding to the index $a$ (the $x$-coordinate) are labeled $\sigma_j'$. The $A^{[j]}$ tensors correspond to the plane waves on the $y$-register, the $B^{[j]}$ tensors correspond to the plane waves on the $x$-register, and $C$ contains the Fourier modes---see also App.~\ref{app:2d_fourier_to_mps} for the notation. (a)~The MPS representation of the state using row-by-row indexing has bond dimension $\chi$. (b)~By reordering the qubits we obtain the hierarchical indexing. Note that all the MPS tensors $B^{[j]}$ are diagonal (see App.~\ref{app:2d_fourier_to_mps}), and so their order can be exchanged. This gives an MPS with bond dimension $\chi^2$. (c)~The snake indexing can be obtained by applying a \cnot gate with one control and several target qubits. This gate can be written as a matrix-product operator with bond dimension equal to two, so the bond dimension of the resulting MPS is $2\chi$.}
\end{figure}

In two dimensions, a sum of Fourier modes given by a $\chi\times\chi$ matrix of Fourier coefficients can be similarly written as a bond-dimension-$\chi$ MPS---the qubits corresponding to the $x$- and $y$-coordinates each contain plane waves with $\chi$ different frequencies, and the two qubit registers are connected by the $\chi\times\chi$ matrix of Fourier coefficients. The detailed derivation can be found in App.~\ref{app:2d_fourier_to_mps}. See Fig.~\ref{fig:MPS}a for a graphical representation.

Consider the previous example of a Fourier series with a cutoff $\Lambda$ in Eq.~\eqref{eq:Fourier_with_cutoff}; there is only a ${(2\Lambda+1)\times(2\Lambda+1)}$ matrix of Fourier modes that contributes to the Fourier series, so we can express the state as an MPS with bond dimension $\chi=2\Lambda+1$. Since the number of terms in the Fourier series does not increase with increasing pixel number, the bond dimension of the MPS will be constant as we increase the number of pixels.

\textbf{Error bounds for decaying Fourier coefficients.}
Generically, the Fourier spectrum of relevant images will not have a cutoff. Rather, the magnitude of the Fourier coefficients decays with increasing frequency (see also Fig.~\ref{fig:Fourier_coefficients}). Here, we consider the approximation error when approximating an image by retaining only Fourier modes up to some cutoff---the resulting state can then be described by an MPS with fixed bond dimension, even if the number of pixels goes to infinity.

In the following, to simplify the notation, we introduce three index sets. First, the set of all frequency pairs contributing to the Fourier transform of a ${2^n\times2^n}$-pixel image, $I_{\text{all}} = \{-2^n\!/2, -2^n\!/2+1,\allowbreak \ldots, 2^n\!/2-1\}^2$, secondly, the set of all frequency pairs contained in the approximation, $I_{\text{appr}} = \{-\Lambda, -\Lambda+1,\allowbreak \ldots, \Lambda\}^2$, and finally, the set of frequency pairs discarded in the approximation, $I_{\text{disc}} = I_{\text{all}} \backslash I_{\text{appr}}$. Using this, the pixel values $f_{ab}$ can be expressed in terms of their Fourier modes as
\begin{equation}
    \hspace{-0.5em}f_{ab}
    = \hspace{-0.8em}\sum_{(p,q)\in I_{\text{all}}}\hspace{-0.6em}\hat{f}_{pq}\:e^{i2\pi pa/2^n}e^{i2\pi qb/2^n}\!
    = 2^n (U \hat{f} U^{\transpose})_{ab},\hspace{-0.5em}
    \label{eq:full_DFT_matrix_notation}
\end{equation}
viewing $\hat{f}_{pq}$ as a matrix and introducing the unitary matrix $U_{ap} = \frac{1}{\sqrt{2^n}} e^{i2\pi pa/2^n}$. Note the factor $2^n$ in Eq.~\eqref{eq:full_DFT_matrix_notation}, which appears due to the convention of the Fourier transform used in Eq.~\eqref{eq:relating_DFT_and_FT_sampling}. In the same way, we can write the pixel values $g_{ab}$ obtained from approximating $f_{ab}$ by truncating the Fourier spectrum; this gives
\begin{equation}
    \hspace{-0.5em}g_{ab}
    = \hspace{-1em}\sum_{(p,q)\in I_{\text{appr}}}\hspace{-0.85em}\hat{g}_{pq}\:e^{i2\pi pa/2^n}e^{i2\pi qb/2^n}\!
    = 2^n (U \hat{g} U^{\transpose})_{ab},\hspace{-0.5em}
    \label{eq:truncated_DFT_matrix_notation}
\end{equation}
where $\hat{g}$ is the matrix of the truncated Fourier coefficients, i.e., for $|p|\leq\Lambda$ and $|q|\leq\Lambda$ we have $\hat{g}_{pq} = \hat{f}_{pq}$ but if either $|p|>\Lambda$ or $|q|>\Lambda$ then $\hat{g}_{pq} = 0$.

The quantum state corresponding to the exact image is
\begin{equation}
    \ket{f} = \frac{1}{\norm{f}_F} \sum_{a,b=0}^{2^n-1} f_{ab} \ket{b}\ket{a},
\end{equation}
where $\norm{f}_F = \sqrt{\sum_{a,b=0}^{2^n-1} |f_{ab}|^2}$ denotes the Frobenius norm of the matrix of pixel values, and the quantum state corresponding to the approximation of the image is
\begin{equation}
    \ket{g} = \frac{1}{\norm{g}_F} \sum_{a,b=0}^{2^n-1} g_{ab} \ket{b}\ket{a}.
\end{equation}
The error of the approximation is given by the norm of the difference of the two states, $\norm{\ket{f} - \ket{g}}_2$, using the usual $2$-norm $\norm{\ket{\psi}}_2 = \sqrt{\braket{\psi}{\psi}}$.%
\footnote{
    Previously, we have used the fidelity $\mathcal{F}=|\!\braket{f}{g}\!|^2$ (or the related infidelity $1-\mathcal{F}$) as a measure of closeness between two states $\ket{f}$ and $\ket{g}$. The two quantities can be related by rewriting the norm of the difference of the two states as
    \begin{equation*}
        \norm{\ket{f} - \ket{g}}_2^2
        = \big(\bra{f} - \bra{g}\big)\big(\ket{f} - \ket{g}\big)
        = 2\big(1-\Re\braket{f}{g}\big).
    \end{equation*}
    By allowing a change in the global phase of $\ket{g}$, we can make the overlap $\braket{f}{g}$ purely real, and can relate the two quantities as
    \begin{equation*}
        \norm{\ket{f} - \ket{g}}_2 = \sqrt{2\left(1-\sqrt{\mathcal{F}}\right)}
        \quad\text{or}\quad
        \mathcal{F} = \left(1 - \frac{1}{2}\norm{\ket{f} - \ket{g}}_2^2\right)^{\!2}\!.\hspace{-1em}
    \end{equation*}
    If the $2$-norm distance is a small parameter $\delta = \norm{\ket{f} - \ket{g}}_2 \ll 1$, we have the relation $\mathcal{F} = 1 - \delta^2 + \bigO{\delta^4}$, which allows to directly relate the infidelity shown in Fig.~\ref{fig:scaling_resolution} to the $2$-norm distance discussed here as $1-\mathcal{F} = \delta^2$ with good accuracy.
}
This expression can then be bounded by
\begin{equation}
    \norm{\ket{f} - \ket{g}}_2 \leq \frac{2\norm{f-g}_F}{\norm{g}_F},
    \label{eq:approx_err_normalized}
\end{equation}
which is proved in App.~\ref{app:proof_appr_err_bounds}. Expressing this in terms of the Fourier coefficients gives
\begin{equation}
\begin{aligned}
    &\norm{\ket{f} - \ket{g}}_2 \leq \frac{2\norm{f-g}_F}{\norm{g}_F}\\
    &\quad= \frac{2\,2^n\|U(\hat{f}-\hat{g})U^{\transpose}\|_F}{2^n\|U\hat{g}U^{\transpose}\|_F}
    = \frac{2\|\hat{f}-\hat{g}\|_F}{\|\hat{g}\|_F},
    \label{eq:approx_err_normalized_fourier}
\end{aligned}
\end{equation}
where we have used that the Frobenius norm is invariant under multiplication by a unitary matrix.
Thus, to get the approximation error we just need to bound the norm of the difference of the Fourier coefficients. Since $\hat{g}$ has the same values as $\hat{f}$ inside the cutoff, but is zero outside the cutoff, the norm of the difference amounts to calculating the weight of the discarded Fourier coefficients, i.e.,
\begin{equation}
    \|\hat{f}-\hat{g}\|_F = \sqrt{\sum_{(p,q)\in I_{\text{disc}}} \!|\hat{f}_{pq}|^2}.
    \label{eq:norm_diff}
\end{equation}
This sum converges if the Fourier coefficients decay fast enough.

In particular, if we assume that the Fourier coefficients $\hat{F}(k,\ell)$ of the function $F(x,y)$ decay exponentially as
\begin{equation}
    |\hat{F}(k,\ell)| \leq C e^{-\alpha|k|} e^{-\beta|\ell|},
    \label{eq:Fourier_exp}
\end{equation}
with some constants $C,\alpha,\beta>0$, we can bound the absolute value of the Fourier coefficients $\hat{f}_{pq}$ of the $2^n\times2^n$-pixel image via Eq.~\eqref{eq:relating_DFT_and_FT_coeff} and obtain a bound for the approximation error $\epsilon = \norm{\ket{f}-\ket{g}}_2$ as
\begin{equation}
\begin{aligned}
    \epsilon &\leq \frac{2C}{\norm{\hat{g}}_F}
    \bigg(\frac{\coth(\beta)}{\sinh(\alpha)} e^{-\alpha\chi}
    + \frac{\coth(\alpha)}{\sinh(\beta)} e^{-\beta\chi}\\[0.1em]
    &\quad+ \frac{e^{-(\alpha+\beta)\chi}}{\sinh(\alpha)\sinh(\beta)}\bigg)^{\!\frac{1}{2}}\!\!
    + \bigO{2^n e^{-\alpha2^n}\!\!, 2^n e^{-\beta2^n}}\!,\hspace{-2em}
    \label{eq:approx_err_exp}
\end{aligned}
\end{equation}
see App.~\ref{app:proof_appr_err_bounds_exp} for the detailed calculation. Importantly, this bound does not increase with the number of pixels as the image resolution is increased, and corrections to this bound decay exponentially with the image size. Note that the factor $1/\norm{\hat{g}}_F$ does not change the asymptotic scaling, as it is dominated by the $(2\Lambda+1)\times(2\Lambda+1)$ smallest-frequency Fourier modes $\hat{F}(k,\ell)$. The corrections due to a finite size of the $2^n\times2^n$-pixel image are exponentially small in $2^n$, see Eq.~\eqref{eq:app_bound_fpq_exp}. Since $\norm{\hat{g}}_F$ cannot become arbitrarily small,%
\footnote{
    This is the case unless we consider edge cases where all Fourier coefficients $\hat{F}(k,\ell)$ with $|k|,|\ell|\leq\Lambda$ are exactly zero, in which case one should rather consider a frequency range with nonzero Fourier coefficients for the approximation, e.g., $\Lambda<|k|,|\ell|\leq2\Lambda$.
}
its inverse cannot induce a factor in the error bound that diverges asymptotically, and thus does not change the scaling. The approximation error decreases exponentially as $\bigO{e^{-\min(\alpha,\beta)\chi/2}}$ with the bond dimension $\chi$, or conversely, if we want to describe the image with a fixed error $\epsilon$ as the image size is increased, the bond dimension scales as $\bigO{\log(1/\epsilon)}$, which is independent of $n$.

The more relevant case for images is when the Fourier coefficients $\hat{F}(k,\ell)$ decay algebraically as
\begin{equation}
    |\hat{F}(k,\ell)| \leq C \frac{1}{(|k|+1)^{\alpha}} \frac{1}{(|\ell|+1)^{\beta}},
    \label{eq:Fourier_alg}
\end{equation}
with some constants $C>0$ and $\alpha,\beta>1$.%
\footnote{
    The additional `$+1$' terms in the denominators of Eq.~\eqref{eq:Fourier_alg} are only included to ensure that we do not divide by zero if one of the Fourier frequencies is zero.
}
We need $\alpha,\beta>1$ so that the Fourier series converges absolutely and we can use Eq.~\eqref{eq:relating_DFT_and_FT_coeff} to bound the absolute values of the Fourier coefficients $\hat{f}_{pq}$ of the $2^n\times2^n$-pixel image. In this case, we can obtain an asymptotic bound for the approximation error as
\begin{equation}
\begin{aligned}
    \epsilon &\leq\sqrt{\bigO{\Lambda^{1-2\alpha}} + \bigO{\Lambda^{1-2\beta}}}\\
    &\hspace{4em}+ \bigO{(2^n)^{1-2\alpha}, (2^n)^{1-2\beta}}.
    \label{eq:approx_err_alg}
\end{aligned}
\end{equation}
The derivation and a more detailed expression for the bound are given in App.~\ref{app:proof_appr_err_bounds_alg}. This means that to achieve an approximation error $\epsilon$ for a $2^n\times2^n$-pixel image the bond dimension needs to scale as $\bigO{\left(1/\epsilon\right)^{1/(\min(\alpha,\beta)-\frac{1}{2})}}$, which---notably---is independent of $n$.

Usually, one considers an MPS approximation efficient even if the bond dimension has to grow polynomially with the system size for a constant approximation error~\cite{Verstraete2006, Schuch2008}. Thus, asymptotically, the error bound above is a strong result, since it implies that the bond dimension does not need to grow with the resolution at all. However, if we compare the quantitative values of the bound with the actual approximation errors we find numerically~(e.g., in Fig.~\ref{fig:scaling_resolution}), then the bound usually overestimates the approximation error by up to an order of magnitude. This is because numerically we use an SVD for the truncation instead of the Fourier transform, which can adaptively choose the best basis for the truncation depending on the input data, while the Fourier transform always truncates in the same basis. Since images typically have more structure than just decaying Fourier coefficients, the Fourier basis is rarely optimal for truncation. We discuss this further and show some numerical comparisons in App.~\ref{app:proof_appr_err_bounds_tightness}.

\textbf{Extensions of the results.}
The results presented here were for states using amplitude encoding with the row-by-row indexing. However, they can easily be extended to the hierarchical and snake indexing, as well as to the FRQI. We will also discuss why the NEQR generically does not fit into this framework.

First, we consider the hierarchical indexing. For hierarchical indexing, the first two qubits label the quadrant of the image in which the pixel can be found. For row-by-row indexing, these two qubits correspond to the most significant bit of the $x$- and $y$-coordinates. The next two qubits for hierarchical indexing label the subquadrant in which the pixel lies. This corresponds to the next-most significant bits of the $x$- and $y$-coordinates of the row-by-row indexing. Thus, we can obtain the state using hierarchical indexing from the state using row-by-row indexing by alternatingly interleaving qubits corresponding to the $x$- and $y$-coordinate, ordered from most significant to least significant bit. Fig.~\ref{fig:MPS}a shows the MPS of the state using row-by-row indexing, and Fig.~\ref{fig:MPS}b shows the result of swapping the qubits into an alternating pattern to construct the state using hierarchical indexing. Thus, from the results of the previous section we can conclude that also states using hierarchical indexing can be approximated by MPS with a bond dimension that does not scale with the image resolution. Note that while this construction of the MPS increases the bond dimension from $\chi$ to $\chi^2$, from the numerical results in Fig.~\ref{fig:scaling_resolution}b we see that an MPS approximation with bond dimension $\chi$ of a state using hierarchical indexing is about as good or better than an MPS approximation with bond dimension $\chi$ of a state using row-by-row indexing.

Next, we consider snake indexing. To obtain the snake indexing from the row-by-row indexing, we need to mirror the $x$-coordinate for every other $y$-coordinate. Mirroring the $x$-coordinate can be done by applying Pauli-$X$ gates on the whole register corresponding to the $x$-coordinate. Doing this only for every other row means making these Pauli-$X$ gates conditional on the least significant bit of the $y$-coordinate. This means we need to apply a \cnot gate with one control and several target qubits, which is shown graphically in Fig.~\ref{fig:MPS}c. Such a gate can be written as a matrix-product operator (MPO), where the two-dimensional virtual index carries the information of whether the control qubit was in the state $\ket{0}$ or $\ket{1}$, and depending on that the tensor connecting the two physical indices either forms an identity matrix or a Pauli-$X$ gate. Such an MPO can change the bond dimension of the MPS at most by a factor of two. Thus, also states using the snake indexing can be expressed as MPS where the bond dimension does not need to grow with increasing image resolution.%
\footnote{
    One could have also considered an ordering of the pixels as given by the Gray code, where the binary integers labeling the $x$- and $y$-coordinate are ordered such that subsequent integers differ only by a single bit. A circuit implementing such a transform is given by a staircase of \cnot gates~\cite{Zhou2015}, so one can connect the row-by-row indexing to the Gray-code indexing (similarly to how the snake indexing can be connected to the row-by-row indexing) by applying the appropriate unitary gate: in this case, a staircase of \cnot gates on both the $x$- and $y$-register. Such a gate cannot change the bond dimension of the MPS by more than a factor of two. While we have not plotted the results for the Gray-code indexing in Fig.~\ref{fig:scaling_resolution} to avoid overcrowding the figure, plotting the results would yield results similar to the other indexing variants.
}

We can also extend the MPS approximability to the FRQI, by writing out the state as
\begin{equation}
\begin{aligned}
    \ket{\psi} = &\frac{1}{2^n} \sum_{a,b=0}^{2^n-1}\ket{c(f_{ab})}\otimes\ket{b}\ket{a}\\
    = &\sum_{a,b=0}^{2^n-1} \frac{1}{2^n}\cos(\frac{\pi}{2}f_{ab})\ket{0}\otimes\ket{b}\ket{a}\\
    + &\sum_{a,b=0}^{2^n-1} \frac{1}{2^n}\sin(\frac{\pi}{2}f_{ab})\ket{1}\otimes\ket{b}\ket{a}.
\end{aligned}
\end{equation}
This can be viewed as the sum of two states using amplitude encoding, one encoding $\frac{1}{2^n}\cos(\frac{\pi}{2}f_{ab})$ and the other one encoding $\frac{1}{2^n}\sin(\frac{\pi}{2}f_{ab})$. Since the decay of the Fourier coefficients can be related to the smoothness of the function~\cite{Grafakos2014}, and the sine and cosine functions are analytic, the newly encoded functions are also smooth and should have similarly decaying Fourier coefficients.%
\footnote{
    This statement can be made more precise if the function $F(\vec{x})$, $\vec{x}\equiv(x,y)$, or its $m$th derivatives, are $\mu$-Hölder continuous. Then, the Fourier coefficients $\hat{F}(\vec{k})$, $\vec{k}\equiv(k,\ell)$, decay as $\bigO{1/|\vec{k}|^{\mu+m}}$~\cite[Theorem~3.3.9]{Grafakos2014}. Since $\sin(x)$ and $\cos(x)$ are analytic functions, also $\sin(F(\vec{x}))$ and $\cos(F(\vec{x}))$, or their $m$th derivatives, are $\mu$-Hölder continuous and their Fourier coefficients also decay as $\bigO{1/|\vec{k}|^{\mu+m}}$. Note that this bound is not necessarily asymptotically tight, at least in one dimension this decay can be further improved, e.g., to a faster decay as $\bigO{1/|k|^{1+\mu+m}}$ if the function is not infinitely oscillatory~\cite[Theorem~1.1]{Nissila2018}, or to include functions whose $m$th derivative is only piecewise continuously differentiable leading to a decay $\bigO{1/|k|^{1+m}}$~\cite[Theorem~5]{Herbert2022}.
}
This is also observed numerically in Fig.~\ref{fig:scaling_resolution}b, where the approximation error of the states using the FRQI is even smaller than that of those using amplitude encoding.

While it seems like this trick can also be applied to the NEQR, now with $2^q$ copies of amplitude encoding states instead of just two, this is not quite the case. For simplicity, let us assume $q=1$; then, the NEQR would correspond to two copies of amplitude encoding states. The amplitudes of one copy would be zero for each black pixel and $1/2^n$ for each white pixel, and the other way around for the other copy. Since the amplitudes are either zero or a constant, they can be represented as the sum of several rectangle functions. Taking the Fourier transform of this gives the sum of several scaled functions of the form $\text{sinc}(k) = \sin(k)/k$, which asymptotically decay as $\bigO{1/k}$. This means the decay of the Fourier coefficients is too slow and we cannot apply the arguments of the previous section to argue for a good MPS approximation. We thus expect a much larger approximation error, which is what we observed numerically in Fig.~\ref{fig:scaling_resolution}b.

As a side note, instead of Fourier transforms we could also consider cosine transforms, which are typically more commonly used in image processing (e.g., in the JPEG compression algorithm) than the usual Fourier transform. Since the cosine transforms can be related to Fourier transforms, we can apply the same error bounds---we make this connection precise in App.~\ref{app:cosine_transforms}.

Finally, we briefly want to comment on the entanglement entropy of these states. For states using amplitude encoding or the FRQI, the error of the MPS approximation decays algebraically with exponent larger than one. From this, one can estimate the decay of the Schmidt values, which is the same asymptotically. This means that the Schmidt values decay fast enough so that even as the system size tends to infinity the bipartite entanglement remains bounded, which explains the behavior observed in Fig.~\ref{fig:scaling_resolution}a; since the NEQR does not admit such an efficient MPS representation, but still has more structure than a random state, it makes sense that the entanglement entropy is somewhere in between the two cases (see Fig.~\ref{fig:scaling_resolution}a). The small entanglement entropy of these states is a useful property for machine learning tasks, besides making them efficiently preparable as input states on a quantum computer. Consider a classification task, where we first prepare an initial state that encodes the input data, we then run a circuit that implements a classification algorithm, and finally we measure some few-qubit observables (e.g., the probability of measuring a certain bitstring) to assign a classification label based on the largest expectation value (e.g., the most likely bitstring outcome). To distinguish this observable from the rest with only a few shots on a quantum computer, we want there to be a big difference between the largest and the second largest expectation value. If the qubits we measure are highly entangled with the rest of the system, their reduced density matrix will be close to an identity matrix, and so all expectation values are close to the trace of the observables. Unless the observables are already biased towards a particular label, the trace will be the same for all of them. Since this is exactly the case we want to avoid, we want the measured qubits to only be weakly entangled with the rest of the system. This means the classification algorithm must disentangle the qubits that are to be measured from the rest of the system. As unitarily disentangling a state is generically a hard problem if the circuit that created the state is unknown~\cite{Chamon2014, Shaffer2014, Morral-Yepes2024}, and we want the classification circuit to be efficient, i.e., to use as few gates as possible, this is only possible if the input state is already only weakly entangled. In this sense, the states discussed here are well-suited as input states for machine learning tasks. Of course, for there to be any hope of quantum advantage, some intermediate state or the unmeasured subsystem of the classification circuit must be highly entangled, so that it is no longer efficiently classically simulable.

\section{\label{sec:image_compression_circuits}Image compression with circuits}
In the previous section we discussed that typical quantum states representing classical image data can be well-approximated by MPS. This means they are quite different from generic quantum states in the Hilbert space, for which an exponentially deep circuit is needed for the preparation on a quantum computer~\cite{Plesch2011, Iten2016}. Here, we discuss how this special structure of the quantum states can lead to more efficient linear-depth circuits for state preparation, and how to obtain them.

\subsection{Sequential circuits from MPS}
\begin{figure}[t]
    \begin{tabular}{l}
        {\sffamily(a)}\\[-1.5em]
        \hspace{0.09\linewidth}\includegraphics[width=0.8\linewidth]{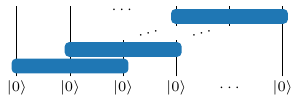}\hspace{0.09\linewidth}\\[1em]
        {\sffamily(b)}\\[-1.5em]
        \hspace{0.09\linewidth}\includegraphics[width=0.8\linewidth]{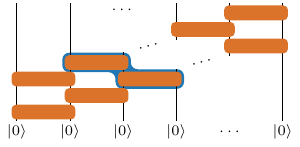}\hspace{0.09\linewidth}
    \end{tabular}
    \caption{\label{fig:1d_sequential_circuit}%
    \textbf{(a)~MPS circuit and (b)~sparse sequential circuit.} (a)~A circuit consisting of gates that act on $\log_2(\chi)+1$ qubits arranged in a sequential pattern is in one-to-one correspondence with an MPS of bond dimension $\chi$---see e.g. Refs.~\cite{Lubasch2020, Lin2021, Barratt2021, Astrakhantsev2023} for details of the mapping. The example shown here is for $\chi=4$. (b)~Instead of using multi-qubit gates, one can also repeat several layers of a circuit corresponding to a $\chi=2$ MPS, shown here for two layers. The two gates framed by the blue box correspond to a sequentially repeated multi-qubit gate---in that sense, sparse sequential circuits can be viewed as a sparsely parametrized subclass of MPS circuits.}
\end{figure}

We can directly turn an MPS with bond dimension $\chi$ into a sequential quantum circuit with $\bigO{\chi^2n}$ two-qubit gates~\cite{Schön2005, Schön2007, Lubasch2020, Smith2022, Lin2021, Barratt2021}. Such a circuit is shown in Fig.~\ref{fig:1d_sequential_circuit}a for $\chi=4$, where each $\chi\times2\times\chi$ MPS tensor gets mapped to a $\log_2(\chi)+1$-qubit gate, with one incoming qubit in the state $\ket{0}$, one outgoing qubit corresponding to the physical leg of the MPS, and the remaining $\log_2(\chi)$ quantum wires corresponding to the virtual legs of the MPS. Since we have seen in the previous section that the bond dimension scales as $\bigO{\log(1/\epsilon)}$ or $\bigO{\left(1/\epsilon\right)^{1/(\min(\alpha,\beta)-\frac{1}{2})}}$, depending on whether the Fourier coefficients decay exponentially or algebraically, the scaling of the number of gates becomes $\bigO{n \log(1/\epsilon)^2}$ or $\bigO{n\left(1/\epsilon\right)^{2/(\min(\alpha,\beta)-\frac{1}{2})}}$, respectively. In both cases the complexity of the circuit reduces from a scaling exponential in $n$ to a scaling linear in $n$, compared to exact state preparation.

The question remains, however, how to obtain the MPS in practice. A popular algorithm to obtain an MPS from a quantum state that yields good results is to iteratively perform an SVD on each bond, and only keep the $\chi$ dominant singular values~\cite{Schollwöck2011}. Since the SVD of an $\ell\times m$ matrix has a computational complexity of $\bigO{\ell m\min(\ell,m)}$, and we need to do SVDs of $\min(2^{i}, \chi, 2^{2n-i+1}) \times 2^{2n-i}$ matrices, where the index $i$ runs over each of the $\bigO{n}$ bonds, the cost of this method scales as $\bigO{\chi 2^{2n}}$.\footnotemark{}
As the bond dimension does not scale asymptotically with $n$, this is asymptotically slightly better than the classical preprocessing of $\bigO{n2^{2n}}$ needed to obtain the exact state preparation circuit with the fewest gates (see Sec.~\ref{sec:image_encodings}). More importantly, asymptotically the circuit preparing the approximate state needs exponentially fewer gates.

Instead of performing the SVD, one could also try to actually apply the Fourier transform, truncate the spectrum of Fourier coefficients, and construct the MPS from that. The full DFT, however, will scale as $\bigO{n2^{2n}}$, which is slightly worse than the SVD and typically results in a worse MPS approximation. Instead of doing the full Fourier transform, one could instead perform a sparse two-dimensional Fourier transform (as suggested in Ref.~\cite{Moosa2023}) to extract only the dominant Fourier modes~\cite{Gilbert2005, Ghazi2013, Gilbert2014}. There are several different implementations for different cases, but with one such implementation estimating the $\bigO{\chi^2}$ dominant Fourier coefficients of an $2^n\times2^n$-pixel image with success-probability $1-\delta$ and error $\epsilon$ scales as $\bigO{\chi^2\,\poly{\log(1/\delta), n, 1/\epsilon}}$~\cite{Gilbert2005}. In principle this could get rid of the exponential scaling entirely, however, it should be examined carefully if real images satisfy the assumptions about the sparsity of the Fourier spectrum that is assumed in the proofs of these algorithms.%
\footnotetext[\value{footnote}]{
    More explicitly, let us assume that $\chi=2^{\eta}$ is some power of two and we have a $2n$-qubit state, then the total cost for turning the state into an MPS by successive SVDs is given by
    \begin{equation*}
    \begin{aligned}
        &\sum_{i=1}^{2n-1} \bigO{\min(2^{i}, \chi, 2^{2n-i+1})^2 \, 2^{2n-i}}\\
        =&\sum_{i=1}^{\eta} \bigO{2^{2i} 2^{2n-i}}
        + \!\sum_{i=\eta+1}^{2n-1-\eta}\! \bigO{\chi^2 2^{2n-i}}\\[-0.5em]
        &\qquad+ \!\sum_{i=2n-\eta}^{2n-1}\! \bigO{(2^{2n-i+1})^2 2^{2n-i}}\\
        =&\,\bigO{\chi 2^{2n}} + \bigO{\chi^2 2^{2n}/\chi} + \bigO{\chi^3} = \bigO{\chi2^{2n}}.%
    \end{aligned}%
    \end{equation*}%
}

\subsection{Sparse circuits}
Directly turning the MPS approximation of an image state into a circuit yields a circuit which needs $\bigO{\chi^2n}$ gates. This leads to a very efficient linear scaling asymptotically, as the bond dimension $\chi$ tends to a constant; however, in practice, the prefactor $\chi^2$ can become very large and can render this approach unsuitable for the currently available noisy intermediate-scale quantum (NISQ) devices~\cite{Preskill2018}. We can make the circuits more NISQ-friendly by choosing a `sparse' circuit structure, consisting only of two-qubit gates so that no further gate decomposition is necessary, and optimize the gates in the circuit directly. There are many potential circuit ansätze that one could choose. Guided by the intuition of the previous section that images are efficiently captured by classical tensor networks due to their limited entanglement entropy, we consider three different circuit ansätze inspired from tensor networks: namely one-dimensional sequential circuits inspired by MPS~\cite{Lin2021, Dilip2022, Astrakhantsev2023}, two-dimensional sequential circuits inspired by two-dimensional isometric tensor networks~\cite{Wei2022, Zalatel2020} and the circuit version of the multi-scale entanglement renormalization ansatz~\cite{Vidal2008, Cong2019, Grant2018}. These tensor-network states also describe states with limited entanglement entropy, but the latter two tensor networks capture a broader range of states than MPS.

To optimize the circuits in the following, we take the fidelity between the target state and the state produced by the circuit as a cost function, and maximize it using the Riemannian implementation of the Adam optimizer from the Python library QGOpt~\cite{qgopt, Luchnikov2021, Oza2009}. Adam is a gradient-based optimizer, for which we can obtain the gradient via automatic differentiation using TensorFlow~\cite{tensorflow}. The Riemannian version optimizes directly over the Riemannian manifold of unitary matrices, allowing each two-qubit gate in the circuit ansatz to be a general gate in $U(4)$ without explicit parametrization. The code for the optimization is available on Zenodo upon reasonable request~\cite{Zenodo}.

To keep the simulation of the circuits tractable on a classical computer, we limit ourselves to images with a resolution of $32\times32$ pixels. This resolution would constitute a significant reduction for the images from the original `Imagenette'~\cite{imagenette} and DIV2K~\cite{DIV2K} datasets, which could introduce unwanted artifacts. Therefore, we switch to using the first $1000$ images from the Fashion-MNIST dataset~\cite{FashionMNIST}. The Fashion-MNIST dataset consists of $28\times28$-pixel grayscale images of fashion articles, which we resize to $32\times32$ pixels using bilinear interpolation. This also allows for direct comparison with other studies using the same dataset, which report classification accuracies on the compressed dataset or infidelities for the compression when using different circuit ansätze~\cite{Dilip2022, Shen2024}.

\textbf{One-dimensional sequential circuits.}
A simple idea to reduce the cost of the multi-qubit gate decompositions in the MPS circuit is to instead use several layers of a sequential circuit consisting of only two-qubit gates, eliminating the need for multi-qubit gate decompositions entirely. An example for two layers is shown in Fig.~\ref{fig:1d_sequential_circuit}b. In the figure, two gates are highlighted in blue, which act like a multi-qubit gate that is applied sequentially, just like in Fig.~\ref{fig:1d_sequential_circuit}a. Therefore, we can equivalently view the circuit as replacing each dense multi-qubit gate, which would require $\bigO{\chi^2}$ two-qubit gates to implement, by a series of $\log_2(\chi)$ two-qubit gates---in that sense, the circuit is a sparse version of the full MPS circuit~\cite{Lin2021}. Numerically, it has been observed that for certain applications these sparse circuits work just as well as MPS~\cite{Lin2021, Dilip2022, Astrakhantsev2023}. The circuits can be obtained either variationally~\cite{Lin2021, Dilip2022} or by repeatedly disentangling the original MPS~\cite{Ran2020}. Using variational methods, one could, in principle, find the best sequential circuit with a fixed number of layers to approximate the target state, as long as one does not get stuck in local minima during the optimization. The optimization could also be done in a hybrid quantum-classical setup, where the circuit is evaluated on a quantum computer and the parameters are updated using some classical optimization scheme. In contrast, constructing the sequential circuit with a fixed number of layers using repeated disentangling gates will not generally converge to the best approximation; however, the computational cost on a classical computer is much lower and one can avoid variational optimization where the computational cost is hard to estimate.

Here, we focus on the variational approach to compress states encoding images (similarly to Ref.~\cite{Dilip2022}). We take the first $1000$ images of the Fashion-MNIST dataset~\cite{FashionMNIST}, and map them to quantum states using amplitude encoding, the FRQI and the NEQR with $1$--$3$ color qubits with snake indexing. We then optimize one-dimensional sequential circuits (as in Fig.~\ref{fig:1d_sequential_circuit}b) with $1$--$3$ layers to approximate the states. The average infidelities of the optimized circuits and the original states are plotted against the number of variational parameters of the circuit in Fig.~\ref{fig:comparison_approx_err}.%
\footnote{\label{fn:parameter_counting}
    Since we do not explicitly parametrize the two-qubit gates, we count the number of parameters as the number of independent real numbers describing a complex isometric matrix. An isometric matrix is a matrix $W\in\C^{n \times p}$, with $n \geq p$, fulfilling $W^{\dagger} W = \Id_{p \times p}$. (Equivalently, it is a subset of $p$ columns of a $n \times n$ unitary matrix.) It has $2np-p^2$ independent real parameters~\cite{Lin2021}. Hence, a two-qubit gate acting on two qubits fixed in the $\ket{0}$ state has $7$ parameters, a two-qubit gate acting on one $\ket{0}$ state has $12$ parameters, and a two-qubit gate acting on arbitrary input states has $16$ parameters.
}
The blue crosses in the left plot show the results for amplitude encoding, the orange crosses in the central plot show the results for the FRQI and the green crosses in the right plot show the results for the NEQR. The shaded area shows the $25$th--$75$th percentiles. Note that the number of parameters in the variational circuit is proportional to the number of qubits. Thus, the same circuit ansatz with the same number of layers can have a different number of parameters depending on the encoding type, as the number of qubits varies. The image resolution, however, remains $32\times32$ pixels for all encodings.

\begin{figure}[t]
    \includegraphics[width=\linewidth, trim={0.08cm 0 0 0}, clip]{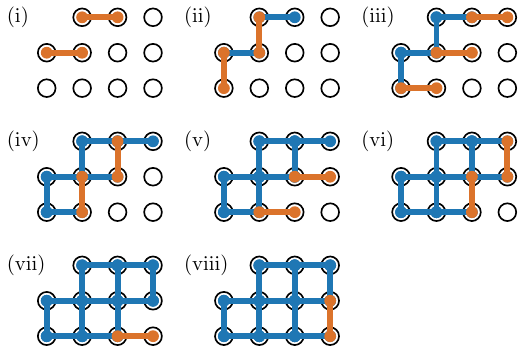}
    \caption{\label{fig:2d_sequential_circuit}%
    \textbf{Constructing a two-dimensional sequential circuit.} Sequential circuits can also be constructed in two dimensions~\cite{Wei2022}. At each step, the orange gates show the newly applied gates, and the blue gates show the ones that have been applied previously. The sequentially repeating pattern here consists of gates along a diagonal, which are applied alternatingly horizontally or vertically.}
\end{figure}

\textbf{Two-dimensional sequential circuits.}
Sequential circuits can not only be implemented in one dimension, but also in higher dimensions~\cite{Wei2022}. Since many superconducting quantum devices have their qubits arranged in a two-dimensional lattice, we will focus on this case. In two dimensions, sequential circuits form a subclass of general two-dimensional tensor networks~\cite{Wei2022, Zalatel2020}, which with a fixed bond dimension can create entanglement entropies following a two-dimensional area law. Conversely, to create a state with an area law in two dimensions using MPS, the bond dimension needs to grow exponentially. Thus, two-dimensional sequential circuits can in principle be a more powerful ansatz than one-dimensional sequential circuits, and we can hope that this also leads to a better approximation of quantum states encoding classical image data.

To test this, we construct a circuit as shown in Fig.~\ref{fig:2d_sequential_circuit}. There, we have eleven qubits arranged in a $3\times4$ square lattice with the top left qubit removed. This corresponds, e.g., to the case of encoding a $32\times32$-pixel image using the FRQI. At each step of the construction, the gates applied in the current step are shown in orange, and the gates applied in previous steps are shown in blue. We start the construction by first applying two-qubit gates horizontally along the top-left-most diagonal, and then applying two-qubit gates vertically along the same diagonal. This procedure is then repeated sequentially on the next diagonals, moving diagonal by diagonal to the bottom right. For different system sizes the construction works analogously. Similar to the case of the one-dimensional sequential circuit, we can view this construction as a single layer, and repeat it several times to obtain sequential circuits with several layers.

As for the one-dimensional sequential circuit, we encode the first $1000$ images of the Fashion-MNIST dataset~\cite{FashionMNIST} using the three encodings discussed in Sec.~\ref{sec:image_encodings} with snake indexing, and optimize the circuit with $1$--$3$ layers to approximate the corresponding quantum states. Since the number of qubits changes for the different encodings, we use different layouts for the qubits. The amplitude encoding uses ten qubits, which we lay out in a $4\times4$ grid, with all qubits above the diagonal from bottom-left to top-right removed; the FRQI and the NEQR with a single color qubit use eleven qubits, which corresponds to the case shown in Fig.~\ref{fig:2d_sequential_circuit}; the NEQR with two color qubits uses twelve qubits, which we can arrange in a $3\times4$ grid without removing any qubits; and the NEQR with three color qubits uses thirteen qubits, which we arrange in a $4\times4$ grid with the three qubits in the top-left corner removed. The resulting infidelities of the approximation are plotted in Fig.~\ref{fig:comparison_approx_err} against the number of variational parameters in the circuit.%
\textsuperscript{\textit{\ref{fn:parameter_counting}}}
The blue pluses in the left plot show the results for the amplitude encoding, the orange pluses in the central plot show the results for the FRQI, and the green pluses in the right plot show the results for the NEQR. The shaded area shows the $25$th--$75$th percentiles.

\begin{figure}[b]
    \includegraphics[width=\linewidth, trim={0.25cm 0 0.25cm 0}, clip]{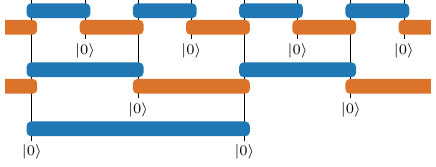}
    \caption{\label{fig:MERA_circuit}%
    \textbf{MERA circuit.} Since the MERA~\cite{Vidal2008} consists of isometric tensors, the tensor network with bond dimension $\chi=2$ can be naturally expressed as a quantum circuit. With each layer, the number of qubits is doubled by inserting a qubit in between the qubits of the previous layer and then entangling it with its nearest neighbors (indicated by the orange and blue gates). This way, entanglement is built up over all length scales of the system.}
\end{figure}

\begin{figure*}[t]
    \centering
    \includegraphics[width=\linewidth]{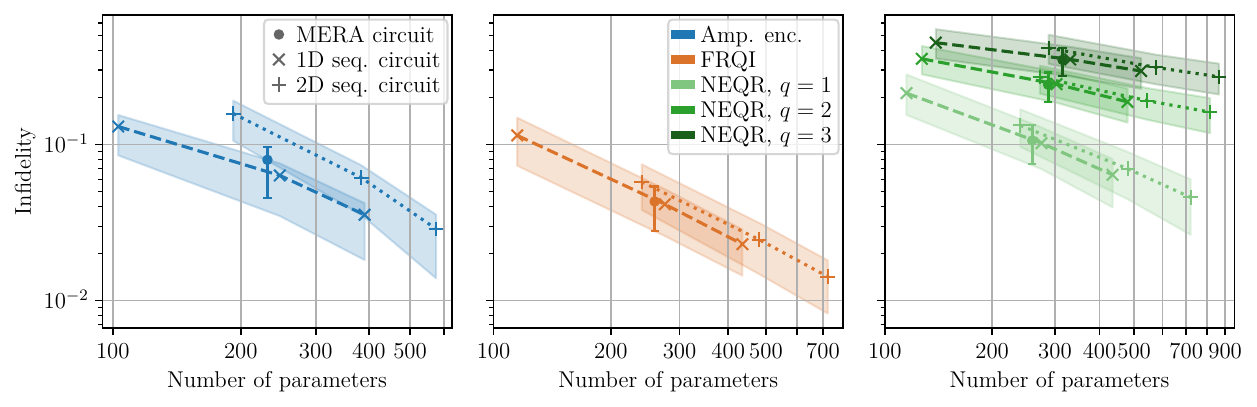}
    \caption{\label{fig:comparison_approx_err}%
    \textbf{Comparison of the approximation error for different circuit types.} The first $1000$ images of the Fashion-MNIST dataset~\cite{FashionMNIST} are encoded using snake indexing in combination with amplitude encoding (left, in blue), the FRQI (center, in orange) and the NEQR with $1$--$3$ color qubits (right, in green). The data shown are the average infidelities of the original state and the variationally optimized circuit (with the infidelity defined as $1 - |\langle\psi_{\text{exact}}|\psi_{\text{circ.}}\rangle|^2$), plotted against the number of parameters in the circuit.\textsuperscript{{\normalfont\textit{\ref{fn:parameter_counting}}}} We show results for a MERA circuit (see Fig.~\ref{fig:MERA_circuit}) as dots, for the one-dimensional sequential circuit (see Fig.~\ref{fig:1d_sequential_circuit}b) as crosses and for the two-dimensional sequential circuit (see Fig.~\ref{fig:2d_sequential_circuit}) as pluses. The error bars and the shaded areas show the $25$th--$75$th percentiles.}
\end{figure*}

\textbf{MERA circuit.}
Another tensor network that can be turned into a quantum circuit is the multi-scale entanglement renormalization ansatz (MERA)~\cite{Vidal2008}. (In the QML literature such a circuit is also known as a quantum convolutional neural network or QCNN~\cite{Cong2019}.) The idea of the MERA is to perform a kind of coarse-graining in the entanglement structure of the state and to subsequently reduce the number of qubits in the system. At each step the qubits are grouped into two-site unit cells. Then, a unitary tensor is applied to all neighboring qubits in different unit cells in order to disentangle the different unit cells. Then, an isometric tensor is applied within each unit cell, mapping each two-qubit state to a single-qubit state. This procedure can be repeated until there are only two qubits left in the system. Running this renormalization in reverse constructs a quantum state, and since all tensors are isometric, it can be written as a quantum circuit. Such a circuit is shown in Fig.~\ref{fig:MERA_circuit}, where the isometries that double the number of qubits are shown in orange, and the entangling operations between the unit cells are shown in blue. Due to its hierarchical structure, the circuit can build up correlations and entanglement at all length scales, and is therefore able to describe states with an entanglement entropy that grows logarithmically in the number of qubits, which one-dimensional sequential circuits with a fixed number of layers are unable to replicate.

Here we view the MERA circuit as a variational ansatz, and optimize the gates in the circuit to best approximate quantum states representing image data. Since the number of qubits of these quantum states is usually not a perfect power of two, we need to slightly adapt the construction of the MERA circuit. If at any layer the number of qubits is even, the construction proceeds as described above and we assume periodic boundary conditions when it comes to disentangling the different unit cells. If at any layer there is an odd number of qubits, we split all qubits into pairs as before except that the last qubit remains without a partner. Then we apply the disentanglers between the unit cells such that the unpaired qubit is disentangled from the last unit cell, and no gate acts on the first qubit. The isometries are applied within the unit cell, which means at this step no gate is applied to the unpaired qubit. This way we have reduced the number of qubits, and can repeat the steps above until only two qubits are left. Running this structure in reverse creates a MERA-like circuit for a number of qubits that is not a power of two. To increase the expressiveness of the MERA with a fixed system size on a classical computer, one usually increases the matrix dimensions of the intermediate unitaries and isometries in the ansatz. However, this is not so straightforward to do on a quantum computer, so we restrict to a fixed number of parameters for the MERA for a given number of qubits.

As before, we variationally optimize the circuits to approximate the states encoding the first $1000$ images of the Fashion-MNIST dataset~\cite{FashionMNIST}, which yields the infidelities against the number of parameters shown in Fig.~\ref{fig:comparison_approx_err}.%
\textsuperscript{\textit{\ref{fn:parameter_counting}}}
The average infidelity for the states using amplitude encoding is shown in the left plot as a blue dot, the average infidelity for the FRQI is shown in the central plot as an orange dot, and the average infidelities for the NEQR with $1$--$3$ color qubits are shown in the right plot as green dots. The error bars show the $25$th--$75$th percentiles.
\medskip

Finally, we can compare the performance of the different circuit ansätze in Fig.~\ref{fig:comparison_approx_err}. For a given encoding (i.e., data represented by the same color and shown in the same subplot), the achievable infidelities appear to depend algebraically on the number of parameters in the ansatz, but not at all on the type of circuit used. Although different encodings use a different number of qubits, and hence the number of parameters for the same circuit type with the same number of layers varies for data shown in different colors, the number of qubits for a given encoding is fixed; thus, for a given encoding the different circuit types can be compared directly. We observe a slight deviation from the power law behavior for states using amplitude encoding and the two-dimensional sequential circuit. This deviation may, however, be an artifact from our specific qubit layout in this case---a $4\times4$ grid with all qubits above the diagonal from bottom-left to top-right removed, leaving only the ten qubits needed. This layout leads to two isolated qubits at the corners that are only connected to a single nearest neighbor, possibly limiting the circuit's ability to fully exploit its two-dimensional structure. For the other encodings with more qubits the layout is more regular and such a deviation is not observed.

It seems a bit surprising at first that all three circuit types perform similarly well, given that in principle the MERA circuit and the two-dimensional sequential circuit are more powerful ansätze---the first being able to describe critical systems with a logarithmically increasing entanglement entropy~\cite{Vidal2008} and the second being able to capture an area law for the entanglement entropy in two dimensions~\cite{Wei2022, Zalatel2020}. The observation that these circuits do not perform significantly better suggests that the entanglement structure of quantum states encoding classical image data is adequately captured by MPS or one-dimensional sequential circuits. This is in agreement with the results we derived in Sec.~\ref{sec:image_compression_MPS}. Essentially, the two-dimensional structure of the image as classical data gets lost once it gets mapped to a quantum state; now the correlations between the qubits correspond to correlations between different length-scales~\cite{Lubasch2018, Garcia-Ripoll2021}, which are well represented by a one-dimensional circuit ansatz.

The algebraic decay of the infidelities in Fig.~\ref{fig:comparison_approx_err} suggests that the number of parameters in the circuit takes over a role similar to that of the bond dimension in the MPS approximation. Consider the case of one-dimensional sequential circuits as a specific example: representing the states generated by arbitrary sequential circuits as an MPS requires the bond dimension of the MPS to grow exponentially with the number of layers in the circuit~\cite{Barratt2021, Lin2021}. However, the inverse relation---how many layers of a sequential circuit are needed to represent an arbitrary MPS as a function of its bond dimension---is generally not known. A naive inversion of the bond dimension scaling would suggest that $\bigO{\log\chi}$ layers of a sequential circuit might suffice to represent a bond-dimension-$\chi$ MPS. However, such an exponential reduction in the number of parameters is only realized in very specific settings, such as time evolution with local one-dimensional Hamiltonians~\cite{Lin2021, Astrakhantsev2023}. The more common scenario is that the number of parameters in the circuit scales as a small power of the MPS bond dimension instead~\cite{Iaconis2023, Bravo-Prieto2020, Haghshenas2022, Jobst2022}. In the case of image compression considered here, we observe that both the number of parameters in the circuit (which is proportional to the number of layers) and the bond dimension of the MPS approximation need to grow algebraically with decreasing approximation error. This suggests that also in this case the number of parameters in the circuit scales algebraically with the MPS bond dimension, rather than logarithmically as the naive expectation predicts. The available data for the circuit approximation is limited to both shallow circuits and low-resolution images; to draw quantitative conclusions about the scaling, in particular to see whether it can be related to the error bounds derived in the previous section, data for circuits with more parameters are needed---corresponding to both deeper circuits and circuits representing higher-resolution images. We plan to explore this scaling with deeper circuits and higher-resolution images in future work.

\section{\label{sec:conclusion}Conclusion}
We have considered three different encodings for mapping two-dimensional classical data to a quantum state, i.e., amplitude encoding, the FRQI and the NEQR. If one can view the data as being equidistant samples from a function with continuous inputs whose Fourier coefficients decay faster than $\bigO{|k|^{-1}|\ell|^{-1}}$, we have shown that the states obtained from amplitude encoding or the FRQI can be efficiently approximated by MPS, in the sense that the bond dimension does not scale asymptotically with the size of the input data. This MPS-representation directly leads to a circuit with a number of gates that, in the worst case, scales as $\bigO{n/\epsilon^4}$, i.e., only linearly in the number of qubits. The exponent of the scaling with the approximation error becomes smaller if the Fourier coefficients decay faster. We have explained why for the NEQR such an MPS approximation is not as efficient, however, numerically we found it to still work better than for generic states. Moreover, we have considered one- and two-dimensional sparse sequential circuits as well as a MERA circuit as variational ansätze to compress quantum states representing classical image data, and found that even though the other circuit ansätze are more powerful in principle, the one-dimensional circuits are capable of achieving the same fidelities with the same number of parameters. This further suggests that the relevant entanglement structure of these states is captured by MPS.

Our results prove a slightly modified version of the conjecture in Refs.~\cite{Garcia-Ripoll2021, Holmes2020}. Ref.~\cite{Garcia-Ripoll2021} showed that when encoding a probability distribution with a bounded derivative in the amplitudes of an $n$-qubit state, adding a single qubit in order to double the resolution only introduces an entanglement entropy of the order $\bigO{2^{-n/2}}$ between the newly added qubit and the previous qubits. Based on this result, and numerical evidence in Refs.~\cite{Garcia-Ripoll2021, Holmes2020}, it was conjectured that also higher-dimensional real-valued functions with a bounded derivative can be efficiently approximated by MPS. Here we proved this for two-dimensional complex-valued functions with sufficiently quickly decaying Fourier coefficients, which also includes functions where the derivative is not bounded. Our results also give theoretical justification for the observation in Ref.~\cite{Dilip2022}, that FRQI states can be well approximated by MPS with a small bond dimension and sequential circuits with only a few layers.

An open question remains how to best obtain the MPS approximation in practice. Using the SVD yields a high-fidelity MPS approximation with a cost of $\bigO{\chi 2^{2n}}$, which is slightly better than the classical preprocessing with cost $\bigO{n2^{2n}}$ needed for finding the optimal circuit that exactly prepares the state~\cite{Amankwah2022} (while also yielding a circuit with exponentially fewer gates); however, the scaling is still exponential in the number of qubits. A full reduction from the exponential scaling seems unlikely if all exponentially many pixels are supposed to be processed to find an approximation. In principle, it might be possible to circumvent the exponential scaling either via stochastic methods like the sparse Fourier transform~\cite{Gilbert2005, Ghazi2013, Gilbert2014} or stochastic SVDs~\cite{Krasulina1969, Shamir2015, Shamir2016}, or via interpolation methods like the tensor cross interpolation~\cite{Savostyanov2011, Savostyanov2014, Dolgov2020, Nunez-Fernandez2022, Ritter2024}. In practice, the classical data we are dealing with still needs to be stored on a classical computer, so the size of the input data cannot be intractably large. For those cases, the usual SVD could suffice for finding good linear-depth circuit approximations, and there might still be a relevant speedup in processing the dataset classically in this way first, before further processing the dataset on a quantum computer. Especially for NISQ devices, this could turn exponential-depth circuits that cannot be run on the quantum hardware into linear-depth circuits that can be run on the quantum hardware and thus allows to retain some of the advantages of a quantum computer.

The overhead of the MPS circuits can be further reduced by optimizing hardware efficient quantum circuits. Here, we have demonstrated the optimization by storing the full wave function classically and optimizing the circuit using a gradient-descent-based method, which limited us to small system sizes. In principle, making use of the MPS structure of the target state allows for a more efficient optimization on a classical computer, and one can explore the use of second-order optimization methods to speed up convergence. These methods would allow one to work with larger circuits, both in system size and the number of parameters, and study the scaling of the approximation error in this regime. This will be instrumental to better estimate the cost of the required circuits in a practical setting. We plan to explore those ideas in future work. Besides the circuits we considered here, several recent proposals have suggested ways to prepare certain MPS states more efficiently---either approximately in $\bigO{\log n}$-depth~\cite{Malz2024} or in constant depth using measurements and unitary feedback based on the measurement outcomes~\cite{Sahay2024a, Sahay2024b, Smith2024, Stephen2024}---which opens another interesting direction to explore.

Here, our focus has been on finding circuits for efficiently preparing quantum states that describe classical input data, but with a clear application to QML in mind. A natural question is thus if we can also find circuits that efficiently classify input data that was prepared in this way? At least for small images, Ref.~\cite{Dilip2022} found that using the same sequential circuit structure for learning the classification task as for the data preparation works well, but the question whether this observation scales to larger system sizes remains open. The successful application of MPS-based algorithms to classical machine learning tasks suggests that the MPS-inspired sequential circuits can also scale to larger systems, however, the types of image encodings used in these applications usually differ significantly from the encodings considered here~\cite{Stoudenmire2017, Bengua2017, Huggins2019, Efthymiou2019}. Other popular circuits for data classification use the structure of a MERA circuit (and are also often called QCNNs in this context)~\cite{Cong2019, Grant2018, Liu2023}. As discussed previously, MERA circuits can in principle be more powerful than one-dimensional sequential circuits in terms of the entanglement entropy they can create; however, they require the implementation of long-range two-qubit gates, and thus the best suited circuit for a classification task may also depend on the underlying quantum hardware. An advantage that both sequential circuits (up to a logarithmic number of layers) and MERA circuits share over the more conventional linear-depth brick wall circuits is that, for local cost functions, they do not suffer from barren plateaus~\cite{McClean2018, Pesah2021, Cervero-Martin2023, Zhang2024}.

\begin{acknowledgments}
    The authors thank Yu-Jie Liu, Michael Lubasch and Adam Gammon-Smith for insightful discussions, and Lukas Lechner for generously granting permission to use their photograph in Fig.~\ref{fig:example_compression}. F.P. thanks Rohit Dilip, Yu-Jie Liu and Adam Gammon-Smith for collaboration on a related previous project.
    The optimization of the variational quantum circuits was implemented using the QGOpt library~\cite{qgopt, Luchnikov2021, Oza2009} and TensorFlow~\cite{tensorflow}.
    This research was funded by the BMW Group. C.A.R. and E.S. are partly funded by the German Ministry for Education and Research (BMB+F) in the Project QAI2-Q-KIS under Grant 13N15583. F.P. acknowledges the support of the Deutsche Forschungsgemeinschaft (DFG, German Research Foundation) under Germany's Excellence Strategy EXC-2111-390814868, the European Research Council (ERC) under the European Union's Horizon 2020 research and innovation program (grant agreement No. 771537), as well as the Munich Quantum Valley, which is supported by the Bavarian state government with funds from the Hightech Agenda Bayern Plus.
    
    \textbf{Data and materials availability:} Data analysis and simulation codes are available on Zenodo upon reasonable request~\cite{Zenodo}.
\end{acknowledgments}
\onecolumngrid
\clearpage

\begin{figure*}[t]
    \makebox[\linewidth][c]{%
        \includegraphics[width=1.03\linewidth]{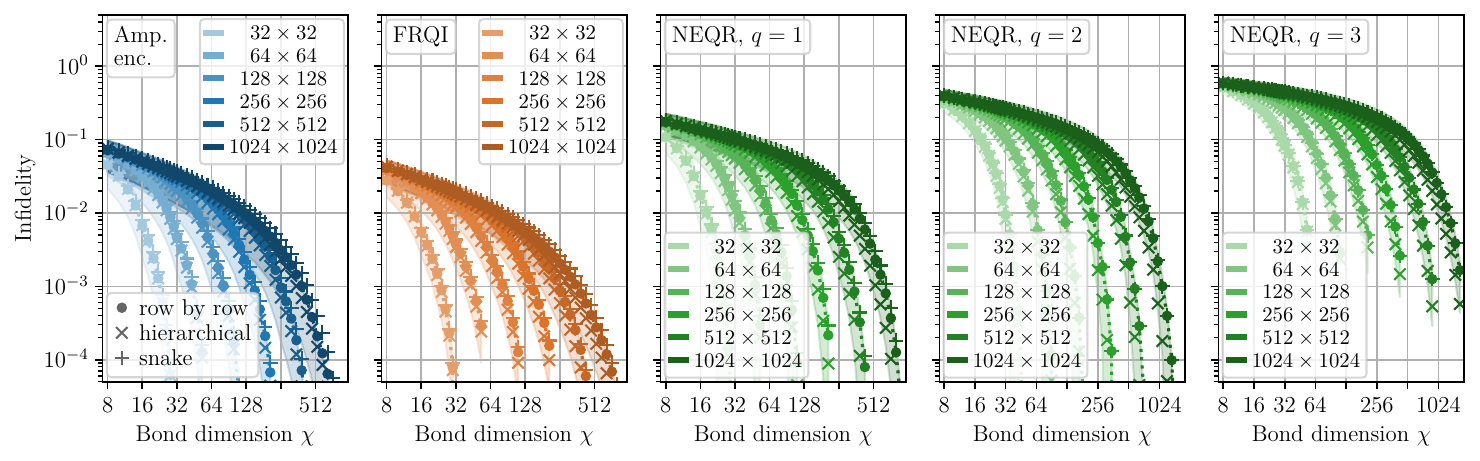}\hspace{0.5cm}}
    \caption{\label{fig:scaling_chi_resolution}%
    \textbf{Scaling of the infidelity of the MPS approximation with increasing bond dimension for different image resolutions.} The infidelity is defined as $1 - |\langle\psi_{\text{exact}}|\psi_{\chi}\rangle|^2$, where $\ket{\psi_{\text{exact}}}$ is the exact state encoding the image and $\ket{\psi_{\chi}}$ is its bond-dimension-$\chi$ approximation. We use the same $1058$ images from the `Imagenette'~\cite{imagenette} and DIV2K~\cite{DIV2K} datasets as in the main text. Each of the five subplots corresponds to one of the encodings, i.e., amplitude encoding, the FRQI or the NEQR with $1$--$3$ color qubits. The different color shades correspond to the different image resolutions. The marker shapes---dots, crosses and pluses---indicate the different indexing variants---respectively, row-by-row indexing, hierarchical indexing and snake indexing (see also Fig.~\ref{fig:pixel_indexing}). The shaded areas show the $25$th--$75$th percentiles for the different encodings using row-by-row indexing. For small bond dimensions, the infidelities seem to follow a straight line independent of the image resolution.}
\end{figure*}
\twocolumngrid
\vspace*{-1.4cm}
\appendix
\section{\label{app:numerics}Additional numerical data}
In this section we present some additional numerical data for the MPS approximation of image states. For all numerical results presented here we use the same $1058$ images from the `Imagenette'~\cite{imagenette} and DIV2K~\cite{DIV2K} datasets as in the main text. In Fig.~\ref{fig:scaling_resolution} in Sec.~\ref{sec:image_compression_MPS}, we only showed the scaling of the infidelity with increasing image resolution for a fixed bond dimension. Here, we also show data for the scaling with increasing bond dimension for different image resolutions. Additionally, we consider a different setting, where we do not scale up the size of the image by increasing the resolution, but where we instead change the size of the image by considering increasingly larger image sections.

\subsection{Scaling with bond dimension}
First, we consider the scaling of the infidelity with increasing bond dimension for different image resolutions. Fig.~\ref{fig:scaling_chi_resolution} shows the average infidelities of the bond-dimension-$\chi$ MPS approximations and the target states, for a range of bond dimensions and image resolutions. As in the main text, we consider different resolutions of the original $1024\times1024$-pixel images by retaining only pixels from every other row and column. The five subplots show the data for each of the five different image encodings we consider, i.e., amplitude encoding, the FRQI and the NEQR with $1$--$3$ color qubits. The different shades of color correspond to the different image resolutions. As before, the different marker shapes---dots, crosses and pluses---denote the different indexing variants---row-by-row, hierarchical and snake indexing. The shaded areas show the $25$th--$75$th percentiles using row-by-row indexing. Similar to Fig.~\ref{fig:scaling_resolution} in the main text, we can see that in terms of the infidelity, the FRQI performs best, showing slightly smaller infidelities than amplitude encoding, while the NEQR performs worst, with the infidelity seemingly further increasing for every color qubit added.

\begin{figure*}[t]
    \begin{tabular}{lclc}
        {\sffamily(a)}\hspace{-1em} && {\sffamily(b)}\hspace{-2em} & \\[-1.5em]
        &  \includegraphics[width=0.45\linewidth]{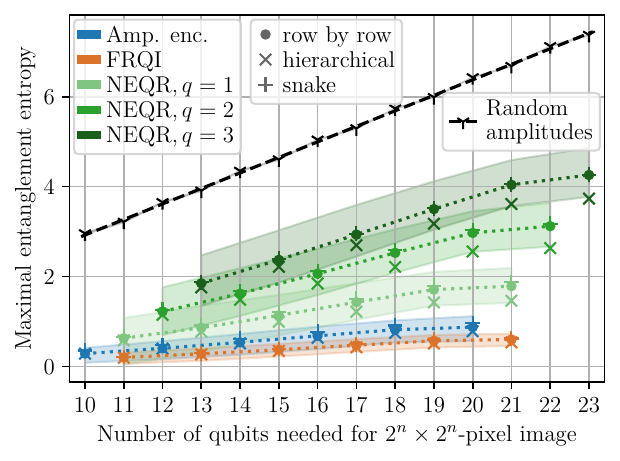}%
        && \includegraphics[width=0.45\linewidth]{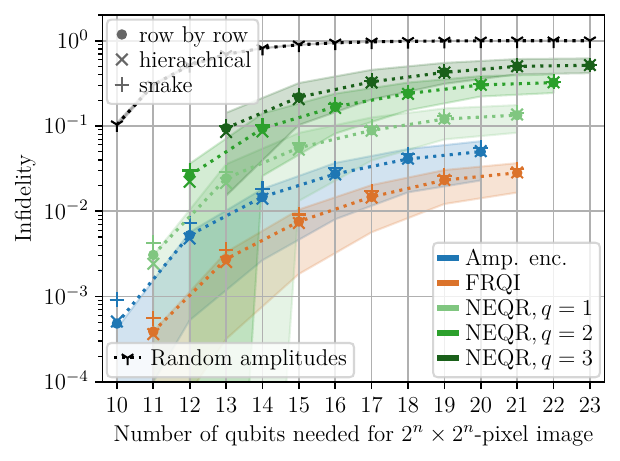}%
    \end{tabular}
    \caption{\label{fig:scaling_zoom}%
    \textbf{Scaling of the entanglement entropy and the approximation error with increasing image section.} In contrast to Fig.~\ref{fig:scaling_resolution} in the main text, here we scale the image size by taking increasingly larger subsections of the original image instead of changing the resolution. The images used are the same $1058$ images from the `Imagenette'~\cite{imagenette} and DIV2K~\cite{DIV2K} datasets as in the main text. (a)~The plot shows the scaling of the maximal entanglement entropy over any bipartition of the state into two contiguous halves. Blue markers show the results using amplitude encoding, orange markers the results using the FRQI, and the different shades of green show the results using the NEQR with $1$--$3$ color qubits. Note that the three different encodings may use a different number of qubits, so that data points corresponding to same-size image subsections do not necessarily align vertically. The marker shapes---dots, crosses and pluses---indicate the different indexing variants---respectively, row-by-row indexing, hierarchical indexing and snake indexing (see also Fig.~\ref{fig:pixel_indexing}). The shaded areas show the $25$th--$75$th percentiles for the different encodings using row-by-row indexing. For reference, we also show the average half-chain entanglement entropy for $200$ states whose amplitudes were drawn from a normal distribution before normalization as black three-pointed stars, the black shaded area shows the $25$th--$75$th percentiles. The black dashed line is a linear fit to the data, showing the generic growth of entanglement with system size~\cite{Page1993}. (b)~The plot shows the infidelity (defined as $1 - |\langle\psi_{\text{exact}}|\psi_{\chi=16}\rangle|^2$) between the exact encoded state and its bond dimension $\chi=16$ approximation. The colors and marker shapes denote the same encodings as before.}
\end{figure*}

Two distinctive features of the MPS approximation are clearly visible in the figure. First, once the image resolution is large enough (i.e., starting with around $64\times64$ pixels), the infidelities for small bond dimensions all seem to fall onto the same line. This is consistent with our expectation from the analytical error bounds. There is an error bound for the MPS representation of the `infinitely high-resolution' image that depends on the bond dimension, and for a resolution with finitely many pixels the finite-size corrections quickly decay with the number of pixels. So, apart from small finite-size effects, we expect the approximations for different image resolutions with the same bond dimension to have the same infidelity; this is indeed what we observe for small bond dimensions in the numerical results. For large bond dimensions, another effect takes over. In the analytic derivation, we only considered asymptotic decays of Fourier coefficients to get the error bounds. The actual approximation error, however, is given by the sum of the (squared) discarded Schmidt values. Since for a state on a finite number of qubits, the number of Schmidt values is finite, the approximation error must go to zero if all Schmidt values are kept. If the sum of discarded Schmidt values decays slowly at first, then it must decay very quickly to zero once one discards only the final few Schmidt values. In Fig.~\ref{fig:scaling_chi_resolution} this shows up as rapidly decaying tails as the bond dimension gets closer to its maximal value.

The second feature visible in Fig.~\ref{fig:scaling_chi_resolution}, is that in the low bond dimension regime, where the infidelity is almost independent of image resolution, the decay of the infidelities approximately follows a power-law (i.e., is linear on the log-log scale in Fig.~\ref{fig:scaling_chi_resolution}). This again is consistent with the expectation from the analytical bounds. However, since we use the SVD here to obtain the MPS representation and not the Fourier transform, the slope will not necessarily be the same as the analytical prediction. The SVD can find a more optimal basis for the truncation compared to the Fourier transform, since the SVD adaptively chooses a basis based on the input state, while the Fourier transform always uses the same basis for the truncation, agnostic to the input state. Therefore, the SVD will typically give much better MPS approximations with a smaller infidelity.

\subsection{Scaling with image sections}
So far we have considered only the scaling to larger system sizes by increasing the resolution of an image, which seems a natural setting for talking about different sizes of the \emph{same} image. In this case, one can make statements about how the cost of the state preparation has to scale with increasing system size, both numerically and analytically, as we have done in the main text. This also could be a practically relevant situation, as one could for example have a high-resolution image on that one wants to perform some computational task. One might then try out the task on a smaller, downscaled version of the image first, to see how well the algorithm performs while investing limited computational resources. Afterwards, one might wonder about how the computational effort increases if one uses the original high-resolution image instead.

\begin{figure*}[t]
    \makebox[\linewidth][c]{%
        \includegraphics[width=1.03\linewidth]{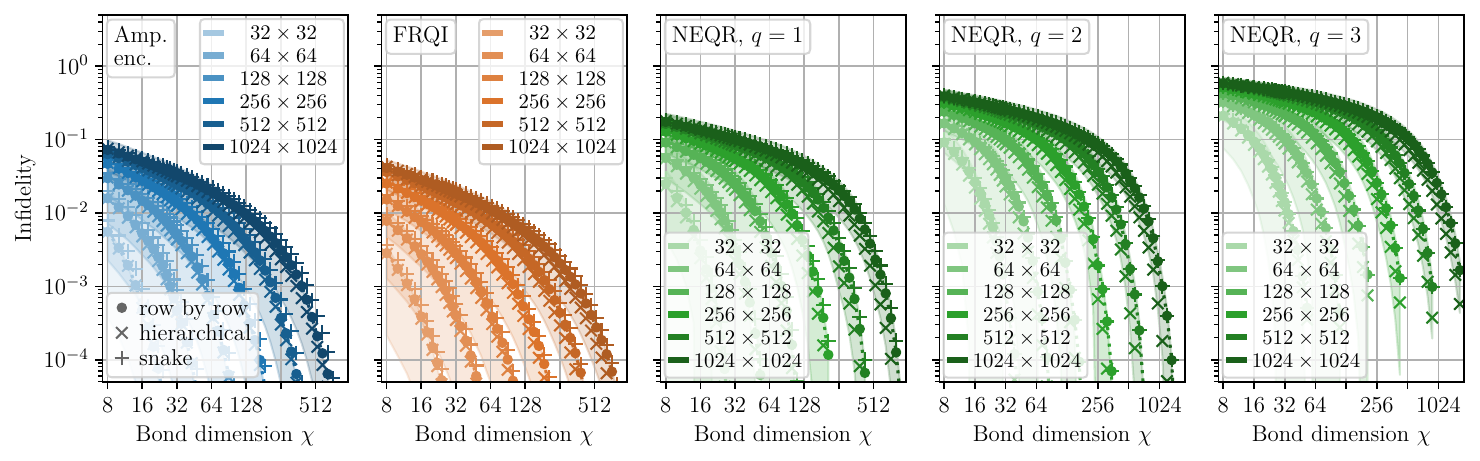}\hspace{0.5cm}}
    \caption{\label{fig:scaling_chi_zoom}%
    \textbf{Scaling of the infidelity of the MPS approximation with increasing bond dimension for differently sized image subsections.} The infidelity is defined as $1 - |\langle\psi_{\text{exact}}|\psi_{\chi}\rangle|^2$, where $\ket{\psi_{\text{exact}}}$ is the exact state encoding the image and $\ket{\psi_{\chi}}$ is its bond-dimension-$\chi$ approximation. We use the same $1058$ images from the `Imagenette'~\cite{imagenette} and DIV2K~\cite{DIV2K} datasets as in the main text, but instead of scaling the resolution of the images, we take increasingly larger subsections of the original image to increase the system size. Each of the five subplots corresponds to one of the encodings, i.e., amplitude encoding, the FRQI or the NEQR with $1$--$3$ color qubits. The different color shades correspond to the different image resolutions. The marker shapes---dots, crosses and pluses---indicate the different indexing variants---respectively, row-by-row indexing, hierarchical indexing and snake indexing (see also Fig.~\ref{fig:pixel_indexing}). The shaded areas show the $25$th--$75$th percentiles for the different encodings using row-by-row indexing. In contrast to scaling the resolution of the image, even for small bond dimensions the infidelities depend on the number of pixels; increasing the number of pixels worsens the infidelity.}
\end{figure*}

However, this is not the only way to increase the image size. Instead one could also zoom out of the original image, and increase the image size by considering a larger field of view than the original image. Then, in principle, each increase in the number of pixels can introduce a lot of new information to the image, and so to keep a good approximation of the image also the amount information kept in the approximation needs to grow. In practice, this case might be less relevant, as for a given image one usually lacks the information about what surrounds the image segment, whereas a given low-resolution image often already contains a lot of the relevant information compared to a high-resolution one. Still, to demonstrate that these two scaling cases are distinct, we present some numerical data here.

For the numerical results, we again use the same $1058$ images from the `Imagenette'~\cite{imagenette} and DIV2K~\cite{DIV2K} datasets as in the main text. To get the different image sizes, we start from the $1024\times1024$-pixel image and only take the central square section of $2^n\times2^n$ pixels of the image, with $n$ going from five to ten. (For $n=10$ this means taking the whole image.) In this way, we get increasingly larger images, which we can then encode in quantum states according to the encodings discussed in Sec.~\ref{sec:image_encodings}. A side-effect of considering these cropped image sections is that most of the time parts of the subject in the image are cut off, and so they are not necessarily in a form anymore as one would typically encounter them in a more natural setting.

The scaling of the entanglement entropy and the approximation error of a bond dimension $\chi=16$ MPS approximation are shown in Fig.~\ref{fig:scaling_zoom}. It is the analogue of the scaling with image resolution shown in the main text in Fig.~\ref{fig:scaling_resolution}, but now scaling the size of the image segment instead. The different colors denote the different encodings, the different marker shapes the different indexing variants and the shaded area shows the $25$th--$75$th percentiles for the row-by-row indexing. The black data shows results for states whose amplitudes are drawn from a normal distribution before normalization. In contrast to the scaling seen in the main text, here we do not see any saturation of the entanglement entropy or the infidelity for any of the image encodings. Since a lot of new information is added with increasing system size, also the bond dimension needs to be increased to keep the approximation quality the same. However, the growth of the entanglement entropy (or the decrease in infidelity) is still slower than that of random states. In the figure, a change in the rate of growth of the entanglement entropy can be seen when reaching the final image size for all image encodings. This can probably be attributed to the fact that for the reduced image sizes some features of the image are cut off, which can introduce sharp features (like discontinuous jumps) that decrease the decay of the Fourier spectrum and make the approximation harder, while much of the smoothness and regularity is expected to be recovered once the whole image is visible.

Finally, we can also study the scaling of the infidelity with increasing bond dimension in the case of scaling the size of the image segment instead of the image resolution. The results are shown in Fig.~\ref{fig:scaling_chi_zoom}. The subplots show the results for the different encodings, i.e., amplitude encoding, the FRQI and the NEQR with $1$--$3$ color qubits; the different marker shapes denote the different indexing variants, i.e., row-by-row, hierarchical and snake indexing; and the different shades of color indicate the size of the image segment. The shaded area shows the $25$th--$75$th percentiles using the row-by-row indexing. In contrast to the scaling of the image resolution, now the decay of the infidelity with increasing bond dimension strongly depends on the size of the image. For a given bond dimension, the infidelity now increases with the image size, while the infidelity for a fixed (small) bond dimension approximation remained roughly the same when increasing the image resolution. Moreover, the decay of the infidelity with increasing bond dimension now becomes slower for larger images, while it remained roughly the same when increasing the image resolution.
\clearpage

\section{\label{app:2d_fourier_to_mps}Constructing an MPS from two-dimensional Fourier modes}
In the main text we claimed that a state whose amplitudes are given by a two-dimensional Fourier transform containing only a $\chi\times\chi$ matrix of Fourier coefficients can be written as an MPS with bond dimension $\chi$. Here we prove this claim.

We consider a $2n$-qubit state whose amplitudes (up to normalization) are given by the $2^n\times2^n$ matrix $f_{ab}$ as
\begin{equation}
    \ket{\psi} \propto \sum_{a,b=0}^{2^n-1} f_{ab}\,\ket{b}\,\ket{a},
\end{equation}
where the basis states $\ket{b}$ live on the first register of $n$ qubits and label the $y$-coordinate of the image, and the basis states $\ket{a}$ live on the second register of $n$ qubits and label the $x$-coordinate. The matrix $f_{ab}$, by assumption, can be written in terms of a $\chi\times\chi$ matrix of Fourier coefficients $\hat{f}_{pq}$ as
\begin{equation}
    f_{ab} = \sum_{p,q=0}^{\chi-1} \hat{f}_{pq}\;e^{i2\pi pa/2^n}\,e^{i2\pi qb/2^n}.
\end{equation}
Expressing the two summation indices in terms of $n$-bit binary integers, ${a=\sum_{j=0}^{n-1}2^{n-1-j}\sigma_j}$ and ${b=\sum_{j=0}^{n-1}2^{n-1-j}\sigma'_j}$, we have
\begin{equation}
\begin{aligned}
    \ket{\psi} &\propto \sum_{a,b=0}^{2^n-1} \sum_{p,q=0}^{\chi-1} \hat{f}_{pq}\;e^{i2\pi pa/2^n}\,e^{i2\pi qb/2^n} \ket{b}\ket{a}\\
    &= \sum_{\{\sigma_j\}}\sum_{\{\sigma'_j\}}\sum_{p,q=0}^{\chi-1} \left(\prod_{j=0}^{n-1} e^{i2\pi q\sigma_j/2^{j+1}}\right)\\
    &\qquad\times\hat{f}_{pq}\left(\prod_{j'=0}^{n-1} e^{i2\pi p\sigma'_{j'}/2^{j'+1}}\right)\\
    &\qquad\times\ket{\sigma_0,\sigma_1,\ldots,\sigma_{n-1}} \ket{\sigma'_0,\sigma'_1,\ldots,\sigma'_{n-1}}.
\end{aligned}
\end{equation}
This defines the following set of MPS tensors: for the first $n$ qubits ($0\leq j \leq n-1$) we have
\begin{equation}
    A^{[j]\sigma_j}_{\alpha_j\alpha_{j+1}}
    = \delta_{\alpha_j,\alpha_{j+1}} e^{i2\pi\alpha_{j+1}\sigma_j/2^{j+1}},
\end{equation}
the states in the two $n$-qubit registers are connected by the matrix
\begin{equation}
    C_{\alpha_{n-1}\beta_0} =
    \sum_{p,q=0}^{\chi-1}\delta_{\alpha_{n-1},q}\,\hat{f}_{pq}\,\delta_{p,\beta_0},
\end{equation}
and for the second $n$ qubits we have
\begin{equation}
    B^{[n+j]\sigma'_j}_{\beta_{j}, \beta_{j+1}} = \delta_{\beta_j,\beta_{j+1}} e^{i2\pi\beta_{j+1}\sigma'_j/2^{j+1}};
\end{equation}
the indices $\alpha_0$ and $\beta_n$ are dummy indices that take only the value $0$, the remaining indices all run from $0$ to $\chi-1$. The amplitudes $\psi_{\sigma_0,\ldots,\sigma_{n-1},\sigma'_0,\ldots,\sigma'_{n-1}}$ of the state (up to normalization) can then be written as an MPS with bond dimension $\chi$ as
\begin{equation}
\begin{aligned}
    &\psi_{\sigma_0,\ldots,\sigma_{n-1},\sigma'_0,\ldots,\sigma'_{n-1}} \propto\\
    &\hspace{1.9em} A^{[0]\sigma_0}\cdots A^{[n-1]\sigma_{n-1}}CB^{[n]\sigma'_0}\cdots B^{[2n-1]\sigma'_{n-1}},\hspace{-1em}
    \label{eq:MPS_amp_enc}
\end{aligned}
\end{equation}
where we omit the virtual indices of the tensors and the matrix product is implied.%
\footnote{
    In the tensor network literature the symbols $A$ and $B$ are often used to denote left- or right-isometric tensors. Here, we use the different symbols purely to differentiate the two halves of the system, and the tensors are not in left- or right-isometric form.
}
See also Fig.~\ref{fig:MPS}a for a diagrammatic representation.

This construction also works if there is not a symmetric cutoff $\Lambda$ in both directions, or if the contributing Fourier coefficients do not form a contiguous region of frequencies---as long as only a finite number of Fourier coefficients $\hat{f}_{pq}$ contribute. The bond dimension $\chi_l$ of the MPS in the left half of the system, i.e., the $\sigma_j$ qubit register, is the number of values $q$ for which there exists a nonzero $\hat{f}_{pq}$. The bond dimension $\chi_r$ of the MPS in the right half of the system, i.e., the $\sigma'_j$ qubit register, is the number of values $p$ for which there exists a nonzero $\hat{f}_{pq}$.

\section{\label{app:proof_appr_err_bounds}Proof of the approximation error bounds}
In this section we give the detailed calculations for the error bounds that were presented in the main text. Afterwards, we show some numerical examples to discuss the tightness of the derived bounds.

First, we need to prove Eq.~\eqref{eq:approx_err_normalized} in the main text; given two quantum states
\begin{equation}
    \ket{f} = \frac{1}{\norm{f}_F} \sum_{a,b=0}^{2^n-1} f_{ab} \ket{b}\ket{a}
\end{equation}
and
\begin{equation}
    \ket{g} = \frac{1}{\norm{g}_F} \sum_{a,b=0}^{2^n-1} g_{ab} \ket{b}\ket{a},
\end{equation}
where $\norm{f}_F = \sqrt{\sum_{a,b=0}^{2^n-1} |f_{ab}|^2}$ denotes the Frobenius norm of the matrix $f$, the norm of the difference between the two states can be written as
\begin{equation}
\begin{aligned}
    &\norm{\ket{f} - \ket{g}}_2=\\
    &\quad= \norm{\sum_{a,b=0}^{2^n-1} \left(\frac{f_{ab}}{\norm{f}_F} - \frac{g_{ab}}{\norm{g}_F}\right) \ket{b}\ket{a}}_2\\
    &\quad= \norm{\frac{f}{\norm{f}_F} - \frac{g}{\norm{g}_F}}_F\\
    &\quad= \norm{\frac{f \norm{g}_F - g \norm{f}_F}{\norm{f}_F \norm{g}_F}}_F\\
    &\quad= \norm{\frac{f (\norm{g}_F - \norm{f}_F) + (f - g) \norm{f}_F}{\norm{f}_F \norm{g}_F}}_F\\
    &\quad\leq \frac{\norm{f}_F \big|\norm{g}_F - \norm{f}_F\big| + \norm{f - g}_F \norm{f}_F}{\norm{f}_F \norm{g}_F}\\
    &\quad\leq \frac{2\norm{f-g}_F}{\norm{g}_F}.
    \label{eq:app_approx_err_normalized_deriv}
\end{aligned}
\end{equation}
To get to the last line we used the reverse triangle inequality $\big|\norm{x} - \norm{y}\big| \leq \norm{x-y}$, which is a consequence of the usual triangle inequality.

\subsection{\label{app:proof_appr_err_bounds_exp}Exponentially decaying Fourier coefficients}
Here, we will assume an exponential decay of the Fourier coefficients $\hat{F}(k,\ell)$ introduced in Eq.~\eqref{eq:Fourier_series}. [Remember that $\hat{F}(k,\ell)$ are the Fourier coefficients of the function $F(x,y)$ which takes continuous inputs $x$ and $y$, from which the pixelated data is obtained via sampling at discrete points, and not the DFT coefficients $\hat{f}_{pq}$ of the pixelated data $f_{ab}$ directly.] First, we will relate them to the Fourier coefficients of a $2^n\times2^n$-pixel image. Then, we can calculate the error we make by only including Fourier coefficients up to some cutoff $\Lambda$, and see how this error scales with the number of pixels $2^{2n}$.

Consider Fourier coefficients that fulfill
\begin{equation}
    |\hat{F}(k,\ell)| \leq C e^{-\alpha|k|} e^{-\beta|\ell|},
    \label{eq:app_Fourier_exp}
\end{equation}
with some constants $C,\alpha,\beta>0$.%
\textsuperscript{\textit{\ref{fn:fourier_within_cutoff}}}
\begin{widetext}
\onecolumngrid
\noindent
Using Eq.~\eqref{eq:relating_DFT_and_FT_coeff}, we can bound the absolute value of the Fourier coefficients of a $2^n\times2^n$-pixel image, which yields\vspace{-2pt}
\begin{equation}
\begin{aligned}
    |\hat{f}_{pq}| &\leq \sum_{i,j\in\Z} |\hat{F}(p+i2^n,q+j2^n)|
    \leq C \left(\sum_{i\in\Z} e^{-\alpha|p+i2^n|}\right) \left(\sum_{j\in\Z} e^{-\beta|q+j2^n|}\right)\\[-0.5em]
    &= C \left(e^{-\alpha |p|} + (e^{\alpha p} + e^{-\alpha p}) \sum_{i=1}^{\infty} e^{-\alpha 2^n i} \right)
    \left(e^{-\beta |q|} + (e^{\beta q} + e^{-\beta q}) \sum_{j=1}^{\infty} e^{-\beta 2^n j} \right)\\
    &= C \left(e^{-\alpha |p|} + (e^{\alpha p} + e^{-\alpha p}) \frac{e^{-\alpha 2^n}}{1 - e^{-\alpha 2^n}} \right)
    \left(e^{-\beta |q|} + (e^{\beta q} + e^{-\beta q}) \frac{e^{-\beta 2^n}}{1 - e^{-\beta 2^n}} \right).
    \label{eq:app_bound_fpq_exp}
\end{aligned}
\end{equation}%
\footnotetext{\label{fn:fourier_within_cutoff}
    This bound on the Fourier coefficients is actually only important for frequencies larger than the cutoff $\Lambda$. The same error bounds hold for any structure of the Fourier coefficients within the cutoff, as the approximation error is given by the sum over the discarded Fourier coefficients and so the coefficients within the cutoff do not contribute.
}

To calculate the error in Eq.~\eqref{eq:approx_err_normalized_fourier} we need to evaluate $\|\hat{f}-\hat{g}\|_F$ as given in Eq.~\eqref{eq:norm_diff}, which sums over the set of discarded Fourier coefficients. We can write this sum over $I_{\text{disc}}$ as the difference of sums over the set of all Fourier coefficients $I_{\text{all}}$ and the set over the kept Fourier coefficients $I_{\text{appr}}$,
\begin{equation}
    \sum_{(p,q)\in I_{\text{disc}}} \!\!|\hat{f}_{pq}|^2 = \!\sum_{(p,q)\in I_{\text{all}}} \!|\hat{f}_{pq}|^2 \, - \hspace{-0.75em}\sum_{(p,q)\in I_{\text{appr}}} \!\!|\hat{f}_{pq}|^2,
    \label{eq:app_sum_disc_fpq}
\end{equation}
and then insert the result from Eq.~\eqref{eq:app_bound_fpq_exp} for $|\hat{f}_{pq}|$. From Eq.~\eqref{eq:app_bound_fpq_exp} we have for the square of $|\hat{f}_{pq}|$ that
\begin{equation}
\begin{aligned}
    |\hat{f}_{pq}|^2
    \leq C^2 &\left(e^{-2\alpha |p|} + 2\,\frac{1 + e^{-2\alpha |p|}}{e^{\alpha 2^n} - 1}
    + \frac{2 + e^{2\alpha p} + e^{-2\alpha p}}{(e^{\alpha 2^n}-1)^2} \right)\\
    \times&\left(e^{-2\beta |p|} + 2\,\frac{1 + e^{-2\beta |p|}}{e^{\beta 2^n} - 1}
    + \frac{2 + e^{2\beta p} + e^{-2\beta p}}{(e^{\beta 2^n}-1)^2} \right).
\end{aligned}
\end{equation}

We first consider the sum over $I_{\text{all}}$ in Eq.~\eqref{eq:app_sum_disc_fpq}, and can perform the sums over $p$ and $q$ individually since $|\hat{f}_{pq}|^2$ factorizes. There are three terms that contribute to the sum over $p$---for the first term we have
\begin{equation}
\begin{aligned}
    \sum_{p=-2^n\!/2}^{2^n\!/2-1} e^{-2\alpha |p|} &= (1+e^{-2\alpha}) \sum_{p=0}^{2^n\!/2-1} e^{-2\alpha p}
    = (1+e^{-2\alpha}) \frac{1-e^{-2\alpha2^n\!/2}}{1-e^{-2\alpha}}
    = \coth(\alpha) \, \big(1 - e^{-\alpha2^n}\big),
\end{aligned}
\end{equation}
where we have used the geometric sum; for the second term we have
\begin{equation}
\begin{aligned}
    \sum_{p=-2^n\!/2}^{2^n\!/2-1} \! 2\,\frac{1 + e^{-2\alpha |p|}}{e^{\alpha 2^n} - 1}
    = 2\,\frac{2^n + \coth(\alpha) \, \big(1 - e^{-\alpha2^n}\big)}{e^{\alpha 2^n} - 1}
    = 2\,\frac{2^n}{e^{\alpha 2^n} - 1} + 2\coth(\alpha)\,e^{-\alpha2^n},
\end{aligned}
\end{equation}
where we could reuse the result from the first term for summing up $e^{-2\alpha|p|}$;
and, finally, for the third term we have
\begin{equation}
\begin{aligned}
    \sum_{p=-2^n\!/2}^{2^n\!/2-1} \frac{2+e^{2\alpha p}+e^{-2\alpha p}}{(e^{\alpha 2^n}-1)^2}
    &= 2\,\frac{2^n}{(e^{\alpha 2^n}-1)^2}
    + \frac{\coth(-\alpha) \, \big(1 - e^{\alpha2^n}\big)}{(e^{\alpha 2^n}-1)^2}
    + \frac{\coth(\alpha) \, \big(1 - e^{-\alpha2^n}\big)}{(e^{\alpha 2^n}-1)^2}\\
    &= 2\,\frac{2^n + \coth(\alpha)\sinh(\alpha2^n)}{(e^{\alpha 2^n}-1)^2},
\end{aligned}
\end{equation}
where we could again reuse the result from the first term for summing up $e^{\pm2\alpha|p|}$. Putting everything together, the result of the sum over $p$ is
\begin{equation}
\begin{aligned}
    \sum_{p=-2^n\!/2}^{2^n\!/2-1}\!
    &e^{-2\alpha|p|}
    + 2\,\frac{1+e^{-2\alpha|p|}}{e^{\alpha 2^n}-1}
    + \frac{2+e^{2\alpha p}+e^{-2\alpha p}}{(e^{\alpha 2^n}-1)^2}=\\
    &= \coth(\alpha) \, \big(1 - e^{-\alpha2^n}\big)
    + 2\,\frac{2^n}{e^{\alpha 2^n} - 1}
    + 2\coth(\alpha) e^{-\alpha2^n}\!
    + 2\,\frac{2^n + \coth(\alpha)\sinh(\alpha2^n)}{(e^{\alpha 2^n}-1)^2}\\
    &= \coth(\alpha) \, \bigg(1 - e^{-\alpha2^n} + 2\,e^{-\alpha2^n} + \frac{\sinh(\alpha2^n)}{(e^{\alpha 2^n}-1)^2}\bigg)
    + 2\,\frac{2^n(e^{\alpha 2^n} - 1)}{(e^{\alpha 2^n} - 1)^2} + 2\,\frac{2^n}{(e^{\alpha 2^n}-1)^2}\\
    &= \coth(\alpha)
    \bigg(1+e^{-\alpha2^n}\! + \frac{\sinh(\alpha2^n)}{(e^{\alpha 2^n}-1)^2}\bigg)
    + \frac{2^n\!/2}{\sinh(\frac{\alpha}{2} 2^n)^2}.
    \label{eq:app_intermediate_sum_2^n_exp}
\end{aligned}
\end{equation}
The results for the sum over $q$ can be obtained in the same way by replacing $\alpha$ with $\beta$.

Now we turn to the sum over $I_{\text{appr}}$; we can again perform the sums over $p$ and $q$ separately, and three terms contribute. The calculation goes analogously to before, only the summation range changes. Defining $\chi=2\Lambda+1$, for the first term we have
\begin{equation}
\begin{aligned}
    \sum_{p=-\Lambda}^{\Lambda} e^{-2\alpha |p|}
    = 1 + 2e^{-2\alpha} \sum_{p=0}^{\Lambda-1} e^{-2\alpha p}
    = 1 + 2e^{-2\alpha} \frac{1-e^{-2\alpha\Lambda}}{1-e^{-2\alpha}}
    = \coth(\alpha) - \frac{e^{-\alpha\chi}}{\sinh(\alpha)};
\end{aligned}
\end{equation}
for the second term we have
\begin{equation}
\begin{aligned}
    \sum_{p=-\Lambda}^{\Lambda} 2\,\frac{1 + e^{-2\alpha |p|}}{e^{\alpha 2^n} - 1}
    = 2\,\frac{\chi+\coth(\alpha)-e^{-\alpha\chi}/\sinh(\alpha)}{e^{\alpha 2^n}-1};
\end{aligned}
\end{equation}
and for the third term we have
\begin{equation}
\begin{aligned}
    \sum_{p=-\Lambda}^{\Lambda} \frac{2+e^{2\alpha p}+e^{-2\alpha p}}{(e^{\alpha 2^n}-1)^2}
    &= 2\,\frac{\chi}{(e^{\alpha 2^n}-1)^2}
    + \frac{\coth(-\alpha) - \frac{e^{\alpha\chi}}{\sinh(-\alpha)}}{(e^{\alpha 2^n}-1)^2}
    + \frac{\coth(\alpha) - \frac{e^{-\alpha\chi}}{\sinh(\alpha)}}{(e^{\alpha 2^n}-1)^2}\\
    &= 2\,\frac{\chi + \sinh(\alpha\chi)/\sinh(\alpha)}{(e^{\alpha 2^n}-1)^2}.
\end{aligned}
\end{equation}
Putting everything together, the result of the sum over $p$ is
\begin{equation}
\begin{aligned}
    \sum_{p=-\Lambda}^{\Lambda}
    &e^{-2\alpha|p|}
    + 2\,\frac{1+e^{-2\alpha|p|}}{e^{\alpha 2^n}-1}
    + \frac{2+e^{2\alpha p}+e^{-2\alpha p}}{(e^{\alpha 2^n}-1)^2}=\\
    &= \coth(\alpha) - \frac{e^{-\alpha\chi}}{\sinh(\alpha)}
    + 2\,\frac{\chi + \coth(\alpha) - e^{-\alpha\chi}/\sinh(\alpha)}{e^{\alpha 2^n}-1}
    + 2\,\frac{\chi + \sinh(\alpha\chi)/\sinh(\alpha)}{(e^{\alpha 2^n}-1)^2}\\
    &= \left(\coth(\alpha) - \frac{e^{-\alpha\chi}}{\sinh(\alpha)}\right)
    + 2\,\frac{\chi (e^{\alpha 2^n}-1)}{(e^{\alpha 2^n}-1)^2}
    + 2\,\frac{\chi + \sinh(\alpha\chi)/\sinh(\alpha)}{(e^{\alpha 2^n}-1)^2}\\
    &= \left(\coth(\alpha) - \frac{e^{-\alpha\chi}}{\sinh(\alpha)}\right)
    \left(1 + \frac{2}{e^{\alpha 2^n}-1}\right)
    + \frac{\chi + e^{-\alpha 2^n} \sinh(\alpha\chi)/\sinh(\alpha)}{2 \sinh(\frac{\alpha}{2} 2^n)^2}.
    \label{eq:app_intermediate_sum_Lambda_exp}
\end{aligned}
\end{equation}
The results for the sum over $q$ can be obtained in the same way by replacing $\alpha$ with $\beta$.

Taking the two results in Eqs.~\eqref{eq:app_intermediate_sum_2^n_exp} and~\eqref{eq:app_intermediate_sum_Lambda_exp} and plugging them into Eq.~\eqref{eq:app_sum_disc_fpq}, we have
\begin{equation}
\begin{aligned}
    \sum_{(p,q)\in I_{\text{disc}}} \!\!|\hat{f}_{pq}|^2
    \leq C^2 &\Bigg(\coth(\alpha)
    \bigg(1 + e^{-\alpha2^n}\! + \frac{\sinh(\alpha2^n)}{(e^{\alpha 2^n}\! - 1)^2}\bigg)
    + \frac{2^n\!/2}{\sinh\!\big(\frac{\alpha}{2} 2^n\big)^2}\Bigg)\\
    \times&\Bigg(\coth(\beta)
    \bigg(1 + e^{-\beta2^n}\! + \frac{\sinh(\beta2^n)}{(e^{\beta 2^n}\! - 1)^2}\bigg)
    + \frac{2^n\!/2}{\sinh\!\big(\frac{\beta}{2} 2^n\big)^2}\Bigg)\\
    - C^2 &\Bigg(\!\bigg(\coth(\alpha) - \frac{e^{-\alpha\chi}}{\sinh(\alpha)}\bigg)
    \left(1 + \frac{2}{e^{\alpha 2^n}-1}\right) + \frac{\chi + e^{-\alpha 2^n} \sinh(\alpha\chi)/\sinh(\alpha)}{2 \sinh\!\big(\frac{\alpha}{2} 2^n\big)^2}\Bigg)\\
    \times&\Bigg(\!\bigg(\coth(\beta) - \frac{e^{-\beta\chi}}{\sinh(\beta)}\bigg)
    \left(1 + \frac{2}{e^{\beta 2^n}-1}\right) + \frac{\chi + e^{-\beta 2^n} \sinh(\beta\chi)/\sinh(\beta)}{2 \sinh\!\big(\frac{\beta}{2} 2^n\big)^2}\Bigg).
    \label{eq:app_exp_bound_nonsimplified}
\end{aligned}
\end{equation}

We can already see that the terms containing only $C^2 \coth(\alpha)\coth(\beta)$ without any $\chi$- or $2^n$-dependence will cancel, and all remaining terms either decay with $\chi$ or with $2^n$. Since we are mostly interested in the asymptotic scaling, we can use that $\sinh(x) = \bigO{e^x}$ for large $x$ to simplify this as
\begin{equation}
\begin{aligned}
    \sum_{(p,q)\in I_{\text{disc}}} \!\!|\hat{f}_{pq}|^2
    \leq C^2 &\left(\coth(\alpha) + \bigO{2^n e^{-\alpha2^n}}\right)
    \left(\coth(\beta) + \bigO{2^n e^{-\beta2^n}}\right)\\[-0.75em]
    - C^2 &\Bigg(\!\bigg(\coth(\alpha) - \frac{e^{-\alpha\chi}}{\sinh(\alpha)}\bigg) + \bigO{\chi e^{-\alpha2^n}}\!\Bigg)
    \Bigg(\!\bigg(\coth(\beta) - \frac{e^{-\beta\chi}}{\sinh(\beta)}\bigg)
    + \bigO{\chi e^{-\beta2^n}}\!\Bigg)\\
    = C^2 &\bigg(\frac{\coth(\alpha)}{\sinh(\beta)} e^{-\beta\chi}
    + \frac{\coth(\beta)}{\sinh(\alpha)} e^{-\alpha\chi}
    + \frac{e^{-(\alpha+\beta)\chi}}{\sinh(\alpha)\sinh(\beta)}\bigg)
    + \bigO{2^n e^{-\alpha2^n}, 2^n e^{-\beta2^n}}.
    \label{eq:app_exp_bound_simplified}
\end{aligned}
\end{equation}
This is the result presented in the main text in Eq.~\eqref{eq:approx_err_exp}.

\subsection{\label{app:proof_appr_err_bounds_alg}Algebraically decaying Fourier coefficients}
The more realistic case is when we assume the Fourier coefficients decay algebraically. The strategy for obtaining the error bounds is the same as for the exponential decay, except that some intermediate results cannot be simplified as nicely now. Consider Fourier coefficients that fulfill
\begin{equation}
    |\hat{F}(k,\ell)| \leq C \frac{1}{(|k|+1)^{\alpha}} \frac{1}{(|\ell|+1)^{\beta}},
    \label{eq:app_Fourier_alg}
\end{equation}
with some constants $C>0$ and $\alpha, \beta > 1$.%
\footnote{
    The additional `$+1$' terms in the denominators of Eq.~\eqref{eq:app_Fourier_alg} are only included to ensure that we do not divide by zero if one of the Fourier frequencies is zero.
}
Using Eq.~\eqref{eq:relating_DFT_and_FT_coeff}, we can bound the absolute value of the Fourier coefficients of a $2^n\times2^n$-pixel image, which yields
\begin{equation}
\begin{aligned}
    \hspace{-0.5em}|\hat{f}_{pq}|
    &\leq\sum_{i,j\in\Z} |\hat{F}(p+i2^n,q+j2^n)|
    \leq C \Bigg(\sum_{i\in\Z} \frac{1}{(|p+i2^n|+1)^{\alpha}}\Bigg)
    \Bigg(\sum_{j\in\Z} \frac{1}{(|q+j2^n|+1)^{\beta}}\Bigg)\\
    &= C \Bigg(\!\frac{1}{(|p|+1)^{\alpha}} + \sum_{i=1}^{\infty} \frac{1}{(i2^n\!-\!p\!+\!1)^{\alpha}} + \frac{1}{(i2^n\!+\!p\!+\!1)^{\alpha}}\!\Bigg)\!
    \Bigg(\!\frac{1}{(|q|+1)^{\beta}} + \sum_{j=1}^{\infty} \frac{1}{(j2^n\!-\!q\!+\!1)^{\beta}} + \frac{1}{(j2^n\!+\!q\!+\!1)^{\beta}}\!\Bigg)\hspace{-0.5em}\\
    &= C
    \Bigg(\!\frac{1}{(|p|+1)^{\alpha}} + \frac{1}{2^{n\alpha}} \sum_{i=1}^{\infty} \frac{1}{\big(i-\frac{p-1}{2^n}\big)^{\!\alpha}} + \frac{1}{\big(i+\frac{p+1}{2^n}\big)^{\!\alpha}}\!\Bigg)\!
    \Bigg(\!\frac{1}{(|q|+1)^{\beta}} + \frac{1}{2^{n\beta}} \sum_{j=1}^{\infty} \frac{1}{\big(j-\frac{q-1}{2^n}\big)^{\!\beta}} + \frac{1}{\big(j+\frac{q+1}{2^n}\big)^{\!\beta}}\!\Bigg).\hspace{-1em}
\end{aligned}
\end{equation}
We can use the Hurwitz zeta function $\zeta(s,a) = \sum_{j=0}^{\infty} \frac{1}{(j+a)^s}$~\cite{Hurwitz_zeta} as a shorthand for the infinite sums, e.g., we have
\begin{equation}
    \sum_{i=1}^{\infty} \frac{1}{\big(i-\frac{p-1}{2^n}\big)^{\!\alpha}}
    = \sum_{i=0}^{\infty} \frac{1}{\big(i+1-\frac{p-1}{2^n}\big)^{\!\alpha}}
    = \zeta(\alpha, 1-{\textstyle\frac{p-1}{2^n}})
\end{equation}
and similar expressions for the other sums. This allows us to write the formula above a bit more succinctly as
\begin{equation}
\begin{aligned}
    |\hat{f}_{pq}|
    &\leq C \Bigg(\frac{1}{(|p|+1)^{\alpha}} + \frac{\zeta\big(\alpha,1-\frac{p-1}{2^n}\big) + \zeta\big(\alpha, 1+\frac{p+1}{2^n}\big)}{2^{n\alpha}}\Bigg)
    \Bigg(\frac{1}{(|q|+1)^{\beta}} + \frac{\zeta\big(\beta,1-\frac{q-1}{2^n}\big) + \zeta\big(\beta, 1+\frac{q+1}{2^n}\big)}{2^{n\beta}}\Bigg)\\
    &\leq C
    \bigg(\frac{1}{(|p|+1)^{\alpha}} + \frac{1}{2^{n\alpha}} \big(\zeta(\alpha,1/2) + \zeta(\alpha, 1)\big)\bigg)
    \bigg(\frac{1}{(|q|+1)^{\beta}} + \frac{1}{2^{n\beta}} \big(\zeta(\beta,1/2) + \zeta(\beta, 1)\big)\bigg).
    \label{eq:app_bound_fpq_alg}
\end{aligned}
\end{equation}
In the second line we used the property of the Hurwitz zeta function that $\zeta(s,a) < \zeta(s,b)$ if $a > b$---in our case for nonnegative $0 \leq p < 2^n/2$ we have $1-\frac{p-1}{2^n} > 1/2$ and $1+\frac{p+1}{2^n} > 1$, while for negative $-2^n/2 \leq p < 0$ we have $1-\frac{p-1}{2^n} > 1$ and $1+\frac{p+1}{2^n} > 1/2$. This covers all possible values of the Fourier frequencies $p\in\{-2^n/2, -2^n/2+1, \ldots, 2^n/2\}$. The same relations hold for $q$.

To calculate the approximation error, we again need to sum $|\hat{f}_{pq}|^2$ over all discarded frequencies, and as in Eq.~\eqref{eq:app_sum_disc_fpq} we split the sum into a difference of the sum over all frequencies and the sum over all retained frequencies. From Eq.~\eqref{eq:app_bound_fpq_alg}, we have for $|\hat{f}_{pq}|^2$ that
\begin{equation}
\begin{aligned}
    |\hat{f}_{pq}|^2
    \leq C^2 &\Bigg(\frac{1}{(|p|+1)^{2\alpha}}
    + 2\,\frac{\zeta(\alpha,1/2) + \zeta(\alpha, 1)}{2^{n\alpha} \, (|p|+1)^{\alpha}}
    + \frac{\big(\zeta(\alpha,1/2) + \zeta(\alpha, 1)\big)^2}{2^{n2\alpha}}\Bigg)\\
    \times&\Bigg(\frac{1}{(|q|+1)^{2\beta}}
    + 2\,\frac{\zeta(\beta,1/2) + \zeta(\beta, 1)}{2^{n\beta} \, (|q|+1)^{\beta}}
    + \frac{\big(\zeta(\beta,1/2) + \zeta(\beta, 1)\big)^2}{2^{n2\beta}}\Bigg).
\end{aligned}
\end{equation}

First, we treat the sum over all frequencies. Since $|\hat{f}_{pq}|^2$ factorizes, we can perform the sums over $p$ and $q$ separately, and three terms contribute to each of the two sums. For the following calculation it is useful to introduce the generalized harmonic numbers $H_s(n) = \sum_{j=1}^n \frac{1}{j^s} = \sum_{j=0}^{n-1} \frac{1}{(j+1)^s}$, which are essentially of the form as the sum we need to calculate. They relate to the Hurwitz zeta function as $H_s(n) = \zeta(s,1) - \zeta(s,n+1)$~\cite[Eq.~$(25.11.4)$]{Hurwitz_zeta}. Then, for the first term in the sum over $p$, we have
\begin{equation}
\begin{aligned}
    \sum_{p=-2^n\!/2}^{2^n\!/2-1} \frac{1}{(|p|+1)^{2\alpha}}
    &= \left(2^n\!/2+1\right)^{-2\alpha} - 1 + 2\sum_{p=0}^{2^n\!/2-1} \frac{1}{(|p|+1)^{2\alpha}}\\
    &= \left(2^n\!/2+1\right)^{-2\alpha} - 1 + 2H_{2\alpha}(2^n\!/2)\\
    &= 2\zeta(2\alpha,1) - 1 + \left(2^n\!/2+1\right)^{-2\alpha}
    - 2\zeta(2\alpha,2^n\!/2+1),
\end{aligned}
\end{equation}
where we have brought the sum into the form of the generalized harmonic numbers in the first line, and then replaced them by their relation to the Hurwitz zeta function. For the second term we have
\begin{equation}
\begin{aligned}
    \sum_{p=-2^n\!/2}^{2^n\!/2-1} 2\,\frac{\zeta(\alpha,1/2) + \zeta(\alpha, 1)}{2^{n\alpha} \, (|p|+1)^{\alpha}}
    &= 2\,\frac{\zeta(\alpha,1/2) + \zeta(\alpha, 1)}{2^{n\alpha}}
    \Big(2\zeta(\alpha,1) - 1 + \left(2^n\!/2+1\right)^{-\alpha}
    - 2\zeta(\alpha,2^n\!/2+1)\Big),
\end{aligned}
\end{equation}
where we could reuse the result from the first term for the sum over $(|p|+1)^{-\alpha}$. Finally, for the third term we have
\begin{equation}
\begin{aligned}
    \sum_{p=-2^n\!/2}^{2^n\!/2-1} \frac{\big(\zeta(\alpha,1/2) + \zeta(\alpha, 1)\big)^2}{2^{n2\alpha}}
    = \frac{\big(\zeta(\alpha,1/2) + \zeta(\alpha, 1)\big)^2}{2^{n(2\alpha-1)}}.\\
\end{aligned}
\end{equation}
Putting everything together, the result of the sum over $p$ is
\begin{equation}
\begin{aligned}
    \sum_{p=-2^n\!/2}^{2^n\!/2-1}\! &\Bigg(\frac{1}{(|p|+1)^{2\alpha}}
    + 2\,\frac{\zeta(\alpha,1/2) + \zeta(\alpha, 1)}{2^{n\alpha} \, (|p|+1)^{\alpha}}
    + \frac{\big(\zeta(\alpha,1/2) + \zeta(\alpha, 1)\big)^2}{2^{n2\alpha}}\Bigg)=\\[0.5em]
    &= 2\zeta(2\alpha,1) - 1 + \left(2^n\!/2 + 1\right)^{-2\alpha} - 2\zeta(2\alpha,2^n\!/2 + 1)\\
    &\quad+2\,\frac{\zeta(\alpha,1/2) + \zeta(\alpha, 1)}{2^{n\alpha}}
    \big(2\zeta(\alpha,1) - 1 + \left(2^n\!/2 + 1\right)^{-\alpha} - 2\zeta(\alpha,2^n\!/2 + 1)\big)\\
    &\quad+\frac{\big(\zeta(\alpha,1/2) + \zeta(\alpha, 1)\big)^2}{2^{n(2\alpha-1)}}.
    \label{eq:app_intermediate_sum_2^n_alg}
\end{aligned}
\end{equation}
The results for the sum over $q$ can be obtained in the same way by replacing $\alpha$ with $\beta$.

Now we treat the sum over all retained frequencies. The sums are essentially the same, only the summation range changes. Again, we can perform the sums over $p$ and $q$ separately, and there are three terms that contribute. For the first term we have
\begin{equation}
\begin{aligned}
    \sum_{p=-\Lambda}^{\Lambda} \frac{1}{(|p|+1)^{2\alpha}}
    = 2\sum_{p=0}^{\Lambda} \frac{1}{(|p|+1)^{2\alpha}} - 1
    = 2H_{2\alpha}(\Lambda+1) - 1
    = 2\zeta(2\alpha,1) - 2\zeta(2\alpha,\Lambda+2) - 1;
\end{aligned}
\end{equation}
for the second term we have
\begin{equation}
\begin{aligned}
    \sum_{p=-\Lambda}^{\Lambda} 2\,\frac{\zeta(\alpha,1/2) + \zeta(\alpha, 1)}{2^{n\alpha} \, (|p|+1)^{\alpha}}
    = 2\,\frac{\zeta(\alpha,1/2) + \zeta(\alpha, 1)}{2^{n\alpha}}
    \big(2\zeta(\alpha,1) - 2\zeta(\alpha,\Lambda+2) - 1\big);
\end{aligned}
\end{equation}
and for the third term we have
\begin{equation}
\begin{aligned}
    \sum_{p=-\Lambda}^{\Lambda}\frac{\big(\zeta(\alpha,1/2) + \zeta(\alpha, 1)\big)^2}{2^{n2\alpha}}
    = (2\Lambda+1)\,\frac{\big(\zeta(\alpha,1/2) + \zeta(\alpha, 1)\big)^2}{2^{n2\alpha}}.
\end{aligned}
\end{equation}
Putting everything together, the result of the sum over $p$ is
\begin{equation}
\begin{aligned}
    \sum_{p=-\Lambda}^{\Lambda} &\Bigg(\frac{1}{(|p|+1)^{2\alpha}}
    + 2\,\frac{\zeta(\alpha,1/2) + \zeta(\alpha, 1)}{2^{n\alpha} \, (|p|+1)^{\alpha}}
    + \frac{\big(\zeta(\alpha,1/2) + \zeta(\alpha, 1)\big)^2}{2^{n2\alpha}}\Bigg)=\\[0.5em]
    &= 2\zeta(2\alpha,1) - 2\zeta(2\alpha,\Lambda+2) - 1\\
    &\quad+ 2\,\frac{\zeta(\alpha,1/2) + \zeta(\alpha, 1)}{2^{n\alpha}}
    \big(2\zeta(\alpha,1) - 2\zeta(\alpha,\Lambda+2) - 1\big)\\
    &\quad+ (2\Lambda+1)\,\frac{\big(\zeta(\alpha,1/2) + \zeta(\alpha, 1)\big)^2}{2^{n2\alpha}}.
    \label{eq:app_intermediate_sum_Lambda_alg}
\end{aligned}
\end{equation}
The results for the sum over $q$ can be obtained in the same way by replacing $\alpha$ with $\beta$.

Taking the two results in Eqs.~\eqref{eq:app_intermediate_sum_2^n_alg} and~\eqref{eq:app_intermediate_sum_Lambda_alg} and plugging them into Eq.~\eqref{eq:app_sum_disc_fpq}, we have
\begin{equation}
\begin{aligned}
    \sum_{(p,q)\in I_{\text{disc}}} \!\!|\hat{f}_{pq}|^2
    \leq C^2 \Bigg(&2\zeta(2\alpha,1) - 1
    + \left(2^n\!/2 + 1\right)^{-2\alpha}\! - 2\zeta(2\alpha,2^n\!/2 + 1)
    + \frac{\big(\zeta(\alpha,1/2) + \zeta(\alpha, 1)\big)^2}{2^{n(2\alpha-1)}}\\[-0.5em]
    &\hspace{4em}+ 2\,\frac{\zeta(\alpha,1/2) + \zeta(\alpha, 1)}{2^{n\alpha}}
    \big(2\zeta(\alpha,1) - 1 + \left(2^n\!/2 + 1\right)^{-\alpha}\! - 2\zeta(\alpha,2^n\!/2 + 1)\big)\!\Bigg)\\
    \times \Bigg(&2\zeta(2\beta,1) - 1
    + \left(2^n\!/2 + 1\right)^{-2\beta}\! - 2\zeta(2\beta,2^n\!/2 + 1)
    + \frac{\big(\zeta(\beta,1/2) + \zeta(\beta, 1)\big)^2}{2^{n(2\beta-1)}}\\[-0.5em]
    &\hspace{4em}+ 2\,\frac{\zeta(\beta,1/2) + \zeta(\beta, 1)}{2^{n\beta}}
    \big(2\zeta(\beta,1) - 1 + \left(2^n\!/2 + 1\right)^{-\beta}\! - 2\zeta(\beta,2^n\!/2 + 1)\big)\!\Bigg)\\
    -C^2 \Bigg(&2\zeta(2\alpha,1) - 1 - 2\zeta(2\alpha,\Lambda+2)
    + (2\Lambda+1)\,\frac{\big(\zeta(\alpha,1/2) + \zeta(\alpha, 1)\big)^2}{2^{n2\alpha}}\\[-0.5em]
    &\hspace{12em}+ 2\,\frac{\zeta(\alpha,1/2) + \zeta(\alpha, 1)}{2^{n\alpha}}
    \big(2\zeta(\alpha,1) - 1 - 2\zeta(\alpha,\Lambda+2)\big)\!\Bigg)\hspace{-2em}\\
    \times \Bigg(&2\zeta(2\beta,1) - 1 - 2\zeta(2\beta,\Lambda+2)
    + (2\Lambda+1)\,\frac{\big(\zeta(\beta,1/2) + \zeta(\beta, 1)\big)^2}{2^{n2\beta}}\\[-0.5em]
    &\hspace{12em}+ 2\,\frac{\zeta(\beta,1/2) + \zeta(\beta, 1)}{2^{n\beta}}
    \big(2\zeta(\beta,1) - 1 - 2\zeta(\beta,\Lambda+2)\big)\!\Bigg).\hspace{-2em}
    \label{eq:app_alg_bound_nonsimplified}
\end{aligned}
\end{equation}
We can see that after expanding all brackets, the term consisting of
\begin{equation}
\begin{aligned}
    C^2 &\left(2\zeta(2\alpha,1) - 1 + 2\,\frac{\zeta(\alpha,1/2) + \zeta(\alpha, 1)}{2^{n\alpha}}
    \big(2\zeta(\alpha,1) - 1\big)\right)\\
    \times &\left(2\zeta(2\beta,1) - 1 + 2\,\frac{\zeta(\beta,1/2) + \zeta(\beta, 1)}{2^{n\beta}}
    \big(2\zeta(\beta,1) - 1\big)\right)
\end{aligned}
\end{equation}
appears twice with opposite signs and will cancel. The remaining terms all either decay with $2^n$ or the cutoff $\Lambda$. Since we are interested in the asymptotic behavior for large $n$, we can use that $\zeta(s,a) = \bigO{a^{1-s}}$ for large $a$~\cite{Hurwitz_zeta} in order to simplify
\begin{equation}
\begin{aligned}
    \hspace{-2.5em}\sum_{(p,q)\in I_{\text{disc}}}\hspace{-1em}|\hat{f}_{pq}|^2\!
    \leq\! 4C^2 &\bigg(
    \zeta(2\alpha,1) - {\textstyle\frac{1}{2}}
    + 2\frac{\zeta(\alpha,1/2) + \zeta(\alpha, 1)}{2^{n\alpha}} \big(\zeta(\alpha,1) - {\textstyle\frac{1}{2}}\big)
    + \bigO{(2^n)^{1-2\alpha}}\bigg)\\
    \times &\bigg(
    \zeta(2\beta,1) - {\textstyle\frac{1}{2}}
    + 2\frac{\zeta(\beta,1/2) + \zeta(\beta, 1)}{2^{n\beta}} \big(\zeta(\beta,1) - {\textstyle\frac{1}{2}}\big)
    + \bigO{(2^n)^{1-2\beta}}\bigg)\\
    -4C^2 &\bigg(
    \zeta(2\alpha,1) \!-\! {\textstyle\frac{1}{2}} \!-\! \zeta(2\alpha,\Lambda\!+\!2)
    + 2\frac{\zeta(\alpha,1/2) \!+\! \zeta(\alpha,1)}{2^{n\alpha}}
    \left(\zeta(\alpha,1) \!-\! {\textstyle\frac{1}{2}} \!-\! \zeta(\alpha,\Lambda\!+\!2)\right)
    + \bigO{\Lambda (2^n)^{-2\alpha}}\!\!\bigg)\hspace{-2em}\\
    \times &\bigg(
    \zeta(2\beta,1) \!-\! {\textstyle\frac{1}{2}} \!-\! \zeta(2\beta,\Lambda\!+\!2)
    + 2\frac{\zeta(\beta,1/2) \!+\! \zeta(\beta,1)}{2^{n\beta}}
    \left(\zeta(\beta,1) \!-\! {\textstyle\frac{1}{2}} \!-\! \zeta(\beta,\Lambda\!+\!2)\right)
    + \bigO{\Lambda (2^n)^{-2\beta}}\!\!\bigg)\hspace{-2em}\\
    =\! 4C^2\! &\Bigg(\!\!
    \bigg(\!\zeta(2\alpha,1) \!-\! {\textstyle\frac{1}{2}}
    \!+\! \frac{\zeta(\alpha,1/2) \!+\! \zeta(\alpha, 1)}{2^{n\alpha-1}}
    \big(\zeta(\alpha,1) \!-\! {\textstyle\frac{1}{2}}\big)\hspace{-0.25em}\bigg)\!
    \bigg(\!\zeta(2\beta,\Lambda\!+\!2)
    \!+\! \frac{\zeta(\beta,1/2) \!+\! \zeta(\beta,1)}{2^{n\beta-1}} \zeta(\beta,\Lambda\!+\!2)\hspace{-0.25em}\bigg)\hspace{-2em}\\
    +\!\!&\phantom{\Bigg(\!\!}
    \bigg(\!\zeta(2\beta,1) \!-\! {\textstyle\frac{1}{2}}
    \!+\! \frac{\zeta(\beta,1/2) \!+\! \zeta(\beta, 1)}{2^{n\beta-1}}
    \big(\zeta(\beta,1) \!-\! {\textstyle\frac{1}{2}}\big)\hspace{-0.25em}\bigg)\!
    \bigg(\!\zeta(2\alpha,\Lambda\!+\!2)
    \!+\! \frac{\zeta(\alpha,1/2) \!+\! \zeta(\alpha,1)}{2^{n\alpha-1}} \zeta(\alpha,\Lambda\!+\!2)\hspace{-0.25em}\bigg)\hspace{-2em}\\
    +\!\!&\phantom{\Bigg(\!\!}
    \bigg(\zeta(2\alpha,\Lambda\!+\!2)
    \!+\! \frac{\zeta(\alpha,1/2) \!+\! \zeta(\alpha,1)}{2^{n\alpha-1}} \zeta(\alpha,\Lambda\!+\!2)\hspace{-0.25em}\bigg)\!
    \bigg(\zeta(2\beta,\Lambda\!+\!2)
    \!+\! \frac{\zeta(\beta,1/2) \!+\! \zeta(\beta,1)}{2^{n\beta-1}} \zeta(\beta,\Lambda\!+\!2)\hspace{-0.25em}\bigg)
    \!\Bigg)\hspace{-4em}\\
    +\!\!&\phantom{\Bigg(\!}
    \bigO{(2^n)^{1-2\alpha}, (2^n)^{1-2\beta}}.
    \label{eq:app_alg_bound_simplified}
\end{aligned}
\end{equation}
\end{widetext}
\twocolumngrid
\noindent%
Notably, this bound does not increase with system size. Since we must have that $2^n>2\Lambda$, for large $\Lambda$ the bound behaves as
\begin{equation}
\begin{aligned}
    \hspace{-0.5em}\sum_{(p,q)\in I_{\text{disc}}} \hspace{-0.5em}|\hat{f}_{pq}|^2
    &\leq \bigO{\Lambda^{1-2\alpha}} + \bigO{\Lambda^{1-2\beta}}\\[-0.75em]
    &\qquad\quad+ \bigO{(2^n)^{1-2\alpha}, (2^n)^{1-2\beta}}.
\end{aligned}
\end{equation}
This is the result presented in the main text in Eq.~\eqref{eq:approx_err_alg}.

\subsection{\label{app:proof_appr_err_bounds_tightness}Tightness of the error bounds}
Having derived the error bounds, we can now investigate their tightness on some example Fourier spectra and compare them both to the actual approximation error from the truncated Fourier series and to the approximation error obtained from successive SVDs, which one would typically use in practice for calculating the MPS representation.

\begin{figure*}
    \raggedright
    \includegraphics[width=\linewidth]{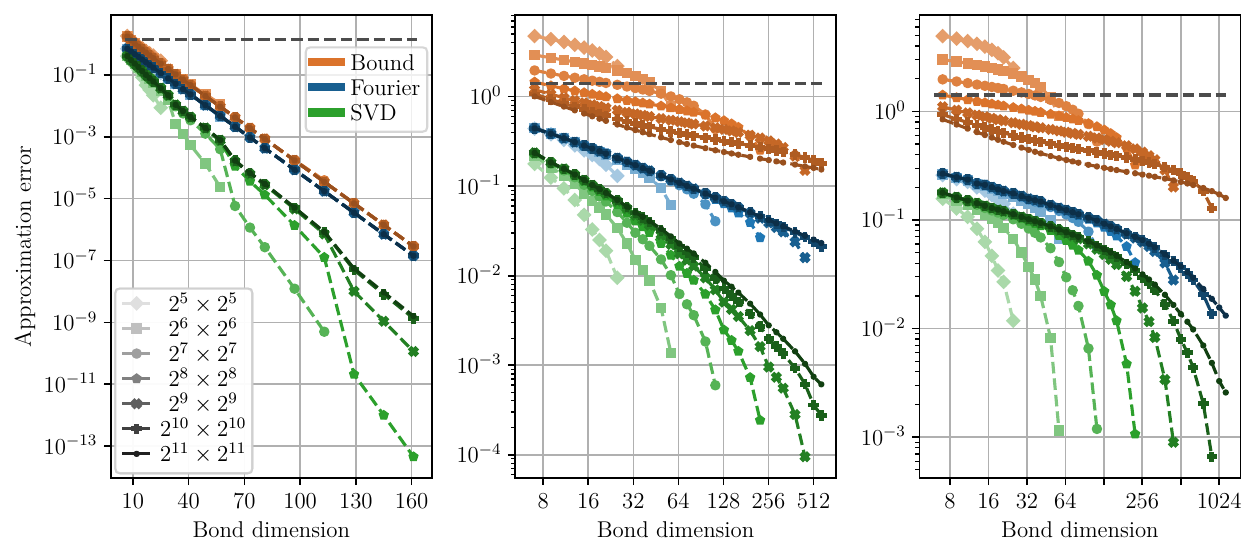}\\[-75.26mm]
    \hspace{7.08mm}{\sffamily(a)}\hspace{50.24mm}{\sffamily(b)}\hspace{50.13mm}{\sffamily(c)}\\[71mm]
    \caption{\label{fig:example_bound_tightness}%
    \textbf{Comparison of the error bounds to the actual approximation errors.} The orange data shows the previously derived error bounds [Eq.~\eqref{eq:approx_err_normalized_fourier} in combination with Eqs.~\eqref{eq:app_exp_bound_nonsimplified} and~\eqref{eq:app_alg_bound_nonsimplified}], the blue and green data show the approximation error $\|\ket{\psi_{\text{exact}}} - \ket{\psi_{\text{approx.}}}\|_2$ for the MPS approximation obtained from truncating the Fourier spectrum and from successive SVDs, respectively. The different color shades and marker shapes show the different image resolutions. The gray dashed line indicates a value of $\sqrt{2}$, which is the approximation error for orthogonal states (and random states in the many-qubit limit). (a)~The data shown in the left plot correspond to an artificial Fourier spectrum, where the absolute values of the Fourier coefficients decay exponentially according to Eq.~\eqref{eq:app_Fourier_exp} with $\alpha=\beta=0.2$ and the phases of the coefficients are drawn randomly as $e^{i\pi x}$, with $x$ drawn from a normal distribution. (b)~The data in the central plot correspond to artificially generated Fourier coefficients with an absolute value saturating the algebraic decay in Eq.~\eqref{eq:app_Fourier_alg} with $\alpha = \beta = 1.2$, and with random phases $e^{i\pi x}$ where $x$ is drawn from a normal distribution. (c)~The data in the right plot shows the results for the example image shown in Fig.~\ref{fig:example_compression}a from the main text. The corresponding Fourier spectrum is shown in Fig.~\ref{fig:example_bound_Fourier_coefficients}.}
\end{figure*}

First, we consider the case of exponentially decaying Fourier coefficients. As an example, we take a spectrum of Fourier coefficients whose absolute values saturate the bound given in Eq.~\eqref{eq:app_Fourier_exp}, with parameters $C=1$ and $\alpha = \beta = 0.2$. In principle, we would need to consider all infinitely many Fourier coefficients, corresponding to all integer pairs of Fourier frequencies, when relating them to a finite-resolution image according to Eq.~\eqref{eq:relating_DFT_and_FT_coeff}. To keep this numerically tractable, we generate a $2^{12}\times2^{12}$ matrix of Fourier coefficients that decay according to Eq.~\eqref{eq:app_Fourier_exp}, and then only consider image resolutions of $2^n\times2^n$ pixels with $n$ going from five to eleven (and not to twelve or above) so that the cutoff has no visible influence on the results. If all Fourier coefficients are positive and saturate the decay in Eq.~\eqref{eq:app_Fourier_exp}, then the matrix is separable and has rank one; to avoid this case we add random phases $e^{i\pi x}$ to the $2^{12}\times2^{12}$ matrix of Fourier coefficients, where $x$ is sampled from a normal distribution with zero mean and a variance of one. We can then obtain the Fourier coefficients of the smaller-resolution images from the $2^{12}\times2^{12}$ matrix via Eq.~\eqref{eq:relating_DFT_and_FT_coeff}. From the Fourier coefficients of the various image resolutions, we can obtain the approximation errors: First, we can calculate the error bounds for the different image resolutions as in Eq.~\eqref{eq:approx_err_normalized_fourier}, using the expression for the bound on the discarded Fourier coefficients in Eq.~\eqref{eq:app_exp_bound_nonsimplified}, where the system-size dependence is still explicit compared to the simplified result concerning the asymptotic behavior in Eq.~\eqref{eq:app_exp_bound_simplified}. Second, we can truncate the Fourier spectrum to a specific bond dimension directly, to calculate the approximation error of an MPS approximation constructed from the Fourier modes. And third, we can transform the Fourier coefficients back to real space via an inverse DFT, and obtain an MPS approximation via successive SVDs. The results are shown in Fig.~\ref{fig:example_bound_tightness}a. There, the orange data points show the values of the bound, the blue data points show the approximation error when truncating the Fourier series and the green data points show the approximation error of the MPS obtained from successive SVDs. The different shades and marker shapes denote the different image resolutions. The gray dashed line shows a reference value of $\sqrt{2}$, which corresponds to the error of an approximation by an orthogonal state or a random state, since a state drawn uniformly at random from the full Hilbert space becomes orthogonal to any reference state as the Hilbert space dimension is increased. This is in contrast to the infidelity we considered for numerical results before (e.g., in Fig.~\ref{fig:scaling_resolution}), where a value of one corresponded to orthogonal states or random states in a high-dimensional Hilbert space.
In the figure, we can see that the bound on the approximation error and the two approximation errors obtained from truncating in the Fourier or the SVD basis all decay exponentially, with only very small differences between the varying image resolutions. In this case, the derived error bound seems tight, since it is close to the actual error obtained from truncating the Fourier series, and both decay asymptotically at the same rate. The approximation error from the SVD is only a little smaller, and only decays slightly faster than the truncation-error in the Fourier basis. Since the coefficients decay exponentially in the Fourier basis, their weight is already highly concentrated and we expect the Fourier basis to be close to optimal for truncation. In general, there are two sources that can contribute to the error bound not being tight compared to the actual approximation error: first, bounding the difference of two normalized states with different normalization constants by the difference of the unnormalized states as in Eq.~\eqref{eq:approx_err_normalized}, or as derived in Eq.~\eqref{eq:app_approx_err_normalized_deriv}, can make the bound loose; and second, the bound on the weight of the discarded Fourier coefficients in Eq.~\eqref{eq:norm_diff}, or explicitly in Eq.~\eqref{eq:app_exp_bound_nonsimplified}, can make the bound loose. In the case at hand, we expect the former effect to contribute more, as the calculation of the weight of the discarded Fourier coefficients is almost exact: for purely positive coefficients applying the triangle inequality in Eq.~\eqref{eq:app_bound_fpq_exp} is exact, and for complex coefficients there is no strong cancellation because the higher-frequency components decay very quickly.

Next, we consider algebraically decaying Fourier coefficients as in Eq.~\eqref{eq:app_Fourier_alg}. We consider an artificial example where the decay of the absolute values of the Fourier coefficients saturates the decay in Eq.~\eqref{eq:app_Fourier_alg} with $C=1$ and $\alpha = \beta = 1.2$. As for the example of the exponential decay before, we generate a $2^{12}\times2^{12}$ matrix of Fourier coefficients, and then add random phases by multiplying each Fourier coefficient by $e^{i\pi x}$, where $x$ is drawn from a normal distribution with zero mean and unit variance. Using Eq.~\eqref{eq:relating_DFT_and_FT_coeff}, we can obtain the different $2^n\times2^n$-pixel image resolutions, with $n$ ranging from five to eleven. With the Fourier coefficients of the different image resolutions, we can calculate the bound on the approximation error using Eq.~\eqref{eq:approx_err_normalized_fourier} in combination with the previously derived bound on the weight of the discarded Fourier coefficients in Eq.~\eqref{eq:app_alg_bound_nonsimplified}; note that the system-size dependence is still explicit in Eq.~\eqref{eq:app_alg_bound_nonsimplified} compared to the simplified result considering the asymptotic behavior in Eq.~\eqref{eq:app_alg_bound_simplified}. Further, we can calculate the approximation error by directly truncating the Fourier spectrum to a specific bond dimension and by calculating the MPS representation with a given bond dimension via successive SVDs. The resulting approximation errors are shown against the bond dimension in Fig.~\ref{fig:example_bound_tightness}b, where the orange data points show the error bounds, the blue data points show the errors from the truncated Fourier spectrum and the green data points show the errors of the SVD-based MPS approximation. The different color shades and marker shapes denote the different image resolutions, the gray dashed line shows a reference value of $\sqrt{2}$.
Compared to the case of exponentially decaying Fourier coefficients, the approximation error obtained from truncating the Fourier spectrum now decays algebraically instead of exponentially. The derived error bound shows stronger finite-size effects than before, and also decays approximately algebraically. We can see that the error bound is less tight than for the exponential decay; in particular, the asymptotic decay of the bound seems slightly slower than that of the approximation error. The approximation error using successive SVDs is smaller than directly truncating in the Fourier basis, the difference is especially significant for large bond dimensions. This is somewhat expected, as an algebraic decay of the Fourier spectrum means their weight is not as concentrated as it was in the case of an exponential decay, making the Fourier basis less optimal for truncation. Meanwhile, the SVD can evidently find and exploit more residual structure in the Fourier spectrum. For the case of an exponentially decaying Fourier spectrum, we argued that of the two causes that can contribute to the looseness of the error bound---i.e., bounding the difference of two normalized states by their unnormalized counterparts in Eqs.~\eqref{eq:approx_err_normalized} and~\eqref{eq:app_approx_err_normalized_deriv}, and bounding the weight of the discarded Fourier coefficients in Eq.~\eqref{eq:norm_diff}---the former played a dominant role while the latter was insignificant. This was because the bound was derived from an exact calculation for positive Fourier coefficients, and complex phases were not expected to change the behavior significantly due to the quick decay of the coefficients. For the case of algebraically decaying Fourier coefficients, even for purely positive Fourier coefficients the weight of the discarded modes cannot be calculated exactly, and differences due to interference with higher-frequency modes become more important as they decay more slowly. Thus, also the bound on the weight of the discarded Fourier coefficients in Eq.~\eqref{eq:app_alg_bound_nonsimplified} becomes less tight, and is one of the reasons why the error bound is observed to deviate more.

\begin{figure}[!t]
    \includegraphics[width=\linewidth]{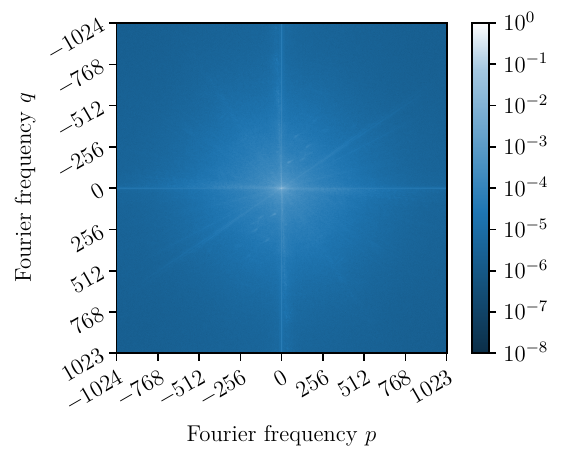}\\[-66.77mm]
    \raggedright{\sffamily(a)}\\[58.77mm]
    \raggedright{\sffamily(b)}\\
    \includegraphics[width=\linewidth]{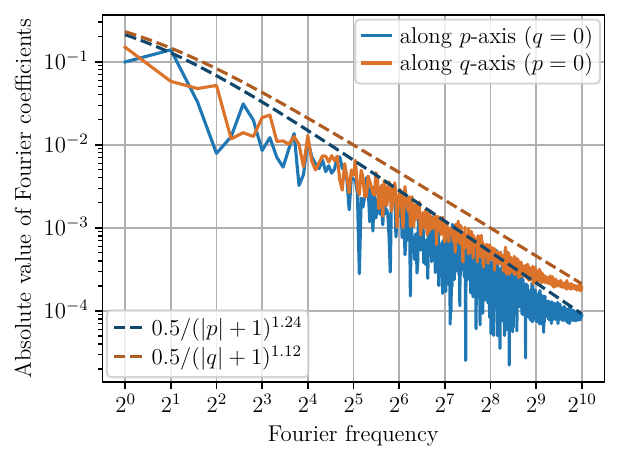}
    \caption{\label{fig:example_bound_Fourier_coefficients}%
    \textbf{Absolute values of the Fourier coefficients of the $\bm{2048\times2048}$-pixel version of the image in Fig.~\ref{fig:example_compression}a.} (a)~The color plot shows the absolute values of the full spectrum of Fourier coefficients. As in Fig.~\ref{fig:Fourier_coefficients}, most of the weight is concentrated at the bright white spot in the center and the two axes corresponding to zero-frequency modes in either direction. (b)~The plot below shows the decay of the absolute value of the Fourier coefficients along the zero-frequency modes---along the $p$-axis where $q=0$ in blue and along the $q$-axis where $p=0$ in orange. To calculate the error bound, we need to estimate the asymptotic decay of the Fourier coefficients in the infinite-resolution limit in accordance with Eq.~\eqref{eq:app_Fourier_alg} from the finite-resolution Fourier coefficients shown here. The dashed lines show the algebraic decay following Eq.~\eqref{eq:app_Fourier_alg} with the estimated parameters $C=0.5$, $\alpha=1.24$ and $\beta=1.12$.}
\end{figure}

The final example we consider is the image shown in Fig.~\ref{fig:example_compression}a in the main text. In this case, the decay of the Fourier coefficients can still be bounded by an algebraic decay. To estimate the values of the parameters appearing in Eq.~\eqref{eq:app_Fourier_alg}, we show the Fourier spectrum in Fig.~\ref{fig:example_bound_Fourier_coefficients}. Fig.~\ref{fig:example_bound_Fourier_coefficients}a shows the absolute values of the full spectrum of Fourier coefficients of the image with a $2048\times2048$-pixel resolution. Similar to the behavior of the averaged Fourier spectrum shown in Fig.~\ref{fig:Fourier_coefficients} in the main text, most of the weight of the Fourier coefficients is concentrated in the bright white spot in the center and the two axes where one of the frequencies is zero. Fig.~\ref{fig:example_bound_Fourier_coefficients}b shows the absolute values of the Fourier coefficients along these two axes: the solid blue line shows the Fourier coefficients for the nonzero frequencies in $x$-direction (along the $p$-axis where $q=0$) and the solid orange line the Fourier coefficients for the nonzero frequencies in $y$-direction (along the $q$-axis where $p=0$). From their decay we can estimate the values of the parameters in Eq.~\eqref{eq:app_Fourier_alg} needed to bound the Fourier coefficients. Note that technically we want to bound the decay of the Fourier coefficients in the limit of infinite pixels, but we only have access to the $2048\times2048$-pixel version. In practice, that means we assume that the resolution is already high enough that further increasing the resolution will not drastically change the decay of the Fourier spectrum, and that the flattening of the decay observed in Fig.~\ref{fig:example_bound_Fourier_coefficients}b for large frequencies is only a finite-size effect---consistent with the behavior of the derived bound on the finite-size Fourier coefficients in Eq.~\eqref{eq:app_bound_fpq_alg}. We estimate the parameters to be $C=0.5$, $\alpha=1.24$ and $\beta=1.12$, and show the corresponding decays as dashed lines in the figure. With these parameters, we can calculate the error bounds for different image resolutions as before, i.e., using Eqs.~\eqref{eq:approx_err_normalized_fourier} and~\eqref{eq:app_alg_bound_nonsimplified}. Also, we can get the actual approximations errors for the different image resolutions from both truncating the Fourier spectrum and successive truncated SVDs. The results are shown in Fig.~\ref{fig:example_bound_tightness}c, where orange data points show the bounds, blue and green data points show the approximation errors when truncating in the Fourier or SVD basis, and the different color shades and marker shapes denote the different resolutions. The gray dashed line shows the reference value $\sqrt{2}$ of an approximation by orthogonal (or random) states. The approximation error when truncating in the Fourier basis decays approximately algebraically. Also the approximation error from truncating in the SVD basis decays algebraically in roughly the same way, and is only marginally smaller than the truncation error in the Fourier basis. The error bound, while still decaying roughly algebraically, seems to decay slower than both actual approximation errors asymptotically. Compared with the previous example of an artificially generated algebraically decaying Fourier spectrum, the bound is somewhat less tight. This is because in that example the Fourier spectrum saturated the decay in Eq.~\eqref{eq:app_Fourier_alg}, so the only looseness in bounding the weight of the discarded Fourier coefficients originated from bounding the infinite sums that appeared in the derivation or from applying the triangle equality to Fourier coefficients with complex phases. In the case of a Fourier spectrum from a real image, Eq.~\eqref{eq:app_Fourier_alg} already constitutes a bound on the actual magnitude of the Fourier coefficients and that itself can be loose; this is especially the case if there is a strong oscillatory behavior as seen in Fig.~\ref{fig:example_bound_Fourier_coefficients}b.

\section{\label{app:cosine_transforms}Error bounds from cosine transforms}
The error bounds for an MPS approximation we presented in the main text crucially depend on how quickly the Fourier coefficients of the function $F(x,y)$ decay. Since the decay of the Fourier coefficients of a function can be related to the smoothness of the function~\cite{Grafakos2014}, any discontinuous jumps could decrease the efficiency of an MPS approximation. When we wrote down the Fourier series of $F(x,y)$ in Eq.~\eqref{eq:Fourier_series}, we implicitly extended the function periodically from $(x,y)\in[0,1)^2$ to $(x,y)\in\R^2$, which can create discontinuous jumps at the boundary of the original input domain $[0,1)^2$. Because of this, one usually uses cosine transforms instead of Fourier transforms in image processing, which corresponds to first mirroring the original function $F(x,y)$ about the origin to get a larger input domain $[-1,1)$, and then periodically extending this function with the larger input domain by doing a Fourier transform. This way, there are no more discontinuous jumps at the boundary, only discontinuities in the derivatives remain.

There are several different cosine transforms, which can be related to Fourier transforms of input data with a larger size and a certain symmetry. We will review this connection for the type I and type II discrete cosine transforms here, and show what changes in the error bounds.

\subsection{Type-I DCT}
First, let us consider the discrete cosine transform of type I (DCT-I). The forward DCT-I of a matrix $f_{ab}$, $a,b\in\{0,1,\ldots,2^n-1\}$ produces coefficients $\hat{f}_{pq}$, $p,q\in\{0,1,\ldots,2^n-1\}$, which are defined as
\begin{equation}
\begin{aligned}
    \hat{f}_{pq} &= \frac{1}{(2^n\!-\!1)^2}\\
    &\quad\!\times\!\sum_{a,b=0}^{2^n-1}\hspace{-0.2em}
    \frac{f_{ab}\,\cos\left(\frac{\pi p a}{2^n-1}\right)\cos\left(\frac{\pi q b}{2^n-1}\right)}
    {(1\!+\!\delta_{a,0}\!+\!\delta_{a,2^n-1})\hspace{-0.05em}(1\!+\!\delta_{a,0}\!+\!\delta_{a,2^n-1})}\hspace{-2em}\\
    &= \frac{1}{\big(2(2^n\!-\!1)\big)^2} \!\sum_{a,b=-2^n+1}^{2^n-2}\!\!\!
    h_{ab}\,e^{-i\frac{2\pi pa}{2(2^n-1)}}\,e^{-i\frac{2\pi qb}{2(2^n-1)}}\hspace{-2em}
    \label{eq:DCT-I}
\end{aligned}
\end{equation}
defining a new matrix of larger dimension $h_{ab} = f_{|a|,|b|}$ for $a,b\in\{-2^n+1,-2^n+2,\ldots,2^n-2\}$.
The inverse transform is defined as
\begin{equation}
\begin{aligned}
    f_{ab} &= \sum_{p,q=0}^{2^n-1} \hat{f}_{pq}
    (2\!-\!\delta_{p,0}\!-\!\delta_{p,2^n-1}) (2\!-\!\delta_{q,0}\!-\!\delta_{q,2^n-1})\\[-0.5em]
    &\hspace{4em}\times\cos\left(\frac{\pi p a}{2^n-1}\right)\cos\left(\frac{\pi q b}{2^n-1}\right)\\
    &= \sum_{p,q=-2^n+1}^{2^n-2}
    \hat{h}_{pq}\,e^{i\frac{2\pi pa}{2(2^n-1)}}\,e^{i\frac{2\pi qb}{2(2^n-1)}},
    \label{eq:iDCT-I}
\end{aligned}
\end{equation}
defining a new matrix of Fourier coefficients $\hat{h}_{pq} = \hat{f}_{|p|,|q|}$ for $p,q\in\{-2^n+1,-2^n+2,\ldots,2^n-2\}$. From this we can see that the DCT-I of the ${2^n\!\times\!2^n}$ matrix $f_{ab}$ is equivalent to the DFT of the ${2(2^n\!-\!1)\times2(2^n\!-\!1)}$ matrix $h_{ab}$, the coefficients of which are just $f_{ab}$ arranged to have a mirror symmetry about the origin.

This can be extended to the cosine series of the function $F(x,y)$ with continuous arguments $(x,y)\in[0,1]^2$; the coefficients of the cosine series are given by
\begin{equation}
\begin{aligned}
    \hspace{-0.25cm}\hat{F}(k,\ell) &= \int_{0}^{1}\d x \int_{0}^{1}\d y\;
    \frac{F(x,y)\cos\left(\pi k x\right)\cos\left(\pi\ell y\right)}
    {(1+\delta_{k,0})(1+\delta_{\ell,0})}\hspace{-0.5cm}\\
    &= \frac{1}{2^2}  \int_{-1}^{1}\d x \int_{-1}^{1}\d y\;
    H(x,y)\,e^{-i2\pi\frac{k}{2}x}\,e^{-i2\pi\frac{\ell}{2}y},\hspace{-0.5cm}
    \label{eq:Cosine_coefficientsI}
\end{aligned}
\end{equation}
defining the new function $H(x,y) = F(|x|,|y|)$ on the extended domain $(x,y)\in[-1,1]^2$;
the cosine series is
\begin{equation}
\begin{aligned}
    F(x,y) &= \sum_{k,\ell=0}^{\infty} \hat{F}(k,\ell) (2-\delta_{k,0}) (2-\delta_{\ell,0})\\[-0.5em]
    &\hspace{4em}\times\cos\left(\pi k x\right)\cos\left(\pi\ell y\right)\\[0.25em]
    &= \sum_{k,\ell\in\Z} \hat{H}(k,\ell)\,e^{i2\pi\frac{k}{2}x}\,e^{i2\pi\frac{\ell}{2}y}
    \label{eq:Cosine_seriesI}
\end{aligned}
\end{equation}
defining the Fourier coefficients $\hat{H}(k,\ell) = \hat{F}(|k|,|\ell|)$ for $k,\ell\in\Z$. Hence, the cosine series of the function $F(x,y)$ defined on $(x,y)\in[0,1]^2$ is equivalent to the Fourier series of the function $H(x,y)$ defined on $(x,y)\in[-1,1]^2$, which is just the function $F$ mirrored about the origin. The advantage here is that if $F(x,y)$ was continuous on $(x,y)\in[0,1]^2$, then the periodically extended function $H(x,y)$ is continuous for all $(x,y)\in\R^2$---in particular, there are no discontinuities at the boundary of the original domain. This should lead to a quicker decay of the Fourier coefficients $\hat{H}(k,\ell)$ compared to $\hat{F}(k,\ell)$.

Sampling $F(x,y)$ on the $2^n\times2^n$ grid $(x,y)=\big(\frac{a}{2^n-1},\frac{b}{2^n-1}\big)$, $a,b\in\{0,1,\ldots,2^n-1\}$, and relating the coefficients $\hat{F}(k,\ell)$ to the coefficients of the DCT-I of $f_{ab} = \big(\frac{a}{2^n-1},\frac{b}{2^n-1}\big)$ is equivalent to sampling the function $H(x,y)$ on the $2(2^n-1)\times2(2^n-1)$ grid $(x,y)=\big(\frac{a}{2^n-1},\frac{b}{2^n-1}\big)$, $a,b\in\{-2^n+1,-2^n+2,\ldots,2^n-2\}$, and relating the coefficients of the Fourier series to the DFT. This is the case considered in the main text, and using Eq.~\eqref{eq:relating_DFT_and_FT_coeff} in this context yields
\begin{equation}
    \hat{h}_{pq} = \sum_{i,j\in\Z} \hat{H}\big(p+i2(2^n-1),q+j2(2^n-1)\big)
\end{equation}
for $p,q\in\{-2^n+1,-2^n+2,\ldots,2^n-2\}$.

With the representation as a Fourier transform, we can use the same derivation for an MPS representation as in App.~\ref{app:2d_fourier_to_mps}. A difference arises in the calculation of the error bounds, as now the Fourier transform includes more frequencies, i.e, values of $p,q\in\{{-2^n+1,}\allowbreak -2^n+2, \ldots, 2^n-2\}$, than there are pixels, i.e., values of $a,b\in\{0,1,\ldots,2^n-1\}$. Still, we can proceed similarly as in the main text. For notational convenience, we define the three index sets $I_{\text{all}} = \{-2^n+1, -2^n+2, \ldots,\allowbreak {2^n-2}\}^2$, containing all possible frequency pairs; $I_{\text{appr}} = \{-\Lambda,-\Lambda+1,\ldots,\Lambda\}^2$, containing all frequency pairs retained in the approximation; and $I_{\text{disc}} = I_{\text{all}}\backslash I_{\text{appr}}$, containing all frequency pairs discarded in the approximation. Using this, we can express $f_{ab}$ as
\begin{equation}
\begin{aligned}
    f_{ab}
    &=\!\sum_{(p,q)\in I_{\text{all}}}\! \hat{h}_{pq}\,
    e^{i\frac{2\pi pa}{2(2^n-1)}}\,e^{i\frac{2\pi qb}{2(2^n-1)}}\\
    &= 2(2^n-1) (U \hat{h} U^{\transpose})_{ab},
\end{aligned}
\end{equation}
where in contrast to the main text $\hat{h}_{pq}$ is a $2(2^n-1)\times2(2^n-1)$ matrix and $U_{ap} = \frac{1}{\sqrt{2(2^n-1)}} e^{i\frac{2\pi pa}{2(2^n-1)}}$ is now a $2^n\times2(2^n-1)$ isometric matrix instead of a unitary one.%
\footnote{
    An isometric matrix $U$ is one for which $UU^{\dagger} = \Id$ but $U^{\dagger}U=P\neq\Id$, with $P = P^2$ a projector matrix.
}
As before, we obtain the approximation $g_{ab}$ to $f_{ab}$ by retaining only Fourier modes up to a cutoff $\Lambda$, which gives
\begin{equation}
\begin{aligned}
    g_{ab}
    &=\!\!\sum_{(p,q)\in I_{\text{appr}}}\!\! \hat{f}_{pq}\,
    e^{i\frac{2\pi pa}{2(2^n-1)}}\,e^{i\frac{2\pi qb}{2(2^n-1)}}\\
    &= 2(2^n-1) (U \hat{g} U^{\transpose})_{ab},
\end{aligned}
\end{equation}
where $\hat{g}$ is the matrix of the truncated Fourier coefficients, i.e., for $|p|\leq\Lambda$ and $|q|\leq\Lambda$ we have $\hat{g}_{pq} = \hat{h}_{pq}$ but if either $|p|>\Lambda$ or $|q|>\Lambda$ then $\hat{g}_{pq} = 0$. The expression for the approximation error in Eq.~\eqref{eq:approx_err_normalized} still holds, but now expressing it in terms of Fourier coefficients requires a bit of caution. We have
\begin{equation}
\begin{aligned}
    &\norm{f - g}_F =\\
    &\quad= 2(2^n-1)\|U(\hat{h} - \hat{g})U^{\transpose}\|_F\\
    &\quad= 2(2^n-1)\|(U \otimes U)\,\text{vec}(\hat{h} - \hat{g})\|_2\\
    &\quad= 2(2^n-1) \big(\text{vec}(\hat{h} - \hat{g})^{\dagger} (U^{\dagger} \otimes U^{\dagger})\\
    &\quad\hspace{6em}\cdot(U \otimes U)\,\text{vec}(\hat{h} - \hat{g})\big)^{\!\frac{1}{2}}\\
    &\quad= 2(2^n-1) \big(\text{vec}(\hat{h} - \hat{g})^{\dagger} (P \otimes P)\,\text{vec}(\hat{h} - \hat{g})\big)^{\!\frac{1}{2}}%
    \hspace{-1em}\\
    &\quad\leq 2(2^n-1) \big(\text{vec}(\hat{h} - \hat{g})^{\dagger} (\Id \otimes \Id)\,\text{vec}(\hat{h} - \hat{g})\big)^{\!\frac{1}{2}}%
    \hspace{-1em}\\
    &\quad= 2(2^n-1)\|\hat{h} - \hat{g}\|_F,
\end{aligned}
\end{equation}
where $\text{vec}(\hat{h} - \hat{g})$ denotes the vectorized version of the matrix $\hat{h} - \hat{g}$, and
\begin{equation}
\begin{aligned}
    \norm{g}_F
    &= 2(2^n-1)\|U\hat{g}U^{\transpose}\|_F\\
    &= 2(2^n-1) \sqrt{\Tr(U^*\hat{g}^{\dagger}U^{\dagger}U\hat{g}U^{\transpose})}\\
    &= 2(2^n-1) \sqrt{\Tr(P^*\hat{g}^{\dagger}P\hat{g})}.
\end{aligned}
\end{equation}
Thus, Eq.~\eqref{eq:approx_err_normalized_fourier} becomes
\begin{equation}
\begin{aligned}
    &\norm{\ket{f} - \ket{g}}_2 \leq \frac{2\norm{f-g}_F}{\norm{g}_F}
    \leq \frac{2\|\hat{h}-\hat{g}\|_F}{\sqrt{\Tr(P^*\hat{g}^{\dagger}P\hat{g})}}.
\end{aligned}
\end{equation}
For $\|\hat{h}-\hat{g}\|_F = \sqrt{\sum_{(p,q)\in I_{\text{disc}}} |\hat{h}_{pq}|^2}$ we can reuse the results from App.~\ref{app:proof_appr_err_bounds}: if the coefficients of the cosine series $\hat{F}(k,\ell)$ decay exponentially or algebraically, then also the Fourier coefficients $\hat{H}(k,\ell)$ decay this way, however, we must substitute ${2^n\to2(2^n-1)}$ as there are now more Fourier coefficients. Still, this leads to an approximation error independent of the resolution of the image. The factor $1/\sqrt{\Tr(P^*\hat{g}^{\dagger}P\hat{g})}$ generically does not introduce an asymptotic scaling. Since $\hat{g}$ is a $2(2^n-1)\times2(2^n-1)$ matrix where only the central $(2\Lambda+1)\times(2\Lambda+1)$ square is nonzero, the increasing matrix dimension of $P=U^{\dagger}U$ has no effect on the result of $\Tr(P^*\hat{g}^{\dagger}P\hat{g})$. The nonzero values of $\hat{g}$ are the values of $\hat{H}(k,\ell)$ up the cutoff $\Lambda$, plus corrections which are exponentially or algebraically small in the number of pixels, depending on the decay of $|\hat{H}(k,\ell)|$---see also Eqs.~\eqref{eq:app_bound_fpq_exp} and~\eqref{eq:app_bound_fpq_alg}. So as long as not all coefficients of the Fourier series within the cutoff are zero, $\Tr(P^*\hat{g}^{\dagger}P\hat{g})$ will not become zero as the number of pixels is increased. Its inverse can thus not diverge, and hence not change the dominant term in the asymptotic scaling.

\subsection{Type-II DCT}
Now we consider the type-II discrete cosine transform (DCT-II), which is more common than the DCT-I and for example used in the JPEG algorithm for image compression. The forward DCT-II of a matrix $f_{ab}$, $a,b\in\{0,1,\ldots,2^n-1\}$ produces coefficients $\hat{f}_{pq}$, $p,q\in\{0,1,\ldots,2^n-1\}$, which are defined as
\begin{equation}
\begin{aligned}
    \hat{f}_{pq} &= \frac{1}{(2^n)^2}\!\sum_{a,b=0}^{2^n-1}\!f_{ab}\\
    &\hspace{3.5em}\times\cos(\frac{\pi p}{2^n}\!\left(a+{\textstyle\frac{1}{2}}\right)\!)
    \cos(\frac{\pi q}{2^n}\!\left(b+{\textstyle\frac{1}{2}}\right)\!)\hspace{-1em}\\
    &= \frac{1}{(2^{n+1})^2} \hspace{-0.4em}\sum_{a,b=-2^n}^{2^n-1}\hspace{-0.5em}
    h_{ab}\,e^{-i\frac{2\pi p}{2^{n+1}}\left(a+\frac{1}{2}\right)}
    e^{-i\frac{2\pi q}{2^{n+1}}\left(b+\frac{1}{2}\right)}\hspace{-2em}
    \label{eq:DCT-II}
\end{aligned}
\end{equation}
defining a new matrix of larger dimension $h_{ab} = f_{|a+\frac{1}{2}|-\frac{1}{2},|b+\frac{1}{2}|-\frac{1}{2}}$ for $a,b\in\{-2^n,-2^n+1,\ldots,2^n-1\}$.
The inverse transform is defined as
\begin{equation}
\begin{aligned}
    f_{ab} &= \sum_{p,q=0}^{2^n-1} \hat{f}_{pq}
    (2-\delta_{p,0}) (2-\delta_{q,0})\\[-0.6em]
    &\hspace{3.5em}\times\cos(\frac{\pi p}{2^n}\!\left(a+{\textstyle\frac{1}{2}}\right)\!)
    \cos(\frac{\pi q}{2^n}\!\left(b+{\textstyle\frac{1}{2}}\right)\!)\hspace{-1em}\\
    &= \!\!\sum_{p,q=-2^n}^{2^n-1}\!\!
    \hat{h}_{pq}\,e^{i\frac{2\pi p}{2^{n+1}}\left(a+\frac{1}{2}\right)}
    e^{i\frac{2\pi q}{2^{n+1}}\left(b+\frac{1}{2}\right)},
    \label{eq:iDCT-II}
\end{aligned}
\end{equation}
defining a new $2^n\times2^n$ matrix of Fourier coefficients $\hat{h}_{pq}$ with $\hat{h}_{pq} = \hat{f}_{|p|,|q|}$ for $p,q\in\{-2^n+1, \allowbreak -2^n+2, \ldots, 2^n-1\}$ and $\hat{h}_{pq}=0$ if either $p=-2^n$ or $q=-2^n$.%
\footnote{
    The condition $\hat{g}_{pq} = 0$ for $p=-2^n$ or $q=-2^n$ follows from the symmetry $g_{ab} = g_{-a-1,b} = g_{a,-b-1}$ of the inputs $g_{ab}$. This can be seen by explicitly calculating $\hat{g}_{-2^n,q}$ or $\hat{g}_{p,-2^n}$.
}
The DCT-II of the $2^n\times2^n$ matrix $f_{ab}$ is thus equivalent to the quarter-wave DFT of the $2^{n+1}\times2^{n+1}$ matrix $h_{ab}$, where each $2^n\times2^n$ quadrant of $h_{ab}$ is a copy of $f_{ab}$, mirrored such that the matrix is point-symmetric about its center, i.e., such that $h_{ab} = h_{-a-1,b} = h_{a,-b-1}$ for $a,b\in\{-2^n,\allowbreak -2^n+1, \ldots, 2^n-1\}$.

The coefficients of the cosine series of $F(x,y)$, $(x,y)\in[0,1]^2$, are given by
\begin{equation}
\begin{aligned}
    \hspace{-0.25cm}\hat{F}(k,\ell) &= \int_{0}^{1}\d x \int_{0}^{1}\d y\;
    \frac{F(x,y)\cos\left(\pi k x\right)\cos\left(\pi\ell y\right)}
    {(1+\delta_{k,0})(1+\delta_{\ell,0})}\hspace{-0.5cm}\\
    &= \frac{1}{2^2}  \int_{-1}^{1}\d x \int_{-1}^{1}\d y\;
    H(x,y)\,e^{-i2\pi\frac{k}{2}x}\,e^{-i2\pi\frac{\ell}{2}y},\hspace{-0.5cm}
    \label{eq:Cosine_coefficientsII}
\end{aligned}
\end{equation}
defining the new function $H(x,y) = F(|x|,|y|)$ on the extended domain $(x,y)\in[-1,1]^2$;
the cosine series is
\begin{equation}
\begin{aligned}
    F(x,y) &= \sum_{k,\ell=0}^{\infty} \hat{F}(k,\ell) (2-\delta_{k,0}) (2-\delta_{\ell,0})\\[-0.5em]
    &\hspace{4em}\times\cos\left(\pi k x\right)\cos\left(\pi\ell y\right)\\[0.25em]
    &= \sum_{k,\ell\in\Z} \hat{H}(k,\ell)\,e^{i2\pi\frac{k}{2}x}\,e^{i2\pi\frac{\ell}{2}y}
    \label{eq:Cosine_seriesII}
\end{aligned}
\end{equation}
defining the Fourier coefficients $\hat{H}(k,\ell) = \hat{F}(|k|,|\ell|)$ for $k,\ell\in\Z$. This yields the same function $H(x,y)$ encountered for the DCT-I, however, instead of sampling $F(x,y)$ on the $2^n\times2^n$ grid with coordinates $(x,y)=\big(\frac{a}{2^n-1},\frac{b}{2^n-1}\big)$, $a,b\in\{0,1,\ldots,2^n-1\}$, we sample it on the $2^n\times2^n$ grid with coordinates $(x,y)=\big(\frac{a+1/2}{2^n},\frac{b+1/2}{2^n}\big)$, $a,b\in\{0,1,\ldots,2^n-1\}$. Relating the coefficients $\hat{F}(k,\ell)$ of the cosine series to the coefficients of the DCT-II is equivalent to sampling the function $H(x,y)$ on the $2^{n+1}\times2^{n+1}$ grid $(x,y)=\big(\frac{a+1/2}{2^n},\frac{b+1/2}{2^n}\big)$ with $a,b\in\{-2^n,\allowbreak-2^n+1,\ldots,2^n-1\}$ and relating the coefficients of the Fourier series to the quarter-wave DFT. Using $e^{i2\pi(k+j2^{n+1})(a+\frac{1}{2})/2^{n+1}} = (-1)^j e^{i2\pi k(a+\frac{1}{2})/2^{n+1}}$ yields a relation similar to Eq.~\eqref{eq:relating_DFT_and_FT_coeff}:
\begin{equation}
    \hat{h}_{pq} = \sum_{i,j\in\Z} (-1)^{i+j}~\hat{H}\big(p+i2N,q+j2N\big)
    \label{eq:app_relating_qwDFT_and_qwFT_coeff}
\end{equation}
for $p,q\in\{-N,-N+1,\ldots,N-1\}$.

Similarly to the case for the DCT-I, we can now reuse the results from the main text under certain substitutions. The MPS construction follows from the representation of the DCT-II as a quarter-wave DFT, where the extra phase due to the shift of the indices $a$ and $b$ by $1/2$ can be absorbed in the DFT coefficients. The bounds on the approximation error when $\hat{F}(k,\ell)$ decays exponentially or algebraically can be reused if one substitutes $2^n\to2^{n+1}$, as then also $\hat{H}(k,\ell)$ decays in the same way, however there are now twice as many Fourier coefficients. The additional phases in Eq.~\eqref{eq:app_relating_qwDFT_and_qwFT_coeff} compared to Eq.~\eqref{eq:relating_DFT_and_FT_coeff} do not play a role, as only the absolute value is relevant.

\bibliographystyle{unsrtnat}
\bibliography{references}
\end{document}